
\input psfig
%
%
%
%

\catcode `\@=11 

\def\@version{1.4}
\def\@verdate{22nd Feb 1994}

%
%
%
%


\newif\ifprod@font

\ifx\@typeface\undefined
  \def\@typeface{Comp. Modern}\prod@fontfalse
\else
  \prod@fonttrue 
\fi

\def\newfam{\alloc@8\fam\chardef\sixt@@n} 

\ifprod@font
\font\fiverm=mtr10 at 5pt
\font\fivebf=mtbx10 at 5pt
\font\fiveit=mtti10 at 5pt
\font\fivesl=mtsl10 at 5pt
\font\fivett=mttt10 at 5pt     \hyphenchar\fivett=-1
\font\fivecsc=mtcsc10 at 5pt
\font\fivesf=mtss10 at 5pt
\font\fivei=mtmi10 at 5pt      \skewchar\fivei='177
\font\fivemib=mtmib10 at 5pt   \skewchar\fivemib='177
\font\fivesy=mtsy10 at 5pt     \skewchar\fivesy='60
\font\fivesyb=mtbsy10 at 5pt   \skewchar\fivesyb='60

\font\sixrm=mtr10 at 6pt
\font\sixbf=mtbx10 at 6pt
\font\sixit=mtti10 at 6pt
\font\sixsl=mtsl10 at 6pt
\font\sixtt=mttt10 at 6pt      \hyphenchar\sixtt=-1
\font\sixcsc=mtcsc10 at 6pt
\font\sixsf=mtss10 at 6pt
\font\sixi=mtmi10 at 6pt       \skewchar\sixi='177
\font\sixmib=mtmib10 at 6pt    \skewchar\sixmib='177
\font\sixsy=mtsy10 at 6pt      \skewchar\sixsy='60
\font\sixsyb=mtbsy10 at 6pt    \skewchar\sixsyb='60

\font\sevenrm=mtr10 at 7pt
\font\sevenbf=mtbx10 at 7pt
\font\sevenit=mtti10 at 7pt
\font\sevensl=mtsl10 at 7pt
\font\seventt=mttt10 at 7pt     \hyphenchar\seventt=-1
\font\sevencsc=mtcsc10 at 7pt
\font\sevensf=mtss10 at 7pt
\font\seveni=mtmi10 at 7pt      \skewchar\seveni='177
\font\sevenmib=mtmib10 at 7pt   \skewchar\sevenmib='177
\font\sevensy=mtsy10 at 7pt     \skewchar\sevensy='60
\font\sevensyb=mtbsy10 at 7pt   \skewchar\sevensyb='60

\font\eightrm=mtr10 at 8pt
\font\eightbf=mtbx10 at 8pt
\font\eightit=mtti10 at 8pt
\font\eighti=mtmi10 at 8pt      \skewchar\eighti='177
\font\eightmib=mtmib10 at 8pt   \skewchar\eightmib='177
\font\eightsy=mtsy10 at 8pt     \skewchar\eightsy='60
\font\eightsyb=mtbsy10 at 8pt   \skewchar\eightsyb='60
\font\eightsl=mtsl10 at 8pt
\font\eighttt=mttt10 at 8pt     \hyphenchar\eighttt=-1
\font\eightcsc=mtcsc10 at 8pt
\font\eightsf=mtss10 at 8pt

\font\ninerm=mtr10 at 9pt
\font\ninebf=mtbx10 at 9pt
\font\nineit=mtti10 at 9pt
\font\ninei=mtmi10 at 9pt      \skewchar\ninei='177
\font\ninemib=mtmib10 at 9pt   \skewchar\ninemib='177
\font\ninesy=mtsy10 at 9pt     \skewchar\ninesy='60
\font\ninesyb=mtbsy10 at 9pt   \skewchar\ninesyb='60
\font\ninesl=mtsl10 at 9pt
\font\ninett=mttt10 at 9pt     \hyphenchar\ninett=-1
\font\ninecsc=mtcsc10 at 9pt
\font\ninesf=mtss10 at 9pt

\font\tenrm=mtr10
\font\tenbf=mtbx10
\font\tenit=mtti10
\font\teni=mtmi10		\skewchar\teni='177
\font\tenmib=mtmib10	\skewchar\tenmib='177
\font\tensy=mtsy10		\skewchar\tensy='60
\font\tensyb=mtbsy10	\skewchar\tensyb='60
\font\tenex=cmex10
\font\tensl=mtsl10
\font\tentt=mttt10		\hyphenchar\tentt=-1
\font\tencsc=mtcsc10
\font\tensf=mtss10

\font\elevenrm=mtr10 at 11pt
\font\elevenbf=mtbx10 at 11pt
\font\elevenit=mtti10 at 11pt
\font\eleveni=mtmi10 at 11pt      \skewchar\eleveni='177
\font\elevenmib=mtmib10 at 11pt   \skewchar\elevenmib='177
\font\elevensy=mtsy10 at 11pt     \skewchar\elevensy='60
\font\elevensyb=mtbsy10 at 11pt   \skewchar\elevensyb='60
\font\elevensl=mtsl10 at 11pt
\font\eleventt=mttt10 at 11pt     \hyphenchar\eleventt=-1
\font\elevencsc=mtcsc10 at 11pt
\font\elevensf=mtss10 at 11pt

\font\twelverm=mtr10 at 12pt
\font\twelvebf=mtbx10 at 12pt
\font\twelveit=mtti10 at 12pt
\font\twelvesl=mtsl10 at 12pt
\font\twelvett=mttt10 at 12pt     \hyphenchar\twelvett=-1
\font\twelvecsc=mtcsc10 at 12pt
\font\twelvesf=mtss10 at 12pt
\font\twelvei=mtmi10 at 12pt      \skewchar\twelvei='177
\font\twelvemib=mtmib10 at 12pt   \skewchar\twelvemib='177
\font\twelvesy=mtsy10 at 12pt     \skewchar\twelvesy='60
\font\twelvesyb=mtbsy10 at 12pt   \skewchar\twelvesyb='60

\font\fourteenrm=mtr10 at 14pt
\font\fourteenbf=mtbx10 at 14pt
\font\fourteenit=mtti10 at 14pt
\font\fourteeni=mtmi10 at 14pt      \skewchar\fourteeni='177
\font\fourteenmib=mtmib10 at 14pt   \skewchar\fourteenmib='177
\font\fourteensy=mtsy10 at 14pt     \skewchar\fourteensy='60
\font\fourteensyb=mtbsy10 at 14pt   \skewchar\fourteensyb='60
\font\fourteensl=mtsl10 at 14pt
\font\fourteentt=mttt10 at 14pt     \hyphenchar\fourteentt=-1
\font\fourteencsc=mtcsc10 at 14pt
\font\fourteensf=mtss10 at 14pt

\font\seventeenrm=mtr10 at 17pt
\font\seventeenbf=mtbx10 at 17pt
\font\seventeenit=mtti10 at 17pt
\font\seventeeni=mtmi10 at 17pt      \skewchar\seventeeni='177
\font\seventeenmib=mtmib10 at 17pt   \skewchar\seventeenmib='177
\font\seventeensy=mtsy10 at 17pt     \skewchar\seventeensy='60
\font\seventeensyb=mtbsy10 at 17pt   \skewchar\seventeensyb='60
\font\seventeensl=mtsl10 at 17pt
\font\seventeentt=mttt10 at 17pt     \hyphenchar\seventeentt=-1
\font\seventeencsc=mtcsc10 at 17pt
\font\seventeensf=mtss10 at 17pt


\newfam\xmfam
\newfam\ymfam

\font\fivexm=mtxm10 at 5pt
\font\sixxm=mtxm10 at 6pt
\font\sevenxm=mtxm10 at 7pt
\font\eightxm=mtxm10 at 8pt
\font\ninexm=mtxm10 at 9pt
\font\tenxm=mtxm10
\font\elevenxm=mtxm10 at 11pt
\font\twelvexm=mtxm10 at 12pt
\font\fourteenxm=mtxm10 at 14pt
\font\seventeenxm=mtxm10 at 17pt

\font\fiveym=mtym10 at 5pt
\font\sixym=mtym10 at 6pt
\font\sevenym=mtym10 at 7pt
\font\eightym=mtym10 at 8pt
\font\nineym=mtym10 at 9pt
\font\tenym=mtym10
\font\elevenym=mtym10 at 11pt
\font\twelveym=mtym10 at 12pt
\font\fourteenym=mtym10 at 14pt
\font\seventeenym=mtym10 at 17pt
\else
\font\fiverm=cmr5
\font\fivei=cmmi5             \skewchar\fivei='177
\font\fivemib=cmmib10 at 5pt  \skewchar\fivemib='177
\font\fivesy=cmsy5            \skewchar\fivesy='60
\font\fivesyb=cmbsy10 at 5pt  \skewchar\fivesyb='60
\font\fivebf=cmbx5

\font\sixrm=cmr6
\font\sixi=cmmi6             \skewchar\sixi='177
\font\sixmib=cmmib10 at 6pt  \skewchar\sixmib='177
\font\sixsy=cmsy6            \skewchar\sixsy='60
\font\sixsyb=cmbsy10 at 6pt  \skewchar\sixsyb='60
\font\sixbf=cmbx6

\font\sevenrm=cmr7
\font\seveni=cmmi7             \skewchar\seveni='177
\font\sevenmib=cmmib10 at 7pt  \skewchar\sevenmib='177
\font\sevensy=cmsy7            \skewchar\sevensy='60
\font\sevensyb=cmbsy10 at 7pt  \skewchar\sevensyb='60
\font\sevenbf=cmbx7

\font\eightrm=cmr8
\font\eightbf=cmbx8
\font\eightit=cmti8
\font\eighti=cmmi8			\skewchar\eighti='177
\font\eightmib=cmmib10 at 8pt	\skewchar\eightmib='177
\font\eightsy=cmsy8			\skewchar\eightsy='60
\font\eightsyb=cmbsy10 at 8pt	\skewchar\eightsyb='60
\font\eightsl=cmsl8
\font\eighttt=cmtt8			\hyphenchar\eighttt=-1
\font\eightcsc=cmcsc10 at 8pt
\font\eightsf=cmss8

\font\ninerm=cmr9
\font\ninebf=cmbx9
\font\nineit=cmti9
\font\ninei=cmmi9			\skewchar\ninei='177
\font\ninemib=cmmib10 at 9pt	\skewchar\ninemib='177
\font\ninesy=cmsy9			\skewchar\ninesy='60
\font\ninesyb=cmbsy10 at 9pt	\skewchar\ninesyb='60
\font\ninesl=cmsl9
\font\ninett=cmtt9			\hyphenchar\ninett=-1
\font\ninecsc=cmcsc10 at 9pt
\font\ninesf=cmss9

\font\tenrm=cmr10
\font\tenbf=cmbx10
\font\tenit=cmti10
\font\teni=cmmi10		\skewchar\teni='177
\font\tenmib=cmmib10	\skewchar\tenmib='177
\font\tensy=cmsy10		\skewchar\tensy='60
\font\tensyb=cmbsy10	\skewchar\tensyb='60
\font\tenex=cmex10
\font\tensl=cmsl10
\font\tentt=cmtt10		\hyphenchar\tentt=-1
\font\tencsc=cmcsc10
\font\tensf=cmss10

\font\elevenrm=cmr10 scaled \magstephalf
\font\elevenbf=cmbx10 scaled \magstephalf
\font\elevenit=cmti10 scaled \magstephalf
\font\eleveni=cmmi10 scaled \magstephalf	\skewchar\eleveni='177
\font\elevenmib=cmmib10 scaled \magstephalf	\skewchar\elevenmib='177
\font\elevensy=cmsy10 scaled \magstephalf	\skewchar\elevensy='60
\font\elevensyb=cmbsy10 scaled \magstephalf	\skewchar\elevensyb='60
\font\elevensl=cmsl10 scaled \magstephalf
\font\eleventt=cmtt10 scaled \magstephalf	\hyphenchar\eleventt=-1
\font\elevencsc=cmcsc10 scaled \magstephalf
\font\elevensf=cmss10 scaled \magstephalf

\font\twelverm=cmr10 scaled \magstep1
\font\twelvebf=cmbx10 scaled \magstep1
\font\twelvei=cmmi10 scaled \magstep1      \skewchar\twelvei='177
\font\twelvemib=cmmib10 scaled \magstep1   \skewchar\twelvemib='177
\font\twelvesy=cmsy10 scaled \magstep1     \skewchar\twelvesy='60
\font\twelvesyb=cmbsy10 scaled \magstep1   \skewchar\twelvesyb='60

\font\fourteenrm=cmr10 scaled \magstep2
\font\fourteenbf=cmbx10 scaled \magstep2
\font\fourteenit=cmti10 scaled \magstep2
\font\fourteeni=cmmi10 scaled \magstep2		\skewchar\fourteeni='177
\font\fourteenmib=cmmib10 scaled \magstep2	\skewchar\fourteenmib='177
\font\fourteensy=cmsy10 scaled \magstep2	\skewchar\fourteensy='60
\font\fourteensyb=cmbsy10 scaled \magstep2	\skewchar\fourteensyb='60
\font\fourteensl=cmsl10 scaled \magstep2
\font\fourteentt=cmtt10 scaled \magstep2	\hyphenchar\fourteentt=-1
\font\fourteencsc=cmcsc10 scaled \magstep2
\font\fourteensf=cmss10 scaled \magstep2

\font\seventeenrm=cmr10 scaled \magstep3
\font\seventeenbf=cmbx10 scaled \magstep3
\font\seventeenit=cmti10 scaled \magstep3
\font\seventeeni=cmmi10 scaled \magstep3	\skewchar\seventeeni='177
\font\seventeenmib=cmmib10 scaled \magstep3	\skewchar\seventeenmib='177
\font\seventeensy=cmsy10 scaled \magstep3	\skewchar\seventeensy='60
\font\seventeensyb=cmbsy10 scaled \magstep3	\skewchar\seventeensyb='60
\font\seventeensl=cmsl10 scaled \magstep3
\font\seventeentt=cmtt10 scaled \magstep3	\hyphenchar\seventeentt=-1
\font\seventeencsc=cmcsc10 scaled \magstep3
\font\seventeensf=cmss10 scaled \magstep3
\fi

\def\hexnumber#1{\ifcase#1 0\or1\or2\or3\or4\or5\or6\or7\or8\or9\or
  A\or B\or C\or D\or E\or F\fi}

\ifprod@font
  \edef\@xm{\hexnumber\xmfam}
  \edef\@ym{\hexnumber\ymfam}
\fi

\def\makestrut{%
  \setbox\strutbox=\hbox{%
    \vrule height.7\baselineskip depth.3\baselineskip width \z@}%
}

\def\baselinestretch{1}
\newskip\tmp@bls

\def\b@ls#1{
  \tmp@bls=#1\relax
  \baselineskip=#1\relax\makestrut
  \normalbaselineskip=\baselinestretch\tmp@bls
  \normalbaselines
}

\def\nostb@ls#1{
  \normalbaselineskip=#1\relax
  \normalbaselines
  \makestrut
}

%

\newfam\mibfam 
\newfam\sybfam 
\newfam\scfam  
\newfam\sffam  

\def\mit{\fam\@ne}

\def\cal{\fam\tw@}

\def\em{\ifdim\fontdimen1\font>\z@ \rm\else\it\fi}

\textfont3=\tenex
\scriptfont3=\tenex
\scriptscriptfont3=\tenex

\setbox0=\hbox{\tenex B} \p@renwd=\wd0 

\def\eightpoint{
  \def\rm{\fam0\eightrm}%
  \textfont0=\eightrm \scriptfont0=\sixrm \scriptscriptfont0=\fiverm%
  \textfont1=\eighti  \scriptfont1=\sixi  \scriptscriptfont1=\fivei%
  \textfont2=\eightsy \scriptfont2=\sixsy \scriptscriptfont2=\fivesy%
  \textfont\itfam=\eightit\def\it{\fam\itfam\eightit}%
  \ifprod@font
    \scriptfont\itfam=\sixit
      \scriptscriptfont\itfam=\fiveit
  \else
    \scriptfont\itfam=\eightit
      \scriptscriptfont\itfam=\eightit
  \fi
  \textfont\bffam=\eightbf%
    \scriptfont\bffam=\sixbf%
      \scriptscriptfont\bffam=\fivebf%
  \def\bf{\fam\bffam\eightbf}%
  \textfont\slfam=\eightsl\def\sl{\fam\slfam\eightsl}%
  \ifprod@font
    \scriptfont\slfam=\sixsl
      \scriptscriptfont\slfam=\fivesl
  \else
    \scriptfont\slfam=\eightsl
      \scriptscriptfont\slfam=\eightsl
  \fi
  \textfont\ttfam=\eighttt\def\tt{\fam\ttfam\eighttt}%
  \ifprod@font
    \scriptfont\ttfam=\sixtt
      \scriptscriptfont\ttfam=\fivett
  \else
    \scriptfont\ttfam=\eighttt
      \scriptscriptfont\ttfam=\eighttt
  \fi
  \textfont\scfam=\eightcsc\def\sc{\fam\scfam\eightcsc}%
  \ifprod@font
    \scriptfont\scfam=\sixcsc
      \scriptscriptfont\scfam=\fivecsc
  \else
    \scriptfont\scfam=\eightcsc
      \scriptscriptfont\scfam=\eightcsc
  \fi
  \textfont\sffam=\eightsf\def\sf{\fam\sffam\eightsf}%
  \ifprod@font
    \scriptfont\sffam=\sixsf
      \scriptscriptfont\sffam=\fivesf
  \else
    \scriptfont\sffam=\eightsf
      \scriptscriptfont\sffam=\eightsf
  \fi
  \textfont\mibfam=\eightmib
    \scriptfont\mibfam=\sixmib
      \scriptscriptfont\mibfam=\fivemib
  \textfont\sybfam=\eightsyb
    \scriptfont\sybfam=\sixsyb
      \scriptscriptfont\sybfam=\fivesyb
  \ifprod@font
    \textfont\xmfam=\eightxm
      \scriptfont\xmfam=\sixxm
        \scriptscriptfont\xmfam=\fivexm
    \textfont\ymfam=\eightym
      \scriptfont\ymfam=\sixym
        \scriptscriptfont\ymfam=\fiveym
  \fi
  \def\oldstyle{\fam\@ne\eighti}%
  \def\boldstyle{\fam\mibfam\eightmib}%
  \b@ls{10pt}\rm%
}

\def\ninepoint{
  \def\rm{\fam0\ninerm}%
  \textfont0=\ninerm \scriptfont0=\sixrm \scriptscriptfont0=\fiverm%
  \textfont1=\ninei  \scriptfont1=\sixi  \scriptscriptfont1=\fivei%
  \textfont2=\ninesy \scriptfont2=\sixsy \scriptscriptfont2=\fivesy%
  \textfont\itfam=\nineit\def\it{\fam\itfam\nineit}%
  \ifprod@font
    \scriptfont\itfam=\sixit
      \scriptscriptfont\itfam=\fiveit
  \else
    \scriptfont\itfam=\nineit
      \scriptscriptfont\itfam=\nineit
  \fi
  \textfont\bffam=\ninebf%
    \scriptfont\bffam=\sixbf%
      \scriptscriptfont\bffam=\fivebf%
  \def\bf{\fam\bffam\ninebf}%
  \textfont\slfam=\ninesl\def\sl{\fam\slfam\ninesl}%
  \ifprod@font
    \scriptfont\slfam=\sixsl
      \scriptscriptfont\slfam=\fivesl
  \else
    \scriptfont\slfam=\ninesl
      \scriptscriptfont\slfam=\ninesl
  \fi
  \textfont\ttfam=\ninett\def\tt{\fam\ttfam\ninett}%
  \ifprod@font
    \scriptfont\ttfam=\sixtt
      \scriptscriptfont\ttfam=\fivett
  \else
    \scriptfont\ttfam=\ninett
      \scriptscriptfont\ttfam=\ninett
  \fi
  \textfont\scfam=\ninecsc\def\sc{\fam\scfam\ninecsc}%
  \ifprod@font
    \scriptfont\scfam=\sixcsc
      \scriptscriptfont\scfam=\fivecsc
  \else
    \scriptfont\scfam=\ninecsc
      \scriptscriptfont\scfam=\ninecsc
  \fi
  \textfont\sffam=\ninesf\def\sf{\fam\sffam\ninesf}%
  \ifprod@font
    \scriptfont\sffam=\sixsf
      \scriptscriptfont\sffam=\fivesf
  \else
    \scriptfont\sffam=\ninesf
      \scriptscriptfont\sffam=\ninesf
  \fi
  \textfont\mibfam=\ninemib
    \scriptfont\mibfam=\sixmib
      \scriptscriptfont\mibfam=\fivemib
  \textfont\sybfam=\ninesyb
    \scriptfont\sybfam=\sixsyb
      \scriptscriptfont\sybfam=\fivesyb
  \ifprod@font
    \textfont\xmfam=\ninexm
      \scriptfont\xmfam=\sixxm
        \scriptscriptfont\xmfam=\fivexm
    \textfont\ymfam=\nineym
      \scriptfont\ymfam=\sixym
        \scriptscriptfont\ymfam=\fiveym
  \fi
  \def\oldstyle{\fam\@ne\ninei}%
  \def\boldstyle{\fam\mibfam\ninemib}%
  \b@ls{\TextLeading plus \Feathering}\rm%
}

\def\tenpoint{
  \def\rm{\fam0\tenrm}%
  \textfont0=\tenrm \scriptfont0=\sevenrm \scriptscriptfont0=\fiverm%
  \textfont1=\teni  \scriptfont1=\seveni  \scriptscriptfont1=\fivei%
  \textfont2=\tensy \scriptfont2=\sevensy \scriptscriptfont2=\fivesy%
  \textfont\itfam=\tenit\def\it{\fam\itfam\tenit}%
  \ifprod@font
    \scriptfont\itfam=\sevenit
      \scriptscriptfont\itfam=\fiveit
  \else
    \scriptfont\itfam=\tenit
      \scriptscriptfont\itfam=\tenit
  \fi
  \textfont\bffam=\tenbf%
    \scriptfont\bffam=\sevenbf%
      \scriptscriptfont\bffam=\fivebf%
  \def\bf{\fam\bffam\tenbf}%
  \textfont\slfam=\tensl\def\sl{\fam\slfam\tensl}%
  \ifprod@font
    \scriptfont\slfam=\sevensl
      \scriptscriptfont\slfam=\fivesl
  \else
    \scriptfont\slfam=\tensl
      \scriptscriptfont\slfam=\tensl
  \fi
  \textfont\ttfam=\tentt\def\tt{\fam\ttfam\tentt}%
  \ifprod@font
    \scriptfont\ttfam=\seventt
      \scriptscriptfont\ttfam=\fivett
  \else
    \scriptfont\ttfam=\tentt
      \scriptscriptfont\ttfam=\tentt
  \fi
  \textfont\scfam=\tencsc\def\sc{\fam\scfam\tencsc}%
  \ifprod@font
    \scriptfont\scfam=\sevencsc
      \scriptscriptfont\scfam=\fivecsc
  \else
    \scriptfont\scfam=\tencsc
      \scriptscriptfont\scfam=\tencsc
  \fi
  \textfont\sffam=\tensf\def\sf{\fam\sffam\tensf}%
  \ifprod@font
    \scriptfont\sffam=\sevensf
      \scriptscriptfont\sffam=\fivesf
  \else
    \scriptfont\sffam=\tensf
      \scriptscriptfont\sffam=\tensf
  \fi
  \textfont\mibfam=\tenmib
    \scriptfont\mibfam=\sevenmib
      \scriptscriptfont\mibfam=\fivemib
  \textfont\sybfam=\tensyb
    \scriptfont\sybfam=\sevensyb
      \scriptscriptfont\sybfam=\fivesyb
  \ifprod@font
    \textfont\xmfam=\tenxm
      \scriptfont\xmfam=\sevenxm
        \scriptscriptfont\xmfam=\fivexm
    \textfont\ymfam=\tenym
      \scriptfont\ymfam=\sevenym
        \scriptscriptfont\ymfam=\fiveym
  \fi
  \def\oldstyle{\fam\@ne\teni}%
  \def\boldstyle{\fam\mibfam\tenmib}%
  \b@ls{11pt}\rm%
}

\def\elevenpoint{
  \def\rm{\fam0\elevenrm}%
  \textfont0=\elevenrm \scriptfont0=\eightrm \scriptscriptfont0=\sixrm%
  \textfont1=\eleveni  \scriptfont1=\eighti  \scriptscriptfont1=\sixi%
  \textfont2=\elevensy \scriptfont2=\eightsy \scriptscriptfont2=\sixsy%
  \textfont\itfam=\elevenit\def\it{\fam\itfam\elevenit}%
  \ifprod@font
    \scriptfont\itfam=\eightit
      \scriptscriptfont\itfam=\sixit
  \else
    \scriptfont\itfam=\elevenit
      \scriptscriptfont\itfam=\elevenit
  \fi
  \textfont\bffam=\elevenbf%
    \scriptfont\bffam=\eightbf%
      \scriptscriptfont\bffam=\sixbf%
  \def\bf{\fam\bffam\elevenbf}%
  \textfont\slfam=\elevensl\def\sl{\fam\slfam\elevensl}%
  \ifprod@font
    \scriptfont\slfam=\eightsl
      \scriptscriptfont\slfam=\sixsl
  \else
    \scriptfont\slfam=\elevensl
      \scriptscriptfont\slfam=\elevensl
  \fi
  \textfont\ttfam=\eleventt\def\tt{\fam\ttfam\eleventt}%
  \ifprod@font
    \scriptfont\ttfam=\eighttt
      \scriptscriptfont\ttfam=\sixtt
  \else
    \scriptfont\ttfam=\eleventt
      \scriptscriptfont\ttfam=\eleventt
  \fi
  \textfont\scfam=\elevencsc\def\sc{\fam\scfam\elevencsc}%
  \ifprod@font
    \scriptfont\scfam=\eightcsc
      \scriptscriptfont\scfam=\sixcsc
  \else
    \scriptfont\scfam=\elevencsc
      \scriptscriptfont\scfam=\elevencsc
  \fi
  \textfont\sffam=\elevensf\def\sf{\fam\sffam\elevensf}%
  \ifprod@font
    \scriptfont\sffam=\eightsf
      \scriptscriptfont\sffam=\sixsf
  \else
    \scriptfont\sffam=\elevensf
      \scriptscriptfont\sffam=\elevensf
  \fi
  \textfont\mibfam=\elevenmib
    \scriptfont\mibfam=\eightmib
      \scriptscriptfont\mibfam=\sixmib
  \textfont\sybfam=\elevensyb
    \scriptfont\sybfam=\eightsyb
      \scriptscriptfont\sybfam=\sixsyb
  \ifprod@font
    \textfont\xmfam=\elevenxm
      \scriptfont\xmfam=\eightxm
       \scriptscriptfont\xmfam=\sixxm
    \textfont\ymfam=\elevenym
      \scriptfont\ymfam=\eightym
        \scriptscriptfont\ymfam=\sixym
   \fi
  \def\oldstyle{\fam\@ne\eleveni}%
  \def\boldstyle{\fam\mibfam\elevenmib}%
  \b@ls{13pt}\rm%
}

\def\fourteenpoint{
  \def\rm{\fam0\fourteenrm}%
  \textfont0\fourteenrm  \scriptfont0\tenrm  \scriptscriptfont0\sevenrm%
  \textfont1\fourteeni   \scriptfont1\teni   \scriptscriptfont1\seveni%
  \textfont2\fourteensy  \scriptfont2\tensy  \scriptscriptfont2\sevensy%
  \textfont\itfam=\fourteenit\def\it{\fam\itfam\fourteenit}%
  \ifprod@font
    \scriptfont\itfam=\tenit
      \scriptscriptfont\itfam=\sevenit
  \else
    \scriptfont\itfam=\fourteenit
      \scriptscriptfont\itfam=\fourteenit
  \fi
  \textfont\bffam=\fourteenbf%
    \scriptfont\bffam=\tenbf%
      \scriptscriptfont\bffam=\sevenbf%
  \def\bf{\fam\bffam\fourteenbf}%
  \textfont\slfam=\fourteensl\def\sl{\fam\slfam\fourteensl}%
  \ifprod@font
    \scriptfont\slfam=\tensl
      \scriptscriptfont\slfam=\sevensl
  \else
    \scriptfont\slfam=\fourteensl
      \scriptscriptfont\slfam=\fourteensl
  \fi
  \textfont\ttfam=\fourteentt\def\tt{\fam\ttfam\fourteentt}%
  \ifprod@font
    \scriptfont\ttfam=\tentt
      \scriptscriptfont\ttfam=\seventt
  \else
    \scriptfont\ttfam=\fourteentt
      \scriptscriptfont\ttfam=\fourteentt
  \fi
  \textfont\scfam=\fourteencsc\def\sc{\fam\scfam\fourteencsc}%
  \ifprod@font
    \scriptfont\scfam=\tencsc
      \scriptscriptfont\scfam=\sevencsc
  \else
    \scriptfont\scfam=\fourteencsc
      \scriptscriptfont\scfam=\fourteencsc
  \fi
  \textfont\sffam=\fourteensf\def\sf{\fam\sffam\fourteensf}%
  \ifprod@font
    \scriptfont\sffam=\tensf
      \scriptscriptfont\sffam=\sevensf
  \else
    \scriptfont\sffam=\fourteensf
      \scriptscriptfont\sffam=\fourteensf
  \fi
  \textfont\mibfam=\fourteenmib
    \scriptfont\mibfam=\tenmib
      \scriptscriptfont\mibfam=\sevenmib
  \textfont\sybfam=\fourteensyb
    \scriptfont\sybfam=\tensyb
      \scriptscriptfont\sybfam=\sevensyb
  \ifprod@font
    \textfont\xmfam=\fourteenxm
      \scriptfont\xmfam=\tenxm
        \scriptscriptfont\xmfam=\sevenxm
   \textfont\ymfam=\fourteenym
      \scriptfont\ymfam=\tenym
        \scriptscriptfont\ymfam=\sevenym
  \fi
  \def\oldstyle{\fam\@ne\fourteeni}%
  \def\boldstyle{\fam\mibfam\fourteenmib}%
  \b@ls{17pt}\rm%
}

\def\seventeenpoint{
  \def\rm{\fam0\seventeenrm}%
  \textfont0\seventeenrm  \scriptfont0\twelverm  \scriptscriptfont0\tenrm%
  \textfont1\seventeeni   \scriptfont1\twelvei   \scriptscriptfont1\teni%
  \textfont2\seventeensy  \scriptfont2\twelvesy  \scriptscriptfont2\tensy%
  \textfont\itfam=\seventeenit\def\it{\fam\itfam\seventeenit}%
  \ifprod@font
    \scriptfont\itfam=\twelveit
      \scriptscriptfont\itfam=\tenit
  \else
    \scriptfont\itfam=\seventeenit
      \scriptscriptfont\itfam=\seventeenit
  \fi
  \textfont\bffam=\seventeenbf%
    \scriptfont\bffam=\twelvebf%
      \scriptscriptfont\bffam=\tenbf%
  \def\bf{\fam\bffam\seventeenbf}%
  \textfont\slfam=\seventeensl\def\sl{\fam\slfam\seventeensl}%
  \ifprod@font
    \scriptfont\slfam=\twelvesl
      \scriptscriptfont\slfam=\tensl
  \else
    \scriptfont\slfam=\seventeensl
      \scriptscriptfont\slfam=\seventeensl
  \fi
  \textfont\ttfam=\seventeentt\def\tt{\fam\ttfam\seventeentt}%
  \ifprod@font
    \scriptfont\ttfam=\twelvett
      \scriptscriptfont\ttfam=\tentt
  \else
    \scriptfont\ttfam=\seventeentt
      \scriptscriptfont\ttfam=\seventeentt
  \fi
  \textfont\scfam=\seventeencsc\def\sc{\fam\scfam\seventeencsc}%
  \ifprod@font
    \scriptfont\scfam=\twelvecsc
      \scriptscriptfont\scfam=\tencsc
  \else
    \scriptfont\scfam=\seventeencsc
      \scriptscriptfont\scfam=\seventeencsc
  \fi
  \textfont\sffam=\seventeensf\def\sf{\fam\sffam\seventeensf}%
  \ifprod@font
    \scriptfont\sffam=\twelvesf
      \scriptscriptfont\sffam=\tensf
  \else
    \scriptfont\sffam=\seventeensf
      \scriptscriptfont\sffam=\seventeensf
  \fi
  \textfont\mibfam=\seventeenmib
    \scriptfont\mibfam=\twelvemib
      \scriptscriptfont\mibfam=\tenmib
  \textfont\sybfam=\seventeensyb
    \scriptfont\sybfam=\twelvesyb
      \scriptscriptfont\sybfam=\tensyb
  \ifprod@font
    \textfont\xmfam=\seventeenxm
      \scriptfont\xmfam=\twelvexm
        \scriptscriptfont\xmfam=\tenxm
    \textfont\ymfam=\seventeenym
      \scriptfont\ymfam=\twelveym
        \scriptscriptfont\ymfam=\tenym
  \fi
  \def\oldstyle{\fam\@ne\seventeeni}%
  \def\boldstyle{\fam\mibfam\seventeenmib}%
  \b@ls{20pt}\rm%
}

\lineskip=1pt      \normallineskip=\lineskip
\lineskiplimit=\z@ \normallineskiplimit=\lineskiplimit



\def\la{\mathrel{\mathchoice {\vcenter{\offinterlineskip\halign{\hfil
$\displaystyle##$\hfil\cr<\cr\sim\cr}}}
{\vcenter{\offinterlineskip\halign{\hfil$\textstyle##$\hfil\cr
<\cr\sim\cr}}}
{\vcenter{\offinterlineskip\halign{\hfil$\scriptstyle##$\hfil\cr
<\cr\sim\cr}}}
{\vcenter{\offinterlineskip\halign{\hfil$\scriptscriptstyle##$\hfil\cr
<\cr\sim\cr}}}}}

\def\ga{\mathrel{\mathchoice {\vcenter{\offinterlineskip\halign{\hfil
$\displaystyle##$\hfil\cr>\cr\sim\cr}}}
{\vcenter{\offinterlineskip\halign{\hfil$\textstyle##$\hfil\cr
>\cr\sim\cr}}}
{\vcenter{\offinterlineskip\halign{\hfil$\scriptstyle##$\hfil\cr
>\cr\sim\cr}}}
{\vcenter{\offinterlineskip\halign{\hfil$\scriptscriptstyle##$\hfil\cr
>\cr\sim\cr}}}}}

\def\getsto{\mathrel{\mathchoice {\vcenter{\offinterlineskip
\halign{\hfil
$\displaystyle##$\hfil\cr\gets\cr\to\cr}}}
{\vcenter{\offinterlineskip\halign{\hfil$\textstyle##$\hfil\cr\gets
\cr\to\cr}}}
{\vcenter{\offinterlineskip\halign{\hfil$\scriptstyle##$\hfil\cr\gets
\cr\to\cr}}}
{\vcenter{\offinterlineskip\halign{\hfil$\scriptscriptstyle##$\hfil\cr
\gets\cr\to\cr}}}}}

\def\lid{\mathrel{\mathchoice {\vcenter{\offinterlineskip\halign{\hfil
$\displaystyle##$\hfil\cr<\cr\noalign{\vskip1.2pt}=\cr}}}
{\vcenter{\offinterlineskip\halign{\hfil$\textstyle##$\hfil\cr<\cr
\noalign{\vskip1.2pt}=\cr}}}
{\vcenter{\offinterlineskip\halign{\hfil$\scriptstyle##$\hfil\cr<\cr
\noalign{\vskip1pt}=\cr}}}
{\vcenter{\offinterlineskip\halign{\hfil$\scriptscriptstyle##$\hfil\cr
<\cr
\noalign{\vskip0.9pt}=\cr}}}}}

\def\gid{\mathrel{\mathchoice {\vcenter{\offinterlineskip\halign{\hfil
$\displaystyle##$\hfil\cr>\cr\noalign{\vskip1.2pt}=\cr}}}
{\vcenter{\offinterlineskip\halign{\hfil$\textstyle##$\hfil\cr>\cr
\noalign{\vskip1.2pt}=\cr}}}
{\vcenter{\offinterlineskip\halign{\hfil$\scriptstyle##$\hfil\cr>\cr
\noalign{\vskip1pt}=\cr}}}
{\vcenter{\offinterlineskip\halign{\hfil$\scriptscriptstyle##$\hfil\cr
>\cr
\noalign{\vskip0.9pt}=\cr}}}}}

\def\grole{\mathrel{\mathchoice {\vcenter{\offinterlineskip\halign{\hfil
$\displaystyle##$\hfil\cr>\cr\noalign{\vskip-1.5pt}<\cr}}}
{\vcenter{\offinterlineskip\halign{\hfil$\textstyle##$\hfil\cr
>\cr\noalign{\vskip-1.5pt}<\cr}}}
{\vcenter{\offinterlineskip\halign{\hfil$\scriptstyle##$\hfil\cr
>\cr\noalign{\vskip-1pt}<\cr}}}
{\vcenter{\offinterlineskip\halign{\hfil$\scriptscriptstyle##$\hfil\cr
>\cr\noalign{\vskip-0.5pt}<\cr}}}}}

\def\leogr{\mathrel{\mathchoice {\vcenter{\offinterlineskip\halign{\hfil
$\displaystyle##$\hfil\cr<\cr\noalign{\vskip-1.5pt}>\cr}}}
{\vcenter{\offinterlineskip\halign{\hfil$\textstyle##$\hfil\cr
<\cr\noalign{\vskip-1.5pt}>\cr}}}
{\vcenter{\offinterlineskip\halign{\hfil$\scriptstyle##$\hfil\cr
<\cr\noalign{\vskip-1pt}>\cr}}}
{\vcenter{\offinterlineskip\halign{\hfil$\scriptscriptstyle##$\hfil\cr
<\cr\noalign{\vskip-0.5pt}>\cr}}}}}

\def\loa{\mathrel{\mathchoice {\vcenter{\offinterlineskip\halign{\hfil
$\displaystyle##$\hfil\cr<\cr\approx\cr}}}
{\vcenter{\offinterlineskip\halign{\hfil$\textstyle##$\hfil\cr
<\cr\approx\cr}}}
{\vcenter{\offinterlineskip\halign{\hfil$\scriptstyle##$\hfil\cr
<\cr\approx\cr}}}
{\vcenter{\offinterlineskip\halign{\hfil$\scriptscriptstyle##$\hfil\cr
<\cr\approx\cr}}}}}

\def\goa{\mathrel{\mathchoice {\vcenter{\offinterlineskip\halign{\hfil
$\displaystyle##$\hfil\cr>\cr\approx\cr}}}
{\vcenter{\offinterlineskip\halign{\hfil$\textstyle##$\hfil\cr
>\cr\approx\cr}}}
{\vcenter{\offinterlineskip\halign{\hfil$\scriptstyle##$\hfil\cr
>\cr\approx\cr}}}
{\vcenter{\offinterlineskip\halign{\hfil$\scriptscriptstyle##$\hfil\cr
>\cr\approx\cr}}}}}

\def\diameter{{\ifmmode\mathchoice
{\ooalign{\hfil\hbox{$\displaystyle/$}\hfil\crcr
{\hbox{$\displaystyle\mathchar"20D$}}}}
{\ooalign{\hfil\hbox{$\textstyle/$}\hfil\crcr
{\hbox{$\textstyle\mathchar"20D$}}}}
{\ooalign{\hfil\hbox{$\scriptstyle/$}\hfil\crcr
{\hbox{$\scriptstyle\mathchar"20D$}}}}
{\ooalign{\hfil\hbox{$\scriptscriptstyle/$}\hfil\crcr
{\hbox{$\scriptscriptstyle\mathchar"20D$}}}}
\else{\ooalign{\hfil/\hfil\crcr\mathhexbox20D}}%
\fi}}

\def\sq{\ifmmode\squareforqed\else{\unskip\nobreak\hfil
\penalty50\hskip1em\null\nobreak\hfil\squareforqed
\parfillskip=0pt\finalhyphendemerits=0\endgraf}\fi}
\def\squareforqed{\hbox{\rlap{$\sqcap$}$\sqcup$}}


\def\bbbc{{\mathchoice {\setbox0=\hbox{$\displaystyle\rm C$}\hbox{\hbox
to0pt{\kern0.4\wd0\vrule height0.9\ht0\hss}\box0}}
{\setbox0=\hbox{$\textstyle\rm C$}\hbox{\hbox
to0pt{\kern0.4\wd0\vrule height0.9\ht0\hss}\box0}}
{\setbox0=\hbox{$\scriptstyle\rm C$}\hbox{\hbox
to0pt{\kern0.4\wd0\vrule height0.9\ht0\hss}\box0}}
{\setbox0=\hbox{$\scriptscriptstyle\rm C$}\hbox{\hbox
to0pt{\kern0.4\wd0\vrule height0.9\ht0\hss}\box0}}}}
\def\bbbq{{\mathchoice {\setbox0=\hbox{$\displaystyle\rm
Q$}\hbox{\raise
0.15\ht0\hbox to0pt{\kern0.4\wd0\vrule height0.8\ht0\hss}\box0}}
{\setbox0=\hbox{$\textstyle\rm Q$}\hbox{\raise
0.15\ht0\hbox to0pt{\kern0.4\wd0\vrule height0.8\ht0\hss}\box0}}
{\setbox0=\hbox{$\scriptstyle\rm Q$}\hbox{\raise
0.15\ht0\hbox to0pt{\kern0.4\wd0\vrule height0.7\ht0\hss}\box0}}
{\setbox0=\hbox{$\scriptscriptstyle\rm Q$}\hbox{\raise
0.15\ht0\hbox to0pt{\kern0.4\wd0\vrule height0.7\ht0\hss}\box0}}}}
\def\bbbt{{\mathchoice {\setbox0=\hbox{$\displaystyle\rm
T$}\hbox{\hbox to0pt{\kern0.3\wd0\vrule height0.9\ht0\hss}\box0}}
{\setbox0=\hbox{$\textstyle\rm T$}\hbox{\hbox
to0pt{\kern0.3\wd0\vrule height0.9\ht0\hss}\box0}}
{\setbox0=\hbox{$\scriptstyle\rm T$}\hbox{\hbox
to0pt{\kern0.3\wd0\vrule height0.9\ht0\hss}\box0}}
{\setbox0=\hbox{$\scriptscriptstyle\rm T$}\hbox{\hbox
to0pt{\kern0.3\wd0\vrule height0.9\ht0\hss}\box0}}}}
\def\bbbs{{\mathchoice
{\setbox0=\hbox{$\displaystyle     \rm S$}\hbox{\raise0.5\ht0\hbox
to0pt{\kern0.35\wd0\vrule height0.45\ht0\hss}\hbox
to0pt{\kern0.55\wd0\vrule height0.5\ht0\hss}\box0}}
{\setbox0=\hbox{$\textstyle        \rm S$}\hbox{\raise0.5\ht0\hbox
to0pt{\kern0.35\wd0\vrule height0.45\ht0\hss}\hbox
to0pt{\kern0.55\wd0\vrule height0.5\ht0\hss}\box0}}
{\setbox0=\hbox{$\scriptstyle      \rm S$}\hbox{\raise0.5\ht0\hbox
to0pt{\kern0.35\wd0\vrule height0.45\ht0\hss}\raise0.05\ht0\hbox
to0pt{\kern0.5\wd0\vrule height0.45\ht0\hss}\box0}}
{\setbox0=\hbox{$\scriptscriptstyle\rm S$}\hbox{\raise0.5\ht0\hbox
to0pt{\kern0.4\wd0\vrule height0.45\ht0\hss}\raise0.05\ht0\hbox
to0pt{\kern0.55\wd0\vrule height0.45\ht0\hss}\box0}}}}
\def\bbbz{{\mathchoice {\hbox{$\sf\textstyle Z\kern-0.4em Z$}}
{\hbox{$\sf\textstyle Z\kern-0.4em Z$}}
{\hbox{$\sf\scriptstyle Z\kern-0.3em Z$}}
{\hbox{$\sf\scriptscriptstyle Z\kern-0.2em Z$}}}}


\ifprod@font
  \mathchardef\la="3\@xm2E
  \mathchardef\getsto="3\@xm1C
  \mathchardef\lid="3\@xm35
  \mathchardef\grole="3\@xm3F
  \mathchardef\loa="3\@xm2F
  \mathchardef\ga="3\@xm26
  \mathchardef\gid="3\@xm3D
  \mathchardef\leogr="3\@xm37
  \mathchardef\goa="3\@xm27
  \mathchardef\sq="0\@xm03
%
%
\def\diameter{{%
  \ifmmode
    \mathchoice
    {\ooalign{\hfil\hbox{$\displaystyle/$}\hfil\crcr
    {\lower.2ex\hbox{$\displaystyle\mathchar"20D$}}}}%
    {\ooalign{\hfil\hbox{$\textstyle/$}\hfil\crcr
    {\lower.2ex\hbox{$\textstyle\mathchar"20D$}}}}%
    {\ooalign{\hfil\hbox{$\scriptstyle/$}\hfil\crcr
    {\lower.1ex\hbox{$\scriptstyle\mathchar"20D$}}}}%
    {\ooalign{\hfil\hbox{$\scriptscriptstyle/$}\hfil\crcr
    {\lower.1ex\hbox{$\scriptscriptstyle\mathchar"20D$}}}}%
  \else
    {\ooalign{\hfil/\hfil\crcr\lower.2ex\hbox{\mathhexbox20D}}}%
  \fi
}}
%
%

\def\bbbc{{\Bbb{C}}}
\def\bbbq{{\Bbb{Q}}}
\def\bbbt{{\Bbb{T}}}
\def\bbbs{{\Bbb{S}}}
\def\bbbz{{\Bbb{Z}}}
\fi


\ifprod@font
\mathchardef\boxdot="2\@xm00
\mathchardef\boxplus="2\@xm01
\mathchardef\boxtimes="2\@xm02
\mathchardef\square="0\@xm03
\mathchardef\blacksquare="0\@xm04
\mathchardef\centerdot="2\@xm05
\mathchardef\lozenge="0\@xm06
\mathchardef\blacklozenge="0\@xm07
\mathchardef\circlearrowright="3\@xm08
\mathchardef\circlearrowleft="3\@xm09
\mathchardef\rightleftharpoons="3\@xm0A
\mathchardef\leftrightharpoons="3\@xm0B
\mathchardef\boxminus="2\@xm0C
\mathchardef\Vdash="3\@xm0D
\mathchardef\Vvdash="3\@xm0E
\mathchardef\vDash="3\@xm0F
\mathchardef\twoheadrightarrow="3\@xm10
\mathchardef\twoheadleftarrow="3\@xm11
\mathchardef\leftleftarrows="3\@xm12
\mathchardef\rightrightarrows="3\@xm13
\mathchardef\upuparrows="3\@xm14
\mathchardef\downdownarrows="3\@xm15
\mathchardef\upharpoonright="3\@xm16

\mathchardef\downharpoonright="3\@xm17
\mathchardef\upharpoonleft="3\@xm18
\mathchardef\downharpoonleft="3\@xm19
\mathchardef\rightarrowtail="3\@xm1A
\mathchardef\leftarrowtail="3\@xm1B
\mathchardef\leftrightarrows="3\@xm1C
\mathchardef\rightleftarrows="3\@xm1D
\mathchardef\Lsh="3\@xm1E
\mathchardef\Rsh="3\@xm1F
\mathchardef\rightsquigarrow="3\@xm20
\mathchardef\leftrightsquigarrow="3\@xm21
\mathchardef\looparrowleft="3\@xm22
\mathchardef\looparrowright="3\@xm23
\mathchardef\circeq="3\@xm24
\mathchardef\succsim="3\@xm25
\mathchardef\gtrsim="3\@xm26
\mathchardef\gtrapprox="3\@xm27
\mathchardef\multimap="3\@xm28
\mathchardef\therefore="3\@xm29
\mathchardef\because="3\@xm2A
\mathchardef\doteqdot="3\@xm2B

\mathchardef\triangleq="3\@xm2C
\mathchardef\precsim="3\@xm2D
\mathchardef\lesssim="3\@xm2E
\mathchardef\lessapprox="3\@xm2F
\mathchardef\eqslantless="3\@xm30
\mathchardef\eqslantgtr="3\@xm31
\mathchardef\curlyeqprec="3\@xm32
\mathchardef\curlyeqsucc="3\@xm33
\mathchardef\preccurlyeq="3\@xm34
\mathchardef\leqq="3\@xm35
\mathchardef\leqslant="3\@xm36
\mathchardef\lessgtr="3\@xm37
\mathchardef\backprime="0\@xm38
\mathchardef\risingdotseq="3\@xm3A
\mathchardef\fallingdotseq="3\@xm3B
\mathchardef\succcurlyeq="3\@xm3C
\mathchardef\geqq="3\@xm3D
\mathchardef\geqslant="3\@xm3E
\mathchardef\gtrless="3\@xm3F
\mathchardef\sqsubset="3\@xm40
\mathchardef\sqsupset="3\@xm41
\mathchardef\vartriangleright="3\@xm42
\mathchardef\vartriangleleft="3\@xm43
\mathchardef\trianglerighteq="3\@xm44
\mathchardef\trianglelefteq="3\@xm45
\mathchardef\bigstar="0\@xm46
\mathchardef\between="3\@xm47
\mathchardef\blacktriangledown="0\@xm48
\mathchardef\blacktriangleright="3\@xm49
\mathchardef\blacktriangleleft="3\@xm4A
\mathchardef\vartriangle="0\@xm4D
\mathchardef\blacktriangle="0\@xm4E
\mathchardef\triangledown="0\@xm4F
\mathchardef\eqcirc="3\@xm50
\mathchardef\lesseqgtr="3\@xm51
\mathchardef\gtreqless="3\@xm52
\mathchardef\lesseqqgtr="3\@xm53
\mathchardef\gtreqqless="3\@xm54
\mathchardef\Rrightarrow="3\@xm56
\mathchardef\Lleftarrow="3\@xm57
\mathchardef\veebar="2\@xm59
\mathchardef\barwedge="2\@xm5A
\mathchardef\doublebarwedge="2\@xm5B
\mathchardef\angle="0\@xm5C
\mathchardef\measuredangle="0\@xm5D
\mathchardef\sphericalangle="0\@xm5E
\mathchardef\varpropto="3\@xm5F
\mathchardef\smallsmile="3\@xm60
\mathchardef\smallfrown="3\@xm61
\mathchardef\Subset="3\@xm62
\mathchardef\Supset="3\@xm63
\mathchardef\Cup="2\@xm64

\mathchardef\Cap="2\@xm65

\mathchardef\curlywedge="2\@xm66
\mathchardef\curlyvee="2\@xm67
\mathchardef\leftthreetimes="2\@xm68
\mathchardef\rightthreetimes="2\@xm69
\mathchardef\subseteqq="3\@xm6A
\mathchardef\supseteqq="3\@xm6B
\mathchardef\bumpeq="3\@xm6C
\mathchardef\Bumpeq="3\@xm6D
\mathchardef\lll="3\@xm6E

\mathchardef\ggg="3\@xm6F

\mathchardef\circledS="0\@xm73
\mathchardef\pitchfork="3\@xm74
\mathchardef\dotplus="2\@xm75
\mathchardef\backsim="3\@xm76
\mathchardef\backsimeq="3\@xm77
\mathchardef\complement="0\@xm7B
\mathchardef\intercal="2\@xm7C
\mathchardef\circledcirc="2\@xm7D
\mathchardef\circledast="2\@xm7E
\mathchardef\circleddash="2\@xm7F
\def\ulcorner{\delimiter"4\@xm70\@xm70 }
\def\urcorner{\delimiter"5\@xm71\@xm71 }
\def\llcorner{\delimiter"4\@xm78\@xm78 }
\def\lrcorner{\delimiter"5\@xm79\@xm79 }
\def\yen{\mathhexbox\@xm55 }
\def\checkmark{\mathhexbox\@xm58 }
\def\circledR{\mathhexbox\@xm72 }
\def\maltese{\mathhexbox\@xm7A }
\mathchardef\lvertneqq="3\@ym00
\mathchardef\gvertneqq="3\@ym01
\mathchardef\nleq="3\@ym02
\mathchardef\ngeq="3\@ym03
\mathchardef\nless="3\@ym04
\mathchardef\ngtr="3\@ym05
\mathchardef\nprec="3\@ym06
\mathchardef\nsucc="3\@ym07
\mathchardef\lneqq="3\@ym08
\mathchardef\gneqq="3\@ym09
\mathchardef\nleqslant="3\@ym0A
\mathchardef\ngeqslant="3\@ym0B
\mathchardef\lneq="3\@ym0C
\mathchardef\gneq="3\@ym0D
\mathchardef\npreceq="3\@ym0E
\mathchardef\nsucceq="3\@ym0F
\mathchardef\precnsim="3\@ym10
\mathchardef\succnsim="3\@ym11
\mathchardef\lnsim="3\@ym12
\mathchardef\gnsim="3\@ym13
\mathchardef\nleqq="3\@ym14
\mathchardef\ngeqq="3\@ym15
\mathchardef\precneqq="3\@ym16
\mathchardef\succneqq="3\@ym17
\mathchardef\precnapprox="3\@ym18
\mathchardef\succnapprox="3\@ym19
\mathchardef\lnapprox="3\@ym1A
\mathchardef\gnapprox="3\@ym1B
\mathchardef\nsim="3\@ym1C
\mathchardef\ncong="3\@ym1D

\mathchardef\varsubsetneq="3\@ym20
\mathchardef\varsupsetneq="3\@ym21
\mathchardef\nsubseteqq="3\@ym22
\mathchardef\nsupseteqq="3\@ym23
\mathchardef\subsetneqq="3\@ym24
\mathchardef\supsetneqq="3\@ym25
\mathchardef\varsubsetneqq="3\@ym26
\mathchardef\varsupsetneqq="3\@ym27
\mathchardef\subsetneq="3\@ym28
\mathchardef\supsetneq="3\@ym29
\mathchardef\nsubseteq="3\@ym2A
\mathchardef\nsupseteq="3\@ym2B
\mathchardef\nparallel="3\@ym2C
\mathchardef\nmid="3\@ym2D
\mathchardef\nshortmid="3\@ym2E
\mathchardef\nshortparallel="3\@ym2F
\mathchardef\nvdash="3\@ym30
\mathchardef\nVdash="3\@ym31
\mathchardef\nvDash="3\@ym32
\mathchardef\nVDash="3\@ym33
\mathchardef\ntrianglerighteq="3\@ym34
\mathchardef\ntrianglelefteq="3\@ym35
\mathchardef\ntriangleleft="3\@ym36
\mathchardef\ntriangleright="3\@ym37
\mathchardef\nleftarrow="3\@ym38
\mathchardef\nrightarrow="3\@ym39
\mathchardef\nLeftarrow="3\@ym3A
\mathchardef\nRightarrow="3\@ym3B
\mathchardef\nLeftrightarrow="3\@ym3C
\mathchardef\nleftrightarrow="3\@ym3D
\mathchardef\divideontimes="2\@ym3E
\mathchardef\varnothing="0\@ym3F
\mathchardef\nexists="0\@ym40
\mathchardef\mho="0\@ym66
\mathchardef\eth="0\@ym67
\mathchardef\eqsim="3\@ym68
\mathchardef\beth="0\@ym69
\mathchardef\gimel="0\@ym6A
\mathchardef\daleth="0\@ym6B
\mathchardef\lessdot="3\@ym6C
\mathchardef\gtrdot="3\@ym6D
\mathchardef\ltimes="2\@ym6E
\mathchardef\rtimes="2\@ym6F
\mathchardef\shortmid="3\@ym70
\mathchardef\shortparallel="3\@ym71
\mathchardef\smallsetminus="2\@ym72
\mathchardef\thicksim="3\@ym73
\mathchardef\thickapprox="3\@ym74
\mathchardef\approxeq="3\@ym75
\mathchardef\succapprox="3\@ym76
\mathchardef\precapprox="3\@ym77
\mathchardef\curvearrowleft="3\@ym78
\mathchardef\curvearrowright="3\@ym79
\mathchardef\digamma="0\@ym7A
\mathchardef\varkappa="0\@ym7B
\mathchardef\hslash="0\@ym7D
\mathchardef\hbar="0\@ym7E
\mathchardef\backepsilon="3\@ym7F


\def\Bbb{\ifmmode\let\next\Bbb@\else
\def\next{\errmessage{Use \string\Bbb\space only in math mode}}\fi\next}
\def\Bbb@#1{{\Bbb@@{#1}}}
\def\Bbb@@#1{\fam\ymfam#1}
\fi


\def\Nulle{0} 
\def\Afe{1}   
\def\Hae{2}   
\def\Hbe{3}   
\def\Hce{4}   
\def\Hde{5}   


\newcount\LastMac       \LastMac=\Nulle

\newskip\half      \half=5.5pt plus 1.5pt minus 2.25pt
\newskip\one       \one=11pt plus 3pt minus 5.5pt
\newskip\onehalf   \onehalf=16.5pt plus 5.5pt minus 8.25pt
\newskip\two       \two=22pt plus 5.5pt minus 11pt

\def\Half{\addvspace{\half}}
\def\One{\addvspace{\one}}
\def\OneHalf{\addvspace{\onehalf}}
\def\Two{\addvspace{\two}}


\def\Raggedright{
  \rightskip=\z@ plus \hsize\relax
}

\def\Fullout{
  \rightskip=\z@\relax
}

\def\Hang#1#2{
  \hangindent=#1%
  \hangafter=#2\relax
}


\newif\ifsp@page
\def\pagestyle#1{\csname ps@#1\endcsname}
\def\thispagestyle#1{\global\sp@pagetrue\gdef\sp@type{#1}}

\def\ps@titlepage{%
  \def\@oddhead{\eightpoint\noindent \the\CatchLine
    \ifprod@font\else\qquad Printed\ \today\fi \hfil}%
  \let\@evenhead=\@oddhead
}

\def\ps@headings{%
  \def\@oddhead{\elevenpoint\it\noindent
    \hfill\the\RightHeader\hskip1.5em\rm\folio}%
  \def\@evenhead{\elevenpoint\noindent
    \folio\hskip1.5em\it\the\LeftHeader\hfill}%
}

\def\ps@plate{%
  \def\@oddhead{\eightpoint\noindent\plt@cap\hfil}%
  \def\@evenhead{\eightpoint\noindent\plt@cap\hfil}%
}



\def\title#1{
  \bgroup
    \vbox to 8pt{\vss}%
    \seventeenpoint
    \Raggedright
    \noindent \strut{\bf #1}\par
  \egroup
}

\def\author#1{
  \bgroup
    \ifnum\LastMac=\Afe \OneHalf\else \vskip 21pt\fi
    \fourteenpoint
    \Raggedright
    \noindent \strut #1\par
    \vskip 3pt%
  \egroup
}

\def\affiliation#1{
  \bgroup
    \vskip -4pt%
    \eightpoint
    \Raggedright
    \noindent \strut {\it #1}\par
  \egroup
  \LastMac=\Afe\relax
}

\def\acceptedline#1{
  \bgroup
    \Two
    \eightpoint
    \Raggedright
    \noindent \strut #1\par
  \egroup
}

\long\def\abstract#1{%
  \bgroup
    \vskip 20pt%
    \everypar{\Hang{11pc}{0}}%
    \noindent{\ninebf ABSTRACT}\par
    \tenpoint
    \Fullout
    \noindent #1\par
  \egroup
}

\long\def\keywords#1{
  \bgroup
    \Half
    \everypar{\Hang{11pc}{0}}%
    \tenpoint
    \Fullout
    \noindent\hbox{\bf Key words:}\ #1\par
  \egroup
}


\def\maketitle{%
  \EndOpening
  \ifsinglecol \else \MakePage\fi
}


\def\pageoffset#1#2{\hoffset=#1\relax\voffset=#2\relax}


\newif\ifAutoNumber \AutoNumberfalse
\newcount\Sec        
\newcount\SecSec
\newcount\SecSecSec

\Sec=\z@

\def\:{\let\@sptoken= } \:  
\def\:{\@xifnch} \expandafter\def\: {\futurelet\@tempc\@ifnch}

\def\@ifnextchar#1#2#3{%
  \let\@tempMACe #1%
  \def\@tempMACa{#2}%
  \def\@tempMACb{#3}%
  \futurelet \@tempMACc\@ifnch%
}

\def\@ifnch{%
\ifx \@tempMACc \@sptoken%
  \let\@tempMACd\@xifnch%
\else%
  \ifx \@tempMACc \@tempMACe%
    \let\@tempMACd\@tempMACa%
  \else%
    \let\@tempMACd\@tempMACb%
  \fi%
\fi%
\@tempMACd%
}

\def\@ifstar#1#2{\@ifnextchar *{\def\@tempMACa*{#1}\@tempMACa}{#2}}

\newskip\@tempskipb

\def\addvspace#1{%
  \ifvmode\else \endgraf\fi%
  \ifdim\lastskip=\z@%
    \vskip #1\relax%
  \else%
    \@tempskipb#1\relax\@xaddvskip%
  \fi%
}

\def\@xaddvskip{%
  \ifdim\lastskip<\@tempskipb%
    \vskip-\lastskip%
    \vskip\@tempskipb\relax%
  \else%
    \ifdim\@tempskipb<\z@%
      \ifdim\lastskip<\z@ \else%
        \advance\@tempskipb\lastskip%
        \vskip-\lastskip\vskip\@tempskipb%
      \fi%
    \fi%
  \fi%
}

\newskip\@tmpSKIP

\def\addpen#1{%
  \ifvmode
    \if@nobreak
    \else
      \ifdim\lastskip=\z@
        \penalty#1\relax
      \else
        \@tmpSKIP=\lastskip
        \vskip -\lastskip
        \penalty#1\vskip\@tmpSKIP
      \fi
    \fi
  \fi
}

\newcount\@clubpen   \@clubpen=\clubpenalty
\newif\if@nobreak    \@nobreakfalse

\def\@noafterindent{%
  \global\@nobreaktrue
  \everypar{\if@nobreak
              \global\@nobreakfalse
              \clubpenalty \@M
              {\setbox\z@\lastbox}%
              \LastMac=\Nulle\relax%
            \else
              \clubpenalty \@clubpen
              \everypar{}%
            \fi}
}

\newcount\gds@cbrk   \gds@cbrk=-300

\def\@nohdbrk{\interlinepenalty \@M\relax}

\let\@par=\par
\def\@restorepar{\def\par{\@par}}

\newif\if@endpe   \@endpefalse
 
\def\@doendpe{\@endpetrue \@nobreakfalse \LastMac=\Nulle\relax%
     \def\par{\@restorepar\everypar{}\par\@endpefalse}%
              \everypar{\setbox\z@\lastbox\everypar{}\@endpefalse}%
}

\def\section{\@ifstar{\@ssection}{\@section}}

\def\@section#1{
  \if@nobreak
    \everypar{}%
    \ifnum\LastMac=\Hae \addvspace{\half}\fi
  \else
    \addpen{\gds@cbrk}%
    \addvspace{\two}%
  \fi
  \bgroup
    \ninepoint\bf
    \Raggedright
    \ifAutoNumber
      \global\advance\Sec \@ne
      \noindent\@nohdbrk\number\Sec\hskip 1pc \uppercase{#1}\par
      \global\SecSec=\z@
    \else
      \noindent\@nohdbrk\uppercase{#1}\par
    \fi
  \egroup
  \nobreak
  \vskip\half
  \nobreak
  \@noafterindent
  \LastMac=\Hae\relax
}

\def\@ssection#1{
  \if@nobreak
    \everypar{}%
    \ifnum\LastMac=\Hae \addvspace{\half}\fi
  \else
    \addpen{\gds@cbrk}%
    \addvspace{\two}%
  \fi
  \bgroup
    \ninepoint\bf
    \Raggedright
    \noindent\@nohdbrk\uppercase{#1}\par
  \egroup
  \nobreak
  \vskip\half
  \nobreak
  \@noafterindent
  \LastMac=\Hae\relax
}

\def\subsection#1{
  \if@nobreak
    \everypar{}%
    \ifnum\LastMac=\Hae \addvspace{1pt plus 1pt minus .5pt}\fi
  \else
    \addpen{\gds@cbrk}%
    \addvspace{\onehalf}%
  \fi
  \bgroup
    \ninepoint\bf
    \Raggedright
    \ifAutoNumber
      \global\advance\SecSec \@ne
      \noindent\@nohdbrk\number\Sec.\number\SecSec \hskip 1pc\relax #1\par
      \global\SecSecSec=\z@
    \else
      \noindent\@nohdbrk #1\par
    \fi
  \egroup
  \nobreak
  \vskip\half
  \nobreak
  \@noafterindent
  \LastMac=\Hbe\relax
}

\def\subsubsection#1{
  \if@nobreak
    \everypar{}%
    \ifnum\LastMac=\Hbe \addvspace{1pt plus 1pt minus .5pt}\fi
  \else
    \addpen{\gds@cbrk}%
    \addvspace{\onehalf}%
  \fi
  \bgroup
    \ninepoint\it
    \Raggedright
    \ifAutoNumber
      \global\advance\SecSecSec \@ne
      \noindent\@nohdbrk\number\Sec.\number\SecSec.\number\SecSecSec
        \hskip 1pc\relax #1\par
    \else
      \noindent\@nohdbrk #1\par
    \fi
  \egroup
  \nobreak
  \vskip\half
  \nobreak
  \@noafterindent
  \LastMac=\Hce\relax
}

\def\paragraph#1{
  \if@nobreak
    \everypar{}%
  \else
    \addpen{\gds@cbrk}%
    \addvspace{\one}%
  \fi%
  \bgroup%
    \ninepoint\it
    \noindent #1\ \nobreak%
  \egroup
  \LastMac=\Hde\relax
  \ignorespaces
}


\let\tx=\relax 


\def\beginlist{%
  \par\if@nobreak \else\addvspace{\half}\fi%
  \bgroup%
    \ninepoint
    \let\item=\list@item%
}

\def\list@item{%
  \par\noindent\hskip 1em\relax%
  \ignorespaces%
}

\def\endlist{\par\egroup\addvspace{\half}\@doendpe}


\def\beginrefs{%
  \par
  \bgroup
    \eightpoint
    \Raggedright
    \let\bibitem=\bib@item
}

\def\bib@item{%
  \par\parindent=1.5em\Hang{1.5em}{1}%
  \everypar={\Hang{1.5em}{1}\ignorespaces}%
  \noindent\ignorespaces
}

\def\endrefs{\par\egroup\@doendpe}


\newtoks\CatchLine

\def\@journal{Mon.\ Not.\ R.\ Astron.\ Soc.\ }  
\def\@pubyear{1994}        
\def\@pagerange{000--000}  
\def\@volume{000}          
\def\@microfiche{}         %

\def\pubyear#1{\gdef\@pubyear{#1}\@makecatchline}
\def\pagerange#1{\gdef\@pagerange{#1}\@makecatchline}
\def\volume#1{\gdef\@volume{#1}\@makecatchline}
\def\microfiche#1{\gdef\@microfiche{and Microfiche\ #1}\@makecatchline}

\def\@makecatchline{%
  \global\CatchLine{%
    {\rm \@journal {\bf \@volume},\ \@pagerange\ (\@pubyear)\ \@microfiche}}%
}

\@makecatchline 

\newtoks\LeftHeader
\def\shortauthor#1{
  \global\LeftHeader{#1}%
}

\newtoks\RightHeader
\def\shorttitle#1{
  \global\RightHeader{#1}%
}

\def\PageHead{
  \begingroup
    \ifsp@page
      \csname ps@\sp@type\endcsname
      \global\sp@pagefalse
    \fi
    \ifodd\pageno
      \let\the@head=\@oddhead
    \else
      \let\the@head=\@evenhead
    \fi
    \vbox to \z@{\vskip-22.5\p@%
      \hbox to \PageWidth{\vbox to8.5\p@{}%
        \the@head
      }%
    \vss}%
  \endgroup
  \nointerlineskip
}

\def\today{%
  \number\day\space
  \ifcase\month\or January\or February\or March\or April\or May\or June\or
    July\or August\or September\or October\or November\or December\fi
  \space\number\year%
}

\def\PageFoot{} 

\def\authorcomment#1{%
  \gdef\PageFoot{%
    \nointerlineskip%
    \vbox to 22pt{\vfil%
      \hbox to \PageWidth{\elevenpoint\noindent \hfil #1 \hfil}}%
  }%
}


\newif\ifplate@page
\newbox\plt@box

\def\beginplatepage{%
  \let\plate=\plate@head
  \let\caption=\fig@caption
  \global\setbox\plt@box=\vbox\bgroup
  \TEMPDIMEN=\PageWidth 
  \hsize=\PageWidth\relax
}

\def\endplatepage{\par\egroup\global\plate@pagetrue}
\def\plate@head#1{\gdef\plt@cap{#1}}


\def\letters{%
  \gdef\folio{\ifnum\pageno<\z@ L\romannumeral-\pageno
    \else L\number\pageno \fi}%
}


\everydisplay{\displaysetup}

\newif\ifeqno
\newif\ifleqno

\def\displaysetup#1$${%
 \displaytest#1\eqno\eqno\displaytest
}

\def\displaytest#1\eqno#2\eqno#3\displaytest{%
 \if!#3!\ldisplaytest#1\leqno\leqno\ldisplaytest
 \else\eqnotrue\leqnofalse\def\eqn{#2}\def\eq{#1}\fi
 \generaldisplay$$}

\def\ldisplaytest#1\leqno#2\leqno#3\ldisplaytest{%
 \def\eq{#1}%
 \if!#3!\eqnofalse\else\eqnotrue\leqnotrue
  \def\eqn{#2}\fi}

\def\generaldisplay{%
\ifeqno \ifleqno 
   \hbox to \hsize{\noindent
     $\displaystyle\eq$\hfil$\displaystyle\eqn$}
  \else
    \hbox to \hsize{\noindent
     $\displaystyle\eq$\hfil$\displaystyle\eqn$}
  \fi
 \else
 \hbox to \hsize{\vbox{\noindent
  $\displaystyle\eq$\hfil}}
 \fi
}


\def\@notice{%
  \par\Two%
  \noindent{\b@ls{11pt}\ninerm This paper has been produced using the
    Blackwell Scientific Publications \TeX\ macros.\par}%
}

\outer\def\bye{\@notice\par\vfill\supereject\end}


\def\start@mess{%
  Monthly notices of the RAS journal style (\@typeface)\space
    v\@version,\space \@verdate.%
}

\everyjob{\Warn{\start@mess}}



\newif\if@debug \@debugfalse  

\def\Print#1{\if@debug\immediate\write16{#1}\else \fi}
\def\Warn#1{\immediate\write16{#1}}
\def\wlog#1{}

\newcount\Iteration 

\def\Single{0} \def\Double{1}                 
\def\Figure{0} \def\Table{1}                  

\def\InStack{0}  
\def\InZoneA{1}
\def\InZoneB{2}
\def\InZoneC{3}

\newcount\TEMPCOUNT 
\newdimen\TEMPDIMEN 
\newbox\TEMPBOX     
\newbox\VOIDBOX     

\newcount\LengthOfStack 
\newcount\MaxItems      
\newcount\StackPointer
\newcount\Point         
\newcount\NextFigure    
\newcount\NextTable     
\newcount\NextItem      

\newcount\StatusStack   
\newcount\NumStack      
\newcount\TypeStack     
\newcount\SpanStack     
\newcount\BoxStack      

\newcount\ItemSTATUS    
\newcount\ItemNUMBER    
\newcount\ItemTYPE      
\newcount\ItemSPAN      
\newbox\ItemBOX         
\newdimen\ItemSIZE      

\newdimen\PageHeight    
\newdimen\TextLeading   
\newdimen\Feathering    
\newcount\LinesPerPage  
\newdimen\ColumnWidth   
\newdimen\ColumnGap     
\newdimen\PageWidth     
\newdimen\BodgeHeight   
\newcount\Leading       

\newdimen\ZoneBSize  
\newdimen\TextSize   
\newbox\ZoneABOX     
\newbox\ZoneBBOX     
\newbox\ZoneCBOX     

\newif\ifFirstSingleItem
\newif\ifFirstZoneA
\newif\ifMakePageInComplete
\newif\ifMoreFigures \MoreFiguresfalse 
\newif\ifMoreTables  \MoreTablesfalse  

\newif\ifFigInZoneB 
\newif\ifFigInZoneC 
\newif\ifTabInZoneB 
\newif\ifTabInZoneC

\newif\ifZoneAFullPage

\newbox\MidBOX    
\newbox\LeftBOX
\newbox\RightBOX
\newbox\PageBOX   

\newif\ifLeftCOL  
\LeftCOLtrue

\newdimen\ZoneBAdjust

\newcount\ItemFits
\def\Yes{1}
\def\No{2}


\MaxItems=15
\NextFigure=\z@        
\NextTable=\@ne

\BodgeHeight=6pt
\TextLeading=11pt    
\Leading=11
\Feathering=\z@      
\LinesPerPage=61     
\topskip=\TextLeading
\ColumnWidth=20pc    
\ColumnGap=2pc       

\newskip\ItemSepamount  
\ItemSepamount=\TextLeading plus \TextLeading minus 4pt

\parskip=\z@ plus .1pt
\parindent=18pt
\widowpenalty=\z@
\clubpenalty=10000
\tolerance=1500
\hbadness=1500
\abovedisplayskip=6pt plus 2pt minus 2pt
\belowdisplayskip=6pt plus 2pt minus 2pt
\abovedisplayshortskip=6pt plus 2pt minus 2pt
\belowdisplayshortskip=6pt plus 2pt minus 2pt

\ninepoint 


\PageHeight=682pt

\PageWidth=2\ColumnWidth
\advance\PageWidth by \ColumnGap

\pagestyle{headings}




\newcount\DUMMY \StatusStack=\allocationnumber
\newcount\DUMMY \newcount\DUMMY \newcount\DUMMY 
\newcount\DUMMY \newcount\DUMMY \newcount\DUMMY 
\newcount\DUMMY \newcount\DUMMY \newcount\DUMMY
\newcount\DUMMY \newcount\DUMMY \newcount\DUMMY 
\newcount\DUMMY \newcount\DUMMY \newcount\DUMMY

\newcount\DUMMY \NumStack=\allocationnumber
\newcount\DUMMY \newcount\DUMMY \newcount\DUMMY 
\newcount\DUMMY \newcount\DUMMY \newcount\DUMMY 
\newcount\DUMMY \newcount\DUMMY \newcount\DUMMY 
\newcount\DUMMY \newcount\DUMMY \newcount\DUMMY 
\newcount\DUMMY \newcount\DUMMY \newcount\DUMMY

\newcount\DUMMY \TypeStack=\allocationnumber
\newcount\DUMMY \newcount\DUMMY \newcount\DUMMY 
\newcount\DUMMY \newcount\DUMMY \newcount\DUMMY 
\newcount\DUMMY \newcount\DUMMY \newcount\DUMMY 
\newcount\DUMMY \newcount\DUMMY \newcount\DUMMY 
\newcount\DUMMY \newcount\DUMMY \newcount\DUMMY

\newcount\DUMMY \SpanStack=\allocationnumber
\newcount\DUMMY \newcount\DUMMY \newcount\DUMMY 
\newcount\DUMMY \newcount\DUMMY \newcount\DUMMY 
\newcount\DUMMY \newcount\DUMMY \newcount\DUMMY 
\newcount\DUMMY \newcount\DUMMY \newcount\DUMMY 
\newcount\DUMMY \newcount\DUMMY \newcount\DUMMY

\newbox\DUMMY   \BoxStack=\allocationnumber
\newbox\DUMMY   \newbox\DUMMY \newbox\DUMMY 
\newbox\DUMMY   \newbox\DUMMY \newbox\DUMMY 
\newbox\DUMMY   \newbox\DUMMY \newbox\DUMMY 
\newbox\DUMMY   \newbox\DUMMY \newbox\DUMMY 
\newbox\DUMMY   \newbox\DUMMY \newbox\DUMMY

\def\wlog{\immediate\write\m@ne}


\def\GetItemAll#1{%
 \GetItemSTATUS{#1}
 \GetItemNUMBER{#1}
 \GetItemTYPE{#1}
 \GetItemSPAN{#1}
 \GetItemBOX{#1}
}

\def\GetItemSTATUS#1{%
 \Point=\StatusStack
 \advance\Point by #1
 \global\ItemSTATUS=\count\Point
}

\def\GetItemNUMBER#1{%
 \Point=\NumStack
 \advance\Point by #1
 \global\ItemNUMBER=\count\Point
}

\def\GetItemTYPE#1{%
 \Point=\TypeStack
 \advance\Point by #1
 \global\ItemTYPE=\count\Point
}

\def\GetItemSPAN#1{%
 \Point\SpanStack
 \advance\Point by #1
 \global\ItemSPAN=\count\Point
}

\def\GetItemBOX#1{%
 \Point=\BoxStack
 \advance\Point by #1
 \global\setbox\ItemBOX=\vbox{\copy\Point}
 \global\ItemSIZE=\ht\ItemBOX
 \global\advance\ItemSIZE by \dp\ItemBOX
 \TEMPCOUNT=\ItemSIZE
 \divide\TEMPCOUNT by \Leading
 \divide\TEMPCOUNT by 65536
 \advance\TEMPCOUNT \@ne
 \ItemSIZE=\TEMPCOUNT pt
 \global\multiply\ItemSIZE by \Leading
}


\def\JoinStack{%
 \ifnum\LengthOfStack=\MaxItems 
  \Warn{WARNING: Stack is full...some items will be lost!}
 \else
  \Point=\StatusStack
  \advance\Point by \LengthOfStack
  \global\count\Point=\ItemSTATUS
  \Point=\NumStack
  \advance\Point by \LengthOfStack
  \global\count\Point=\ItemNUMBER
  \Point=\TypeStack
  \advance\Point by \LengthOfStack
  \global\count\Point=\ItemTYPE
  \Point\SpanStack
  \advance\Point by \LengthOfStack
  \global\count\Point=\ItemSPAN
  \Point=\BoxStack
  \advance\Point by \LengthOfStack
  \global\setbox\Point=\vbox{\copy\ItemBOX}
  \global\advance\LengthOfStack \@ne
  \ifnum\ItemTYPE=\Figure 
   \global\MoreFigurestrue
  \else
   \global\MoreTablestrue
  \fi
 \fi
}


\def\LeaveStack#1{%
 {\Iteration=#1
 \loop
 \ifnum\Iteration<\LengthOfStack
  \advance\Iteration \@ne
  \GetItemSTATUS{\Iteration}
   \advance\Point by \m@ne
   \global\count\Point=\ItemSTATUS
  \GetItemNUMBER{\Iteration}
   \advance\Point by \m@ne
   \global\count\Point=\ItemNUMBER
  \GetItemTYPE{\Iteration}
   \advance\Point by \m@ne
   \global\count\Point=\ItemTYPE
  \GetItemSPAN{\Iteration}
   \advance\Point by \m@ne
   \global\count\Point=\ItemSPAN
  \GetItemBOX{\Iteration}
   \advance\Point by \m@ne
   \global\setbox\Point=\vbox{\copy\ItemBOX}
 \repeat}
 \global\advance\LengthOfStack by \m@ne
}


\newif\ifStackNotClean

\def\CleanStack{%
 \StackNotCleantrue
 {\Iteration=\z@
  \loop
   \ifStackNotClean
    \GetItemSTATUS{\Iteration}
    \ifnum\ItemSTATUS=\InStack
     \advance\Iteration \@ne
     \else
      \LeaveStack{\Iteration}
    \fi
   \ifnum\LengthOfStack<\Iteration
    \StackNotCleanfalse
   \fi
 \repeat}
}


\def\FindItem#1#2{%
 \global\StackPointer=\m@ne 
 {\Iteration=\z@
  \loop
  \ifnum\Iteration<\LengthOfStack
   \GetItemSTATUS{\Iteration}
   \ifnum\ItemSTATUS=\InStack
    \GetItemTYPE{\Iteration}
    \ifnum\ItemTYPE=#1
     \GetItemNUMBER{\Iteration}
     \ifnum\ItemNUMBER=#2
      \global\StackPointer=\Iteration
      \Iteration=\LengthOfStack 
     \fi
    \fi
   \fi
  \advance\Iteration \@ne
 \repeat}
}


\def\FindNext{%
 \global\StackPointer=\m@ne 
 {\Iteration=\z@
  \loop
  \ifnum\Iteration<\LengthOfStack
   \GetItemSTATUS{\Iteration}
   \ifnum\ItemSTATUS=\InStack
    \GetItemTYPE{\Iteration}
   \ifnum\ItemTYPE=\Figure
    \ifMoreFigures
      \global\NextItem=\Figure
      \global\StackPointer=\Iteration
      \Iteration=\LengthOfStack 
    \fi
   \fi
   \ifnum\ItemTYPE=\Table
    \ifMoreTables
      \global\NextItem=\Table
      \global\StackPointer=\Iteration
      \Iteration=\LengthOfStack 
    \fi
   \fi
  \fi
  \advance\Iteration \@ne
 \repeat}
}


\def\ChangeStatus#1#2{%
 \Point=\StatusStack
 \advance\Point by #1
 \global\count\Point=#2
}



\def\Zone{\InZoneA}

\ZoneBAdjust=\z@

\def\MakePage{
 \global\ZoneBSize=\PageHeight
 \global\TextSize=\ZoneBSize
 \global\ZoneAFullPagefalse
 \global\topskip=\TextLeading
 \MakePageInCompletetrue
 \MoreFigurestrue
 \MoreTablestrue
 \FigInZoneBfalse
 \FigInZoneCfalse
 \TabInZoneBfalse
 \TabInZoneCfalse
 \global\FirstSingleItemtrue
 \global\FirstZoneAtrue
 \global\setbox\ZoneABOX=\box\VOIDBOX
 \global\setbox\ZoneBBOX=\box\VOIDBOX
 \global\setbox\ZoneCBOX=\box\VOIDBOX
 \loop
  \ifMakePageInComplete
 \FindNext
 \ifnum\StackPointer=\m@ne
  \NextItem=\m@ne
  \MoreFiguresfalse
  \MoreTablesfalse
 \fi
 \ifnum\NextItem=\Figure
   \FindItem{\Figure}{\NextFigure}
   \ifnum\StackPointer=\m@ne \global\MoreFiguresfalse
   \else
    \GetItemSPAN{\StackPointer}
    \ifnum\ItemSPAN=\Single \def\Zone{\InZoneB}\relax
     \ifFigInZoneC \global\MoreFiguresfalse\fi
    \else
     \def\Zone{\InZoneA}
     \ifFigInZoneB \def\Zone{\InZoneC}\fi
    \fi
   \fi
   \ifMoreFigures\Print{}\FigureItems\fi
 \fi
\ifnum\NextItem=\Table
   \FindItem{\Table}{\NextTable}
   \ifnum\StackPointer=\m@ne \global\MoreTablesfalse
   \else
    \GetItemSPAN{\StackPointer}
    \ifnum\ItemSPAN=\Single\relax
     \ifTabInZoneC \global\MoreTablesfalse\fi
    \else
     \def\Zone{\InZoneA}
     \ifTabInZoneB \def\Zone{\InZoneC}\fi
    \fi
   \fi
   \ifMoreTables\Print{}\TableItems\fi
 \fi
   \MakePageInCompletefalse 
   \ifMoreFigures\MakePageInCompletetrue\fi
   \ifMoreTables\MakePageInCompletetrue\fi
 \repeat
 \ifZoneAFullPage
  \global\TextSize=\z@
  \global\ZoneBSize=\z@
  \global\vsize=\z@\relax
  \global\topskip=\z@\relax
  \vbox to \z@{\vss}
  \eject
 \else
 \global\advance\ZoneBSize by -\ZoneBAdjust
 \global\vsize=\ZoneBSize
 \global\hsize=\ColumnWidth
 \global\ZoneBAdjust=\z@
 \ifdim\TextSize<23pt
 \Warn{}
 \Warn{* Making column fall short: TextSize=\the\TextSize *}
 \vskip-\lastskip\eject\fi
 \fi
}

\def\MakeRightCol{
 \global\TextSize=\ZoneBSize
 \MakePageInCompletetrue
 \MoreFigurestrue
 \MoreTablestrue
 \global\FirstSingleItemtrue
 \global\setbox\ZoneBBOX=\box\VOIDBOX
 \def\Zone{\InZoneB}
 \loop
  \ifMakePageInComplete
 \FindNext
 \ifnum\StackPointer=\m@ne
  \NextItem=\m@ne
  \MoreFiguresfalse
  \MoreTablesfalse
 \fi
 \ifnum\NextItem=\Figure
   \FindItem{\Figure}{\NextFigure}
   \ifnum\StackPointer=\m@ne \MoreFiguresfalse
   \else
    \GetItemSPAN{\StackPointer}
    \ifnum\ItemSPAN=\Double\relax
     \MoreFiguresfalse\fi
   \fi
   \ifMoreFigures\Print{}\FigureItems\fi
 \fi
 \ifnum\NextItem=\Table
   \FindItem{\Table}{\NextTable}
   \ifnum\StackPointer=\m@ne \MoreTablesfalse
   \else
    \GetItemSPAN{\StackPointer}
    \ifnum\ItemSPAN=\Double\relax
     \MoreTablesfalse\fi
   \fi
   \ifMoreTables\Print{}\TableItems\fi
 \fi
   \MakePageInCompletefalse 
   \ifMoreFigures\MakePageInCompletetrue\fi
   \ifMoreTables\MakePageInCompletetrue\fi
 \repeat
 \ifZoneAFullPage
  \global\TextSize=\z@
  \global\ZoneBSize=\z@
  \global\vsize=\z@\relax
  \global\topskip=\z@\relax
  \vbox to \z@{\vss}
  \eject
 \else
 \global\vsize=\ZoneBSize
 \global\hsize=\ColumnWidth
 \ifdim\TextSize<23pt
 \Warn{}
 \Warn{* Making column fall short: TextSize=\the\TextSize *}
 \vskip-\lastskip\eject\fi
\fi
}

\def\FigureItems{
 \Print{Considering...}
 \ShowItem{\StackPointer}
 \GetItemBOX{\StackPointer} 
 \GetItemSPAN{\StackPointer}
  \CheckFitInZone 
  \ifnum\ItemFits=\Yes
   \ifnum\ItemSPAN=\Single
     \ChangeStatus{\StackPointer}{\InZoneB} 
     \global\FigInZoneBtrue
     \ifFirstSingleItem
      \hbox{}\vskip-\BodgeHeight
     \global\advance\ItemSIZE by \TextLeading
     \fi
     \unvbox\ItemBOX\ItemSep
     \global\FirstSingleItemfalse
     \global\advance\TextSize by -\ItemSIZE
     \global\advance\TextSize by -\TextLeading
   \else
    \ifFirstZoneA
     \global\advance\ItemSIZE by \TextLeading
     \global\FirstZoneAfalse\fi
    \global\advance\TextSize by -\ItemSIZE
    \global\advance\TextSize by -\TextLeading
    \global\advance\ZoneBSize by -\ItemSIZE
    \global\advance\ZoneBSize by -\TextLeading
    \ifFigInZoneB\relax
     \else
     \ifdim\TextSize<3\TextLeading
     \global\ZoneAFullPagetrue
     \fi
    \fi
    \ChangeStatus{\StackPointer}{\Zone}
    \ifnum\Zone=\InZoneC \global\FigInZoneCtrue\fi
  \fi
   \Print{TextSize=\the\TextSize}
   \Print{ZoneBSize=\the\ZoneBSize}
  \global\advance\NextFigure \@ne
   \Print{This figure has been placed.}
  \else
   \Print{No space available for this figure...holding over.}
   \Print{}
   \global\MoreFiguresfalse
  \fi
}

\def\TableItems{
 \Print{Considering...}
 \ShowItem{\StackPointer}
 \GetItemBOX{\StackPointer} 
 \GetItemSPAN{\StackPointer}
  \CheckFitInZone 
  \ifnum\ItemFits=\Yes
   \ifnum\ItemSPAN=\Single
    \ChangeStatus{\StackPointer}{\InZoneB}
     \global\TabInZoneBtrue
     \ifFirstSingleItem
      \hbox{}\vskip-\BodgeHeight
     \global\advance\ItemSIZE by \TextLeading
     \fi
     \unvbox\ItemBOX\ItemSep
     \global\FirstSingleItemfalse
     \global\advance\TextSize by -\ItemSIZE
     \global\advance\TextSize by -\TextLeading
   \else
    \ifFirstZoneA
    \global\advance\ItemSIZE by \TextLeading
    \global\FirstZoneAfalse\fi
    \global\advance\TextSize by -\ItemSIZE
    \global\advance\TextSize by -\TextLeading
    \global\advance\ZoneBSize by -\ItemSIZE
    \global\advance\ZoneBSize by -\TextLeading
    \ifFigInZoneB\relax
     \else
     \ifdim\TextSize<3\TextLeading
     \global\ZoneAFullPagetrue
     \fi
    \fi
    \ChangeStatus{\StackPointer}{\Zone}
    \ifnum\Zone=\InZoneC \global\TabInZoneCtrue\fi
   \fi
  \global\advance\NextTable \@ne
   \Print{This table has been placed.}
  \else
  \Print{No space available for this table...holding over.}
   \Print{}
   \global\MoreTablesfalse
  \fi
}


\def\CheckFitInZone{%
{\advance\TextSize by -\ItemSIZE
 \advance\TextSize by -\TextLeading
 \ifFirstSingleItem
  \advance\TextSize by \TextLeading
 \fi
 \ifnum\Zone=\InZoneA\relax
  \else \advance\TextSize by -\ZoneBAdjust
 \fi
 \ifdim\TextSize<3\TextLeading \global\ItemFits=\No
 \else \global\ItemFits=\Yes\fi}
}

\def\BeginOpening{%
  \thispagestyle{titlepage}%
  \global\setbox\ItemBOX=\vbox\bgroup%
    \hsize=\PageWidth%
    \hrule height \z@
    \ifsinglecol\vskip 6pt\fi 
}

\let\begintopmatter=\BeginOpening  

\def\EndOpening{%
  \One
  \egroup
  \ifsinglecol
    \box\ItemBOX%
    \vskip\TextLeading plus 2\TextLeading
    \@noafterindent
  \else
    \ItemNUMBER=\z@%
    \ItemTYPE=\Figure
    \ItemSPAN=\Double
    \ItemSTATUS=\InStack
    \JoinStack
  \fi
}


\newif\if@here  \@herefalse

\def\no@float{\global\@heretrue}
\let\nofloat=\relax 

\def\beginfigure{%
  \@ifstar{\global\@dfloattrue \@bfigure}{\global\@dfloatfalse \@bfigure}%
}

\def\@bfigure#1{%
  \par
  \if@dfloat
    \ItemSPAN=\Double
    \TEMPDIMEN=\PageWidth
  \else
    \ItemSPAN=\Single
    \TEMPDIMEN=\ColumnWidth
  \fi
  \ifsinglecol
    \TEMPDIMEN=\PageWidth
  \else
    \ItemSTATUS=\InStack
    \ItemNUMBER=#1%
    \ItemTYPE=\Figure
  \fi
  \bgroup
    \hsize=\TEMPDIMEN
    \global\setbox\ItemBOX=\vbox\bgroup
      \eightpoint\nostb@ls{10pt}%
      \let\caption=\fig@caption
      \ifsinglecol \let\nofloat=\no@float\fi
}

\def\fig@caption#1{%
  \vskip 5.5pt plus 6pt%
  \bgroup 
    \eightpoint\nostb@ls{10pt}%
    \setbox\TEMPBOX=\hbox{#1}%
    \ifdim\wd\TEMPBOX>\TEMPDIMEN
      \noindent \unhbox\TEMPBOX\par
    \else
      \hbox to \hsize{\hfil\unhbox\TEMPBOX\hfil}%
    \fi
  \egroup
}

\def\endfigure{%
  \par\egroup 
  \egroup
  \ifsinglecol
    \if@here \midinsert\global\@herefalse\else \topinsert\fi
      \unvbox\ItemBOX
    \endinsert
  \else
    \JoinStack
    \Print{Processing source for figure \the\ItemNUMBER}%
  \fi
}


\newbox\tab@cap@box
\def\tab@caption#1{\global\setbox\tab@cap@box=\hbox{#1\par}}

\newtoks\tab@txt@toks
\long\def\tab@txt#1{\global\tab@txt@toks={#1}\global\table@txttrue}

\newif\iftable@txt  \table@txtfalse
\newif\if@dfloat    \@dfloatfalse

\def\begintable{%
  \@ifstar{\global\@dfloattrue \@btable}{\global\@dfloatfalse \@btable}%
}

\def\@btable#1{%
  \par
  \if@dfloat
    \ItemSPAN=\Double
    \TEMPDIMEN=\PageWidth
  \else
    \ItemSPAN=\Single
    \TEMPDIMEN=\ColumnWidth
  \fi
  \ifsinglecol
    \TEMPDIMEN=\PageWidth
  \else
    \ItemSTATUS=\InStack
    \ItemNUMBER=#1%
    \ItemTYPE=\Table
  \fi
  \bgroup
    \eightpoint\nostb@ls{10pt}%
    \global\setbox\ItemBOX=\vbox\bgroup
      \let\caption=\tab@caption
      \let\tabletext=\tab@txt
      \ifsinglecol \let\nofloat=\no@float\fi
}

\def\endtable{%
  \par\egroup 
  \egroup
  \setbox\TEMPBOX=\hbox to \TEMPDIMEN{%
    \hss
    \vbox{%
      \hsize=\wd\ItemBOX
      \ifvoid\tab@cap@box
      \else
        \noindent\unhbox\tab@cap@box
        \vskip 5.5pt plus 6pt%
      \fi
      \box\ItemBOX
      \iftable@txt
        \vskip 10pt%
        \eightpoint\nostb@ls{10pt}%
        \noindent\the\tab@txt@toks
        \global\table@txtfalse
      \fi
    }%
    \hss
  }%
  \ifsinglecol
    \if@here \midinsert\global\@herefalse\else \topinsert\fi
      \box\TEMPBOX
    \endinsert
  \else
    \global\setbox\ItemBOX=\box\TEMPBOX
    \JoinStack
    \Print{Processing source for table \the\ItemNUMBER}%
  \fi
}

\def\UnloadZoneA{%
\FirstZoneAtrue
 \Iteration=\z@
  \loop
   \ifnum\Iteration<\LengthOfStack
    \GetItemSTATUS{\Iteration}
    \ifnum\ItemSTATUS=\InZoneA
     \GetItemBOX{\Iteration}
     \ifFirstZoneA \vbox to \BodgeHeight{\vfil}%
     \FirstZoneAfalse\fi
     \unvbox\ItemBOX\ItemSep
     \LeaveStack{\Iteration}
     \else
     \advance\Iteration \@ne
   \fi
 \repeat
}

\def\UnloadZoneC{%
\Iteration=\z@
  \loop
   \ifnum\Iteration<\LengthOfStack
    \GetItemSTATUS{\Iteration}
    \ifnum\ItemSTATUS=\InZoneC
     \GetItemBOX{\Iteration}
     \ItemSep\unvbox\ItemBOX
     \LeaveStack{\Iteration}
     \else
     \advance\Iteration \@ne
   \fi
 \repeat
}


\def\ShowItem#1{
  {\GetItemAll{#1}
  \Print{\the#1:
  {TYPE=\ifnum\ItemTYPE=\Figure Figure\else Table\fi}
  {NUMBER=\the\ItemNUMBER}
  {SPAN=\ifnum\ItemSPAN=\Single Single\else Double\fi}
  {SIZE=\the\ItemSIZE}}}
}

\def\ShowStack{%
 \Print{}
 \Print{LengthOfStack = \the\LengthOfStack}
 \ifnum\LengthOfStack=\z@ \Print{Stack is empty}\fi
 \Iteration=\z@
 \loop
 \ifnum\Iteration<\LengthOfStack
  \ShowItem{\Iteration}
  \advance\Iteration \@ne
 \repeat
}

\def\B#1#2{%
\hbox{\vrule\kern-0.4pt\vbox to #2{%
\hrule width #1\vfill\hrule}\kern-0.4pt\vrule}
}


\newif\ifsinglecol   \singlecolfalse

\def\onecolumn{%
  \global\output={\singlecoloutput}%
  \global\hsize=\PageWidth
  \global\vsize=\PageHeight
  \global\ColumnWidth=\hsize
  \global\TextLeading=12pt
  \global\Leading=12
  \global\singlecoltrue
  \global\let\onecolumn=\relax
  \global\let\footnote=\sing@footnote
  \global\let\vfootnote=\sing@vfootnote
  \ninepoint 
  \message{(Single column)}%
}

\def\singlecoloutput{%
  \shipout\vbox{\PageHead\pagebody\PageFoot}%
  \advancepageno
  \ifplate@page
    \shipout\vbox{%
      \sp@pagetrue
      \def\sp@type{plate}%
      \global\plate@pagefalse
      \PageHead\vbox to \PageHeight{\unvbox\plt@box\vfil}\PageFoot%
    }%
    \message{[plate]}%
    \advancepageno
  \fi
  \ifnum\outputpenalty>-\@MM \else\dosupereject\fi%
}

\def\ItemSep{\vskip\ItemSepamount\relax}

\def\ItemSepbreak{\par\ifdim\lastskip<\ItemSepamount
  \removelastskip\penalty-200\ItemSep\fi%
}


\let\@@endinsert=\endinsert 

\def\endinsert{\egroup 
  \if@mid \dimen@\ht\z@ \advance\dimen@\dp\z@ \advance\dimen@12\p@
    \advance\dimen@\pagetotal \advance\dimen@-\pageshrink
    \ifdim\dimen@>\pagegoal\@midfalse\p@gefalse\fi\fi
  \if@mid \ItemSep\box\z@\ItemSepbreak
  \else\insert\topins{\penalty100 
    \splittopskip\z@skip
    \splitmaxdepth\maxdimen \floatingpenalty\z@
    \ifp@ge \dimen@\dp\z@
    \vbox to\vsize{\unvbox\z@\kern-\dimen@}
    \else \box\z@\nobreak\ItemSep\fi}\fi\endgroup%
}


\def\gobbleone#1{}
\def\gobbletwo#1#2{}
\let\footnote=\gobbletwo 
\let\vfootnote=\gobbleone

\def\sing@footnote#1{\let\@sf\empty 
  \ifhmode\edef\@sf{\spacefactor\the\spacefactor}\/\fi
  \hbox{$^{\hbox{\eightpoint #1}}$}\@sf\sing@vfootnote{#1}%
}

\def\sing@vfootnote#1{\insert\footins\bgroup\eightpoint\b@ls{9pt}%
  \interlinepenalty\interfootnotelinepenalty
  \splittopskip\ht\strutbox 
  \splitmaxdepth\dp\strutbox \floatingpenalty\@MM
  \leftskip\z@skip \rightskip\z@skip \spaceskip\z@skip \xspaceskip\z@skip
  \noindent $^{\scriptstyle\hbox{#1}}$\hskip 4pt%
    \footstrut\futurelet\next\fo@t%
}

\def\footnoterule{\kern-3\p@ \hrule height \z@ \kern 3\p@}

\skip\footins=19.5pt plus 12pt minus 1pt
\count\footins=1000
\dimen\footins=\maxdimen


\def\landscape{%
  \global\TEMPDIMEN=\PageWidth
  \global\PageWidth=\PageHeight
  \global\PageHeight=\TEMPDIMEN
  \global\let\landscape=\relax
  \onecolumn
  \message{(landscape)}%
  \raggedbottom
}


\output{%
  \ifLeftCOL
    \global\setbox\LeftBOX=\vbox to \ZoneBSize{\box255\unvbox\ZoneBBOX}%
    \global\LeftCOLfalse
    \MakeRightCol
  \else
    \setbox\RightBOX=\vbox to \ZoneBSize{\box255\unvbox\ZoneBBOX}%
    \setbox\MidBOX=\hbox{\box\LeftBOX\hskip\ColumnGap\box\RightBOX}%
    \setbox\PageBOX=\vbox to \PageHeight{%
      \UnloadZoneA\box\MidBOX\UnloadZoneC}%
    \shipout\vbox{\PageHead\box\PageBOX\PageFoot}%
    \advancepageno
    \ifplate@page
      \shipout\vbox{%
        \sp@pagetrue
        \def\sp@type{plate}%
        \global\plate@pagefalse
        \PageHead\vbox to \PageHeight{\unvbox\plt@box\vfil}\PageFoot%
      }%
      \message{[plate]}%
      \advancepageno
    \fi
    \global\LeftCOLtrue
    \CleanStack
    \MakePage
  \fi
}


\Warn{\start@mess}


\catcode `\@=12 






\newif\ifprintcomments
 
 
\printcommentsfalse

\def\xx{XXX}


\def\etal{et al.~}


\def\Msun{{\rm\,M_\odot}}
\def\Lsun{{\rm\,L_\odot}}
\def\kms{{\rm\,km\,s^{-1}}}


\def\Vrms{V_{\rm rms}}


\def\asim{\mathord{\sim}}


\def\spose#1{\hbox to 0pt{#1\hss}}
\def\lta{\mathrel{\spose{\lower 3pt\hbox{$\sim$}}
    \raise 2.0pt\hbox{$<$}}}
\def\gta{\mathrel{\spose{\lower 3pt\hbox{$\sim$}}
    \raise 2.0pt\hbox{$>$}}}


\newdimen\hssize
\hssize=8.4truecm
\newdimen\hdsize
\hdsize=17.7truecm


\def\today{\ifcase\month\or
 January\or February\or March\or April\or May\or June\or
 July\or August\or September\or October\or November\or December\fi
 \space\number\day, \number\year}
 

\newcount\eqnumber
\eqnumber=1
 
\def\new{\hbox{(\the\eqnumber )}\global\advance\eqnumber by 1}
 
\def\first{\hbox{(\the\eqnumber a)}\global\advance\eqnumber by 1}
\def\last#1{\advance\eqnumber by -1 \hbox{(\the\eqnumber#1)}\advance
     \eqnumber by 1}
 
\def\ref#1{\advance\eqnumber by -#1 \the\eqnumber
     \advance\eqnumber by #1}
 
\def\nref#1{\advance\eqnumber by -#1 \the\eqnumber
     \advance\eqnumber by #1}

\def\eqnam#1{\xdef#1{\the\eqnumber}}



\pageoffset{-0.85truecm}{0.45truecm}



\pagerange{}
\pubyear{submitted: \xx}
\volume{}

\begintopmatter

\title{Nuclear stellar discs in early-type galaxies ---\hfill\break\noindent
I.~HST and WHT observations}

\author{Frank C. van den Bosch$^{1,2}$, Walter Jaffe$^1$ and Roeland P. van 
der Marel$^{3,4}$}

\affiliation{$^1$ Sterrewacht Leiden, Postbus 9513, 2300 RA Leiden, 
             The Netherlands}
\vskip 0.2truecm
\affiliation{$^2$ Astronomy Dept., University of Washington,
P.O. Box 351580, Seattle WA 98195-1580, USA}
\vskip 0.2truecm
\affiliation{$^3$ Institute for Advanced Study, Princeton, NJ 08540, USA}
\vskip 0.2truecm
\affiliation{$^4$ Hubble Fellow}

\shortauthor{F.C.~van den Bosch, W.~Jaffe and R.P.~van der Marel}

\shorttitle{Nuclear stellar discs in early-type galaxies --- I.}


\abstract{%
We present multi-colour ($U$, $V$ and $I$) photometry obtained with the
second Wide Field and Planetary Camera (WFPC2) on the Hubble Space
Telescope (HST), and spectra taken with the William Herschel Telescope
(WHT) and the HST Faint Object Spectrograph (FOS), for the early-type
galaxies NGC~4342 and NGC~4570. These galaxies are intermediate
between ellipticals and lenticulars, and they both have a small
nuclear stellar disc in addition to their main outer disc.

Colour images reveal no colour differences between the nuclear discs
and the bulges. Comparison of the $U-V$ and $V-I$ colours with stellar
population models indicates that the central regions of both galaxies
are of intermediate age ($\asim 8$ Gyr) and of high metallicity. For
NGC~4342 this is consistent with the values of the line strengths in
the central region derived from the FOS spectra. For NGC~4570, an 
unusually large H$\beta$ line strength may suggest recent star 
formation.

The long-slit WHT spectra have a high signal-to-noise ratio
($S/N$) and a spatial resolution of $\asim 1''$. They are used to
determine the stellar rotation velocities $V$, the velocity
dispersions $\sigma$, and the deviations of the line-of-sight velocity
profiles from a Gaussian shape, as quantified by the Gauss-Hermite
moments $h_3$ and $h_4$. Both galaxies are rapidly rotating, and they
both have velocity dispersions that increase strongly towards the
centre. The $V$ and $h_3$ profiles clearly reflect radial changes
in the relative contributions of the different structural components
identified photometrically.

The FOS was used to obtain spectra with the $0.26''$ diameter circular
aperture at seven different positions in the central region of
each galaxy. Measurements of $V$ and $\sigma$ yield the stellar
kinematics at four times higher spatial resolution than available from
the WHT spectra. The FOS spectra of NGC~4342 indicate a
central velocity dispersion of $\asim 420\kms$, higher than the $\asim
320\kms$ measured from the WHT spectra. Also, the nuclear rotation
gradient measured with the FOS is steeper than that measured with the
WHT; it reaches $V_{\rm rot} \sim 200\kms$ at $0.25''$. The rapid
stellar motions seen in the centre of NGC~4342 suggest a large central
mass concentration, possibly a massive black hole. The kinematics of
the more massive NGC~4570 are less spectacular, with a central
velocity dispersion of $\sim 250 \kms$ and a central rotation curve
that reaches only $\sim 60 \kms$ at $0.25''$. The stellar kinematical
measurements for both galaxies will be interpreted quantitatively with
detailed dynamical models in a series of companion papers.
}

\keywords{%
galaxies: individual: NGC~4342, NGC~4570 --
galaxies: elliptical and lenticular, cD --
galaxies: nuclei --
galaxies: kinematics and dynamics --
galaxies: abundances
}

\maketitle  


\section{1 Introduction}

\tx The discovery with the Hubble Space Telescope (HST) of a number of 
early-type galaxies with very small stellar discs, with scale lengths
of the order of 20 pc (van den Bosch \etal 1994; Forbes 1994; Lauer
\etal 1995), has opened new windows on galaxy dynamics and
formation. From a dynamical point of view, nuclear stellar discs are
interesting because their kinematics allow an accurate determination
of the central mass density of their host galaxies (van den Bosch \&
de Zeeuw 1996). The existence of a morphologically and kinematically
distinct stellar component in the nucleus of these galaxies raises the
question whether they formed coevally with the host galaxy, or arose
from evolution of the host galaxy in a later stage. One possible form
of evolution would be gas infall to the centre induced by either a bar
or a merger, which after subsequent star-formation could result in a
nuclear disc.


\begintable{1}
\caption{{\bf Table~1.} Parameters of the observed galaxies}
\halign{#\hfil&\quad #\hfil\quad& \hfil#\hfil&
\hfil# \quad& \hfil# \quad& \hfil#\hfil\quad& #\hfil\quad \cr
NGC & RC2 & $ M_{B} $ & $D_{25}$ & $V_{\rm hel}$ &
$S_{100 \mu{\rm m}}$ & $S_{21 {\rm cm}}$ \cr
 & & & (arcsec) & (km/s) & (mJy) & (mJy) \cr
4342 & S0$^{-}$ & -17.47 & 84.8 & 714 & $0 \pm 160$ & $<3$ \cr
4570 & S0 & -19.04 & 244.4 & 1730 & $0 \pm 100$ & $<10$ \cr
}
\tabletext{Column~(1) gives the NGC number of the galaxy. Column~(2) gives
the galaxy type according to the Second Reference Catalogue (RC2; de
Vaucouleurs \etal 1976). The absolute blue magnitude (for a Virgo
distance of 15 Mpc) is listed in column~(3), whereas column~(4) gives
the major axis isophotal diameter at the surface brightness level
$\mu_{B} = 25.0$ mag arcsec$^{-2}$. Column~(5) lists the heliocentric
velocity in $\kms$. Columns~(6) and~(7) give upper limits on the IRAS
flux density at $100 \mu {\rm m}$ (Knapp \etal 1989) and on the flux
density at 21 cm (Wrobel 1991).}
\endtable


Stellar discs with sizes of 0.2--1~kpc were found in several
elliptical galaxies from ground-based observations (e.g., Nieto
\etal 1991). The nuclear discs discussed here are considerably smaller,
and have sizes $\lta 100$~pc. They were discovered in
monochromatic broad-band images that were taken with the HST Planetary
Camera (PC) before any corrections had been made for the spherical
aberration of the HST primary mirror.

To investigate the kinematics of the nuclear discs, and to learn about
their formation, we have taken spectroscopic and improved photometric
data for two early-type galaxies in the Virgo cluster: NGC~4342 and
NGC~4570. These galaxies are intermediate between ellipticals and
lenticulars, and in both cases previous HST images revealed the
presence of bright nuclear discs (van den Bosch \etal 1994). Global
parameters of the two galaxies are listed in Table~1.

We obtained multi-colour ($U$,$V$,$I$) images with the second Wide
Field and Planetary Camera (WFPC2) aboard the HST, in order to
investigate the detailed morphology of the different components in
NGC~4342 and NGC~4570, and to study their differences in colour. We
also obtained long-slit spectra with the William Herschel Telescope
(WHT), and higher spatial-resolution single-aperture spectra with the
HST Faint Object Spectrograph (FOS), to determine the stellar
kinematics of the different components and investigate the stellar
populations. The HST data were obtained after the telescope was
serviced to correct for the spherical aberration of the primary
mirror.


\begintable{2}
\caption{{\bf Table~2.} Log of HST/WFPC2 observations}
\halign{#\hfil&\quad \hfil#\hfil\quad& \hfil#\hfil\quad&
 \hfil#\hfil\quad& \hfil#\hfil\quad& \hfil#\hfil\quad \cr
NGC & Filter & colour & date & $t_{\rm exp}$ & \# exp \cr
 & & & & (sec) & \cr
4342 & F336W & U & 21/01/96 & 3600 & 4 \cr
     & F555W & V & 21/01/96 &  200 & 2 \cr
     & F814W & I & 21/01/96 &  200 & 2 \cr
4570 & F336W & U & 19/04/96 & 5100 & 5 \cr
     & F555W & V & 19/04/96 &  400 & 2 \cr
     & F814W & I & 19/04/96 &  460 & 2 \cr
}
\tabletext{Column~(4) lists the total exposure time per filter. 
Column~(5) gives the number of exposures per filter.}
\endtable


The nuclear discs have a small angular size, even for HST standards,
and are embedded in complex larger structures. Reliable astrophysical
interpretation in terms of dynamics and stellar populations therefore
requires careful reduction and modeling. In this paper we describe the
observations, the reduction, and the parameterization of the
results. The quantity and diversity of the data has led us to
parameterize it into a form suitable for presentation and
modeling. The imaging data is parameterized in terms of both
elliptical isophotal parameters and multi-Gaussian decompositions; the
kinematic data is parameterized through Gauss-Hermite expansions of the
line-of-sight velocity distributions.

A detailed interpretation of the data will be presented in a series of
companion papers. Scorza \& van den Bosch (1997) will discuss the
results of the decomposition of both galaxies into bulge and disc
components. Cretton \& van den Bosch (1997) will present detailed
three-integral modeling of the dynamics of NGC~4342. Van den Bosch \&
Emsellem (1997) will present evidence that the galaxy NGC~4570 has
been shaped under the influence of a rapidly tumbling bar potential.

In Section~2 we present the reduction and parameterization of the
multi-colour HST photometry. Sections~3 and~4 contain the reduction of
the WHT and HST/FOS spectra, respectively. The stellar kinematical
analysis is presented in Section~5. Section~6 discusses the stellar
populations of both galaxies, based on the broad-band colour and line
strength data. We summarize and discuss our conclusions in
Section~7. Throughout this paper we adopt a distance of 15 Mpc for
both galaxies, as appropriate for the Virgo cluster (Jacoby, Ciardullo
\& Ford 1990).

\section{2 HST multi-colour photometry}

In this section we describe the observations (Section~2.1), the
reduction (Section~2.2) and the parameterization of the broad-band
WFPC2 images. We derive multi-Gaussian decompositions (Section~2.3)
and standard isophotal parameters (Sections~2.4 and~2.5). We also
subtract a pure elliptical model from the observations to emphasize
the disc structures (Figure~6).

\subsection{2.1 Observations}

We obtained $U$, $V$ and $I$ band images of NGC~4342 and NGC~4570
using the HST/WFPC2 as part of our GO-project
\#6107. A detailed description of the WFPC2 can be found in the HST
WFPC2 Instrument Handbook (Burrows \etal 1995).  The nuclei of the
galaxies were centred in the Planetary Camera chip (PC1), which
consists of $800 \times 800$ pixels of $0.0455'' \times 0.0455''$
each. Exposures were taken with the broad band filters F336W, F555W
and F814W; these correspond closely to the Johnson~$U$ and~$V$ bands,
and the Cousins~$I$ band, respectively. In each band several separate
exposures were taken. Table~2 lists the log of the observations. All
exposures were taken with the telescope guiding in fine lock, yielding
a nominal pointing stability of $\asim 3$~mas. Since there was no
danger of saturation, the analogue-to-digital gain was set to its low
setting of 7.12 electrons/DN (where DN is the number of counts).  The
CCD read-out noise was 5.24 electrons; the dark rate was only 0.003
electrons pixel$^{-1}$sec$^{-1}$.

\subsection{2.2 Reduction}

The images were calibrated with the standard `pipeline' that is
maintained by the Space Telescope Science Institute (STScI). The
reduction steps, including e.g., bias subtraction, dark current
subtraction and flat-fielding, are described in detail by Holtzman
\etal (1995a).

Subsequent reduction was done using standard {\tt IRAF} tasks.  With
the $V$- and $I$-band filters, two separate frames were obtained for
each galaxy within one HST orbit. These frames were offset from each
other by $11 \times 11$ pixels. This allows removal of chip defects,
such as hot pixels and bad columns, as well as cosmic rays.  After
shifting over an integer $11 \times 11$ pixels, we checked the
alignment of the two frames by measuring the positions of a number of
globular clusters present on the PC1. The alignment was found to be
better than 0.05 pixels. For the $U$-band images, 4-5 separate
exposures were available with different exposure times. Frames
obtained in different orbits where offset from each other by an
integer $11 \times 11$ pixels. Once again we used globular clusters to
determine the offsets, and found them to be accurate at the 0.05-pixel
level. Therefore, we could align the frames by shifting over integer
pixels, without the need for interpolation. Registered frames for the
same galaxy and filter were combined with cosmic-ray rejection. A
constant background was subtracted from all combined frames, as
measured at the boundaries of the WF2 CCD, where the galactic
contribution is negligible.


\beginfigure{1}
\centerline{\psfig{figure=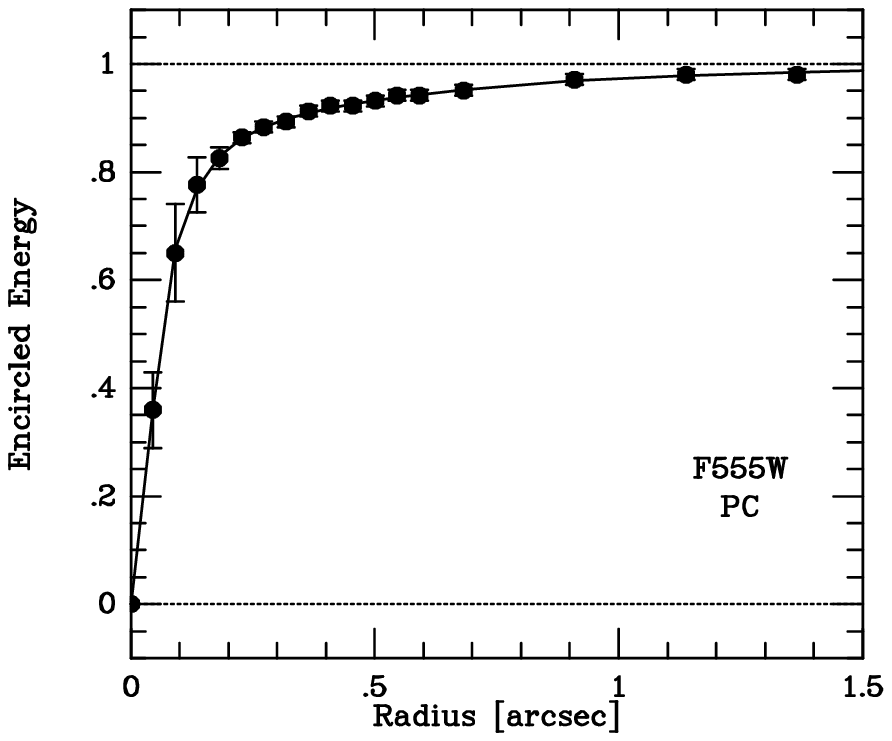,width=\hssize}}\smallskip
\caption{{\bf Figure~1.} The dots show measurements of the 
encircled energy of the PSF, as a function of radius, for the HST
Planetary Camera (PC) with the F555W filter. The solid line is the
encircled energy of the model PSF used in the MGE method. This is 
a sum of 5 circular Gaussians.}
\endfigure



\beginfigure*{2}
\centerline{\psfig{figure=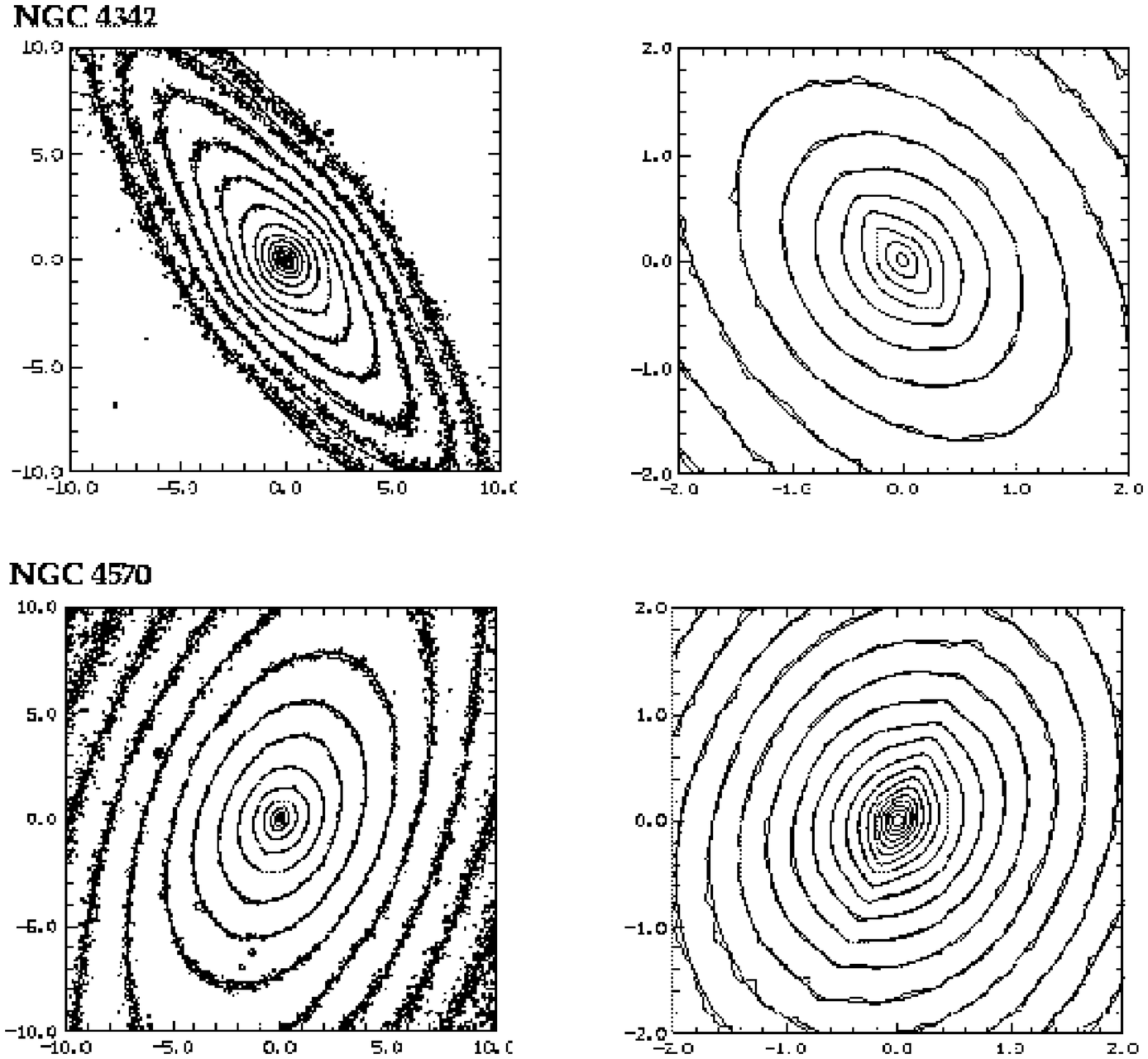,width=\hdsize,clip=}}
\smallskip
\caption{{\bf Figure~2.} Contour maps of the WFPC2 $V$-Band images of NGC~4342
and NGC~4570 (without PSF deconvolution), at two different
scales: $20''\times 20''$ (plots on the left) and $4'' \times 4''$
(plots on the right). Contours of the best-fitting
MGE models are superimposed.}
\endfigure


In order to convert the raw counts in the F336W, F555W and F814W
frames to Johnson $U$ and $V$, and Cousins $I$ band magnitudes,
respectively, we performed a flux calibration following the guidelines
given by Holtzman \etal (1995b). The equations that convert counts to
$U$, $V$ and $I$ surface brightness magnitudes include $U-V$ and $V-I$
colour terms. We approximated those by the average values found for
ellipticals. After the photometric calibration, we calculated the
$U-V$ and $V-I$ colours from our images, and iterated the calibration
until the colours had converged.

For the $U$-band, two further corrections are required.  First,
photometry in the UV is sensitive to the presence of contaminants on
the CCD. There is a linear behaviour between the light being lost due
to those contaminants and the day since the last decontamination of
the CCD. We used the formula of Holtzman \etal (1995b), and applied
corrections of 0.0070 and 0.0119 magnitudes to the zero-points of the
F336W images of NGC~4342 and NGC~4570, respectively.  Secondly, the UV
filters have a considerable red leak. Table~6.5 in the WFPC2
Instrument Handbook gives an estimate for the percentage of red light
leaking through the filter for a number of stellar spectra. Since
early-type galaxies consist mainly of late-type stars, we estimate
that 2--15\% of the light falling through the F336W filter is coming
from wavelengths around 7500{\AA}. We assumed that 8\% of the flux
through the F336W filter is due to the red leak, and increased the
U-band photometric zero-point by 0.0905 magnitudes. We estimate
that our final photometric accuracies are $\lta 0.02$ magnitudes for
the $V$- and $I$-band, and $\lta 0.08$ magnitudes for the $U$-band
(mainly due to uncertainties in the amount of red leak).

\subsection{2.3 Multi-Gaussian fitting}

The HST point-spread-function (PSF) has improved significantly with
the 1993 refurbishment mission. Figure~1 shows measurements of the
encircled energy curve for the F555W filter and the PC CCD (Holtzman
\etal 1994). The FWHM of the PSF is only $\asim 0.1''$. However, the
PSF wings are still very broad; several percent of the light is
scattered more than 1 arcsec away.  Since the luminosity profiles of
NGC 4342 and NGC 4570 are strongly peaked towards the centre, PSF
convolution still has a considerable degrading effect. Deconvolution
therefore remains essential to obtain the maximum amount of
information from our images. We have explored two methods of PSF
deconvolution: the Multi Gaussian Expansion (MGE) method (this
section), and direct Richardson-Lucy deconvolution (Section~2.4).

The MGE method was developed by Monnet, Bacon \& Emsellem (1992). It
builds a model for the galaxy, while deconvolving for the effects of
PSF convolution at the same time. The method assumes that both the PSF
and the deconvolved (i.e., intrinsic) surface brightness distribution
of the galaxy can be approximated by a sum of Gaussians. Each Gaussian
is parameterized by 6 parameters: the centre $(x_i,y_i)$, the position
angle, the flattening $q_i$, the standard deviation $\sigma_i$, and
the central intensity $I_i$. We approximated the HST $V$-band PSF as
the sum of 5 circular (i.e., $q=1$) Gaussians, chosen so as to fit the
observed encircled energy curve (Figure~1). Using this PSF model we
derived the parameters of the $N$ Gaussians that describe the {\it
deconvolved} surface brightness, by fitting to the HST $V$-band galaxy
images. Since both the PSF and the deconvolved surface brightness are
assumed to be sums of Gaussians, the convolution is analytical. In the
fitting we forced each of the $N$ Gaussians to have the same position
angle and centre (i.e., the MGE model is assumed to be
axisymmetric). Therefore, the model is described by $3N + 3$ free
parameters, which are fit simultaneously to the images using a global
bidimensional fitting process. More and more components are added
until convergence is achieved (for details on the MGE method, see
Emsellem, Monnet \& Bacon, 1994). We found our fits to converge for
$N=11$ Gaussian components, for both NGC~4342 and NGC~4570.

The results are shown in Figure~2, which displays contour maps
of the HST $V$-band images, with superimposed the best fitting
MGE-models. In general the fits are excellent. For NGC~4342 a small
discrepancy is seen on the outside. This is due to the fact that the
isophotes of this galaxy twist slightly at large radii (see
Section~2.5), which we ignore by forcing the position angles of the
different Gaussians to be the same. In both galaxies there is a clear
multi-component structure: the isophotes are highly flattened and
discy at the outside (due to the outer disc), less flattened at
intermediate radii $(r \approx 1'')$, where the bulge is dominating the
light, and again very elongated and discy close to the centre, due to
the nuclear disc.  The parameters of the MGE models are listed
in Table~3.


\begintable*{3}
\caption{{\bf Table~3.} Parameters of MGE models for the 
deconvolved $V$-band surface brightness}
\halign{#\hfil&\quad \hfil# \quad& \hfil# \quad& \hfil# \quad&
\hfil#\hfill \quad& \hfil# \quad& \hfil# \quad& \hfil# \quad \cr
    & NGC 4342 &&&& NGC 4570 &&\cr
$i$ & \hfill$I_i$\hfill & \hfill$\sigma_i$\hfill & \hfill$q_i$\hfill & & 
\hfill$I_i$\hfill & \hfill$\sigma_i$\hfill & \hfill$q_i$\hfill \cr
    & ($\Lsun {\rm pc}^{-2}$) & (arcsec) & & & ($\Lsun {\rm pc}^{-2}$) 
    & (arcsec) & \cr
1 & 3136240.0 & 0.02 & 0.119 & & 1755160.0 &  0.02 & 0.158 \cr
2 &   95319.8 & 0.08 & 0.841 & &   61238.0 &  0.09 & 0.800 \cr
3 &   42954.3 & 0.26 & 0.632 & &   21526.4 &  0.23 & 0.748 \cr
4 &   48520.9 & 0.36 & 0.136 & &   21589.3 &  0.26 & 0.140 \cr
5 &   17155.4 & 0.42 & 0.848 & &   11285.3 &  0.51 & 0.780 \cr
6 &    4930.9 & 0.72 & 0.521 & &    5728.7 &  0.60 & 0.120 \cr
7 &    8657.3 & 0.79 & 0.840 & &    7911.9 &  1.11 & 0.809 \cr
8 &    3207.9 & 1.80 & 0.759 & &    3800.5 &  2.73 & 0.635 \cr
9 &    2154.3 & 3.89 & 0.275 & &    1624.8 &  4.20 & 0.700 \cr
10&    1085.9 & 9.11 & 0.270 & &    1095.8 & 12.88 & 0.350 \cr
11&     219.1 & 9.61 & 0.836 & &     334.4 & 17.12 & 0.583 \cr
}
\tabletext{Column (1) gives the index number of each Gaussian.
Columns~(2) and~(5) give its central surface brightness,
columns~(3) and~(6) give its standard deviation, and columns~(4) and~(7)
give its flattening.}
\endtable



\beginfigure*{3}
\centerline{\psfig{figure=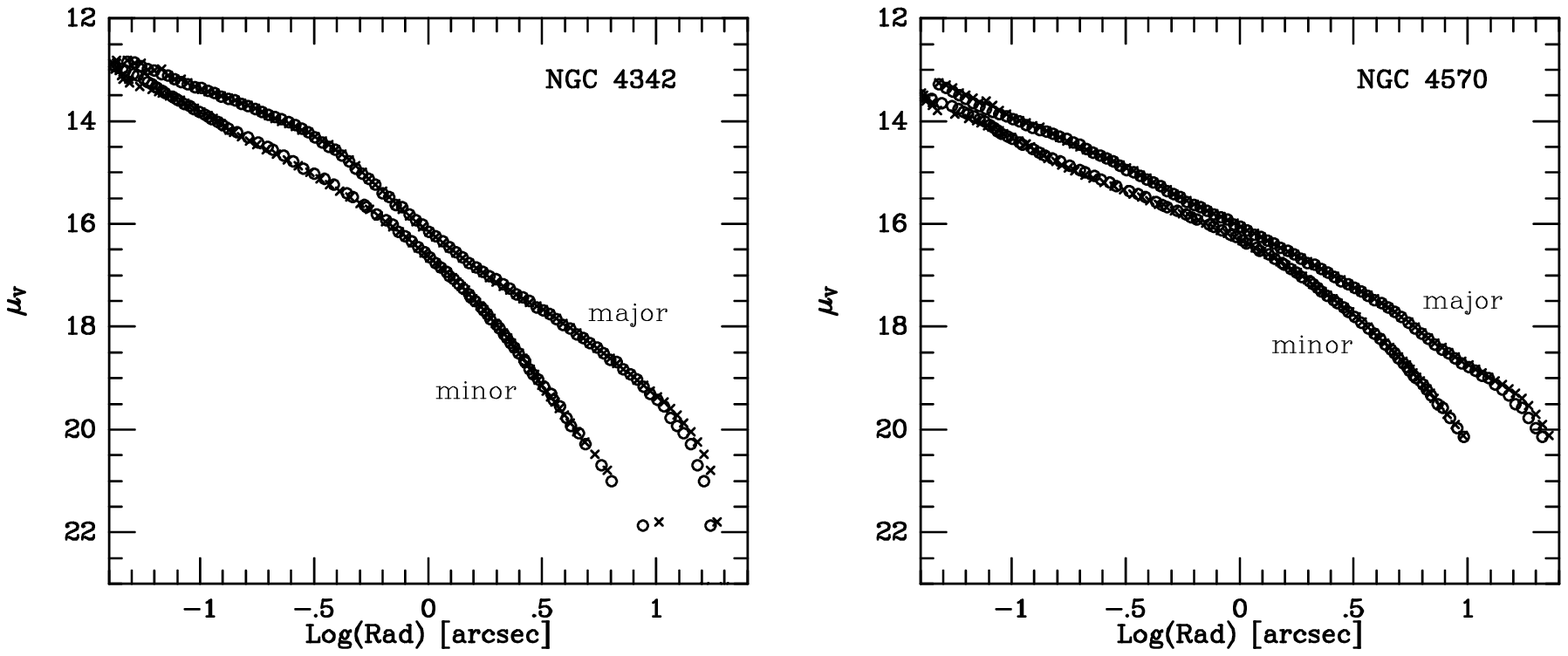,width=\hdsize}}\smallskip
\caption{{\bf Figure~3.} The projected $V$-band surface brightness profiles
(in mag arcsec$^{-2}$) of the Lucy-deconvolved (open circles) and
MGE-deconvolved (crosses) images of NGC~4342 and NGC~4570. 
Agreement between the two methods of deconvolution is excellent.
Profiles along both the major and the minor axes are shown.
The excess light along the major axes clearly reveals the
`double-disc' structure in both galaxies. The minor axis profiles show
that the bulges have a luminosity profile similar to that of
low-luminosity elliptical galaxies, with very steep cusps.}
\endfigure


\subsection{2.4 Luminosity profiles}

Richardson-Lucy iteration (Lucy 1974) provides an alternative PSF
deconvolution method. For this, accurate knowledge of the PSF is
required. We calculated model PSFs appropriate for each given filter
and position of the nucleus on the PC1 CCD, using the TinyTim software
package. Since our observations were made while guiding in fine
lock, no corrections for telescope jitter were necessary.

We used the Richardson-Lucy algorithm to deconvolve the $V$-band
images of NGC~4342 and NGC~4570; 20 iterations were found to be
sufficient for convergence. We subsequently derived the luminosity
profiles along the major and minor axes, using the isophote fitting
procedure described below (Section~2.5). The resulting $V$-band
surface brightness profiles are shown in Figure~3 (open cicrles). 
The difference
between the major and minor axis profiles clearly reveals the excess
light due to the nuclear and outer disc components. The minor axis
profiles, which have a negligible contribution of disc light, show a
double power-law behaviour for the bulge luminosity distribution, with
a steep cusp. Such profiles are characteristic for low luminosity
ellipticals (Gebhardt \etal 1996). The bulge luminosity profiles
will be further discussed by Scorza \& van den Bosch (1997).

The crosses in Figure~3 correspond to the same luminosity profiles but
now determined from the MGE model of the intrinsic (i.e., deconvolved)
surface brightness, again using the isophote-fitting procedure. 
The agreement with the luminosity profiles derived from the Lucy-deconvolved 
images is excellent. Small discrepancies can be seen at the outside. These 
are related to the discrepancies seen in Figure~2 and originate from
neglecting the small amount of isophote twisting, when constructing the MGE 
models.


\beginfigure*{4}
\centerline{\psfig{figure=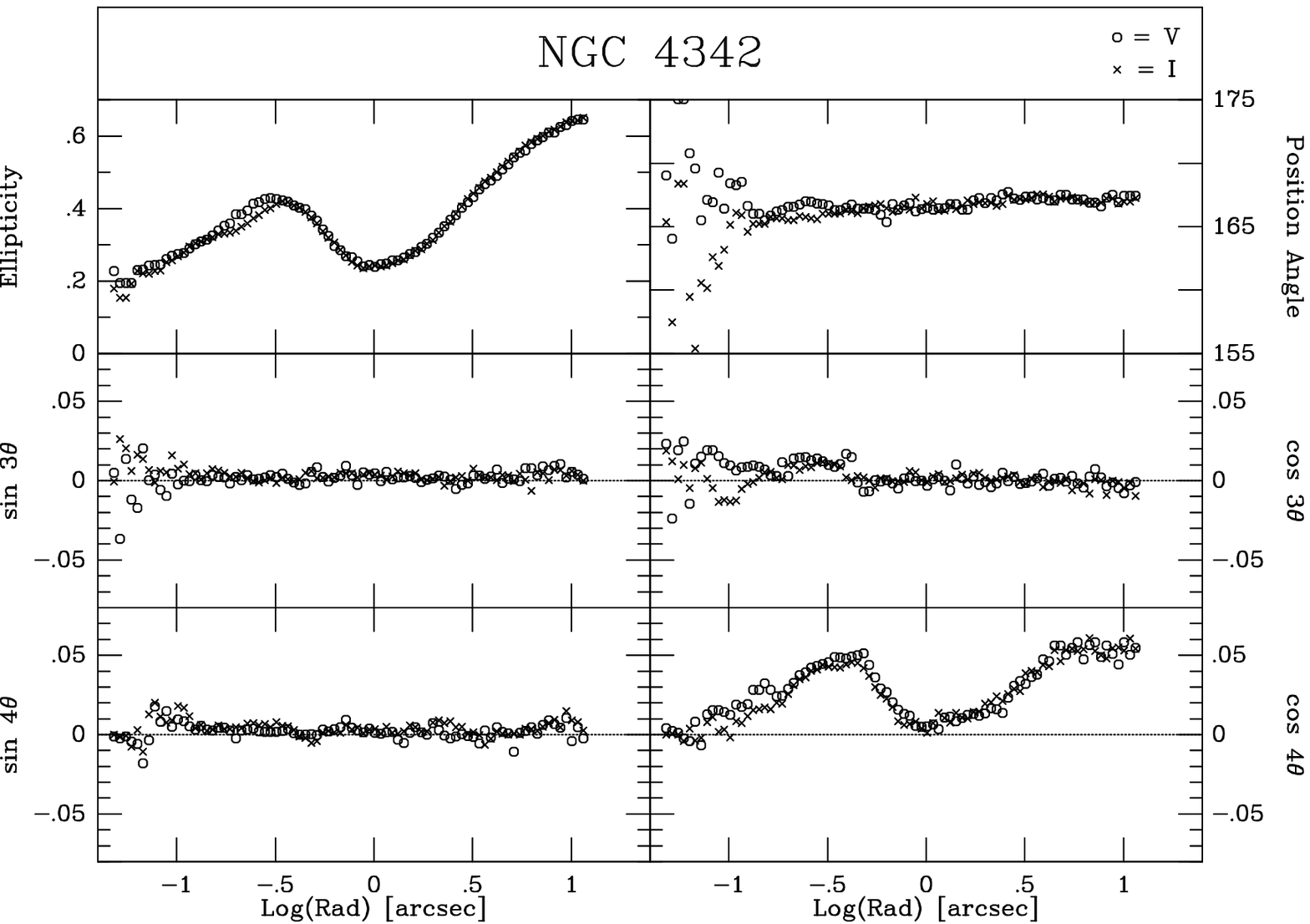,width=0.8\hdsize}}\smallskip
\caption{{\bf Figure~4.} The isophotal parameters as a function 
of log(radius), for the $V$- and $I$-band images of NGC 4342. Both the
ellipticity and the $\cos 4\theta$-term clearly reveal the double-disc
structure.}
\endfigure



\beginfigure*{5}
\centerline{\psfig{figure=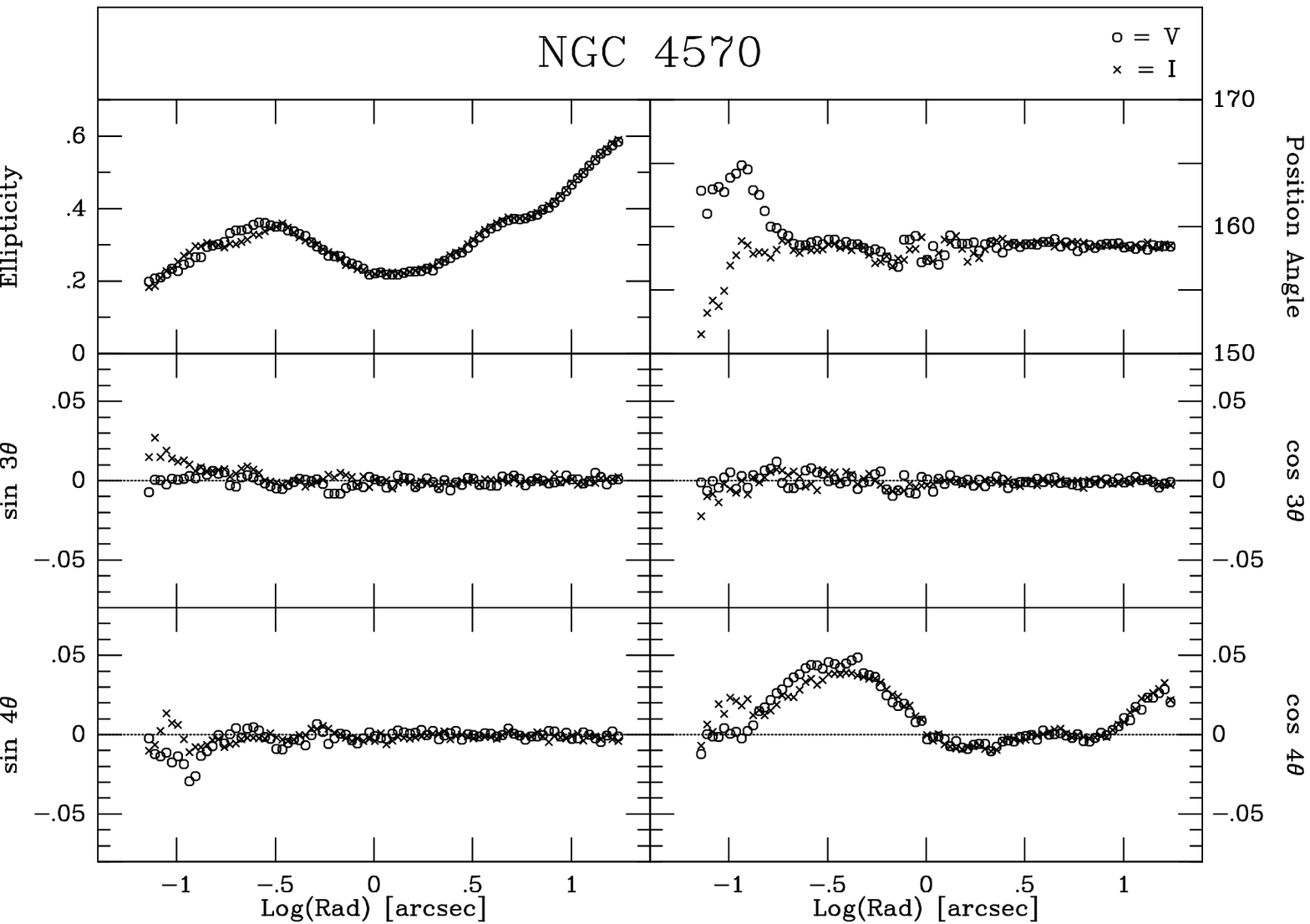,width=0.8\hdsize}}\smallskip
\caption{{\bf Figure~5.} Same as Figure~4, but now for NGC 4570.}
\endfigure



\beginfigure*{6}
\line{\psfig{figure=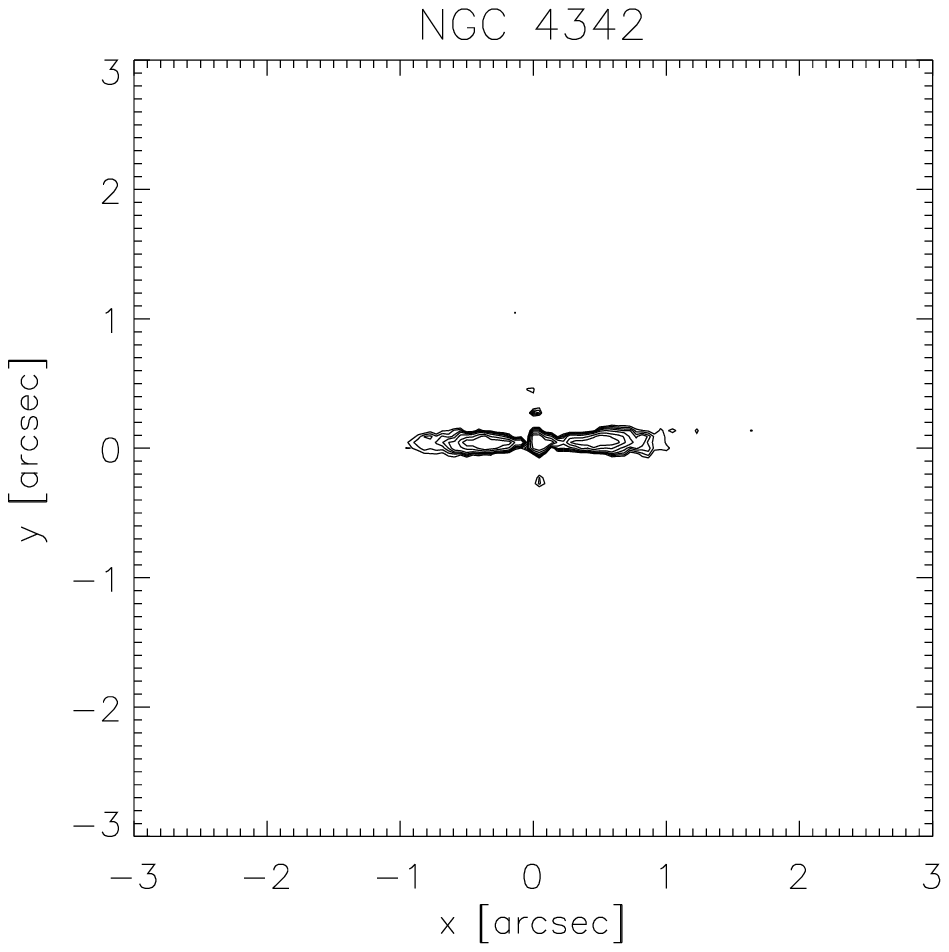,width=\hssize}\hfill
\psfig{figure=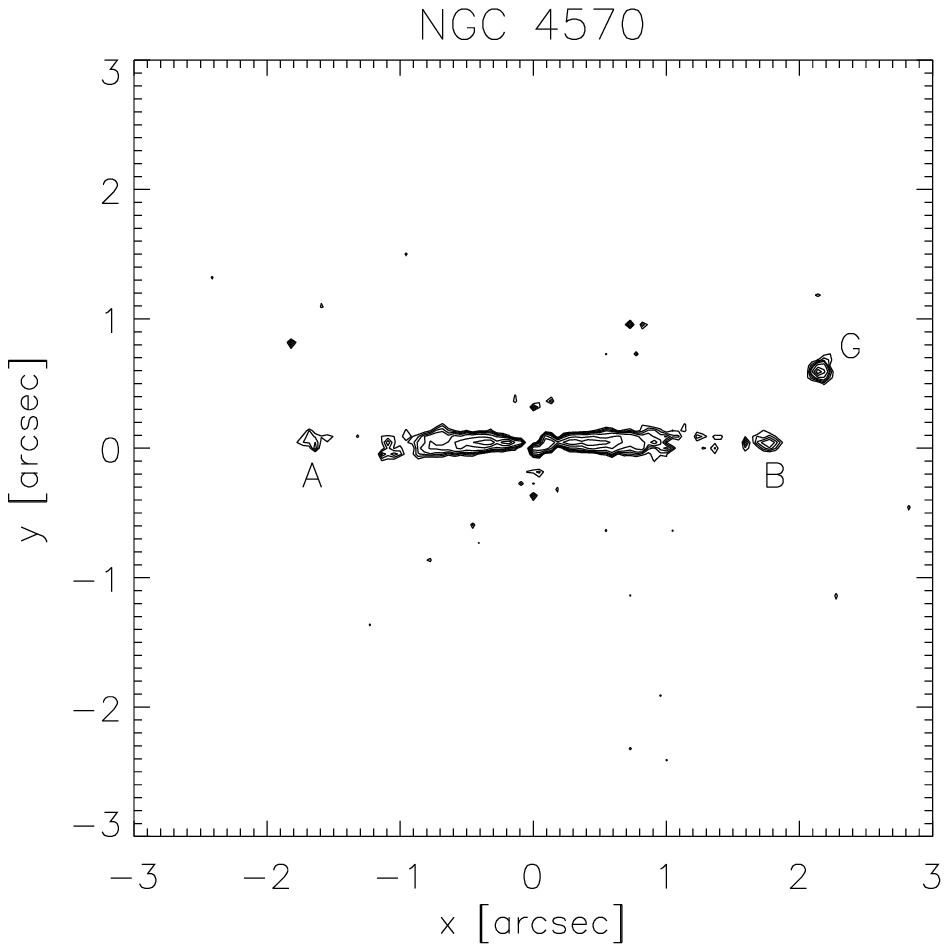,width=\hssize}}\smallskip
\caption{{\bf Figure~6.} Contour maps of the $V$-band residual images 
of NGC~4342 (left) and NGC 4570 (right), obtained after subtraction of
perfectly elliptical galaxy models. The images were rotated to align
the major axis of each galaxy with the $x$-axis. Highly flattened
nuclear discs are clearly visible inside the central arcsec. Only
positive contours are plotted for clarity. The apparent central point
sources are artifacts, due to the limited radial extent of the galaxy
models that were subtracted. The features `A' and `B' in NGC~4570 are
both at $1.7''$ from the centre, and will be discussed elsewhere. The
feature marked `G' is a globular cluster.}
\endfigure


\subsection{2.5 Isophotal analysis}

We derived the ellipticity and position angle of the isophotes, as a
function of radius, for each colour and each galaxy, from the
non-deconvolved images. In addition, the sin and cos $3\theta$
and $4\theta$ terms were derived that describe the high-order
deviations of the isophotes from pure ellipses (e.g., Lauer 1985;
Jedrzejewski 1987; Bender, D\"obereiner \& M\"ollenhoff 1988). For
a pure ellipse, these coefficients are all equal to
zero. Positive $\cos 4\theta$ terms correspond to `discy' isophotes,
whereas `boxy' isophotes give rise to negative values of the $\cos
4\theta$ term.

The results for the $V$- and $I$-band images are shown in Figures~4
and~5. Again, the double-disk structure of both galaxies is clear:
the isophotes are highly flattened and discy at the outside ($r >
1''$), moderately flattened and nicely elliptical at intermediate
radii ($r \approx 1''$), and again strongly elongated and discy inside
$1''$. At radii inside $\asim 0.5''$, the measured parameters are
not a reliable representation of their intrinsic values, due to
the convolution with the HST PSF. 

For both galaxies, there is almost
no difference between the $V$- and $I$-band parameters. The same
appears to be true for the $U$-band parameters (not plotted here),
although these are noisier due to lower $S/N$.
Except for the $\cos 4\theta$ term, all other high-order
terms that express deviations from elliptical isophotes are close
to zero. For both galaxies the position angle is close to
constant, although in NGC~4342 there is a mild, but significant twist
of a few degrees. Comparison with the isophotal parameters derived
from Lucy deconvolved F555W images obtained with the PC before the HST
refurbishment (van den Bosch \etal 1994) generally shows good
agreement, with one exception. The pre-refurbishment images
revealed strange `wiggles' in the higher order terms of the Fourier
expansion, interpreted by van den Bosch \etal (1994) as indicative of
a patchy dust distribution. However, from the WFPC2 images presented
here no such evidence for dust is found. It therefore seems likely
that the pre-refurbishment images suffered from insufficiently
corrected measles due to contaminants.

Michard (1994) showed that the inner isophotes of strongly flattened
galaxies containing a sharp central feature are distorted in the
process of convolution and subsequent deconvolution, in a way that can
mimic the presence of a nuclear disc. Michard therefore suggested that
the nuclear discs inferred from isophotal analysis of deconvolved
images could be merely an artifact of the deconvolution procedure. Van
den Bosch \etal (1994) found similar deconvolution induced distortions
from tests performed on model galaxies and concluded that the
isophotal parameters inside $0.5''$ (e.g., five times the FWHM of the
PSF) indeed could not be trusted. Their evidence for the presence of
nuclear discs was therefore based solely on the photometry outside
$0.5''$. The new data presented here clearly shows the nuclear discs
in NGC~4342 and NGC~4570 even in data that are {\it not} deconvolved
(Figures~2, 4 and~5). This proves incontrovertibly that the nuclear
discs are real structures.

In order to reveal more clearly the nuclear discs, we constructed
residual $V$-band images by subtracting a model image that has the
same luminosity and ellipticity profile as the real image, but is
taken to have perfectly elliptical isophotes. These residual images
reveal the structures that are responsible for the higher-order
deviations from perfectly elliptical isophotes. Contour maps of these
images are shown in Figure~6. We only show the inner $3''
\times 3''$ regions where the nuclear discs clearly stand out. The
models that were subtracted from the images are based on the isophotal
parameters outside $0.1''$; inside that radius no meaningful isophotes
can be fitted. As a resulting artifact, the residual images show a
central point source in addition to the nuclear discs.  The residual
image of NGC~4570 reveals, besides the nuclear disc, two unresolved
features, marked `A' and `B'. These features are both at $1.7''$
offset from the centre, and are perfectly aligned with the nuclear
disc. The nature of these features is discussed in van den Bosch \&
Emsellem (1997).  The feature marked `G' is a globular cluster.


\begintable{4}
\caption{{\bf Table~4.} Log of long-slit WHT observations}
\halign{#\hfil&\quad \hfil#\hfil\quad& \hfil#\hfil\quad&
\hfil#\hfil\quad& \hfil#\hfil\quad& \hfil#\hfil\quad& 
\hfil#\hfil\quad& \hfil#\hfil \cr
NGC & slit pos. & PA & slit & $S$ & exp & airmass & off \cr
 & & & ($''$) & ($''$) & (min) & & ($''$) \cr
4342 & major axis & 347 & 1.0 & 0.80 & 90 & 1.10 & 0.15 \cr
         & minor axis & 257 & 1.0 & 0.95 & 90 & 1.52 & 0.24 \cr
4570 & major axis & 159 & 1.0 & 1.10 & 80 & 1.12 & -0.22 \cr
         & offset axis & 159 & 1.0 & 1.05 & 70 & 1.38 & 0.94 \cr
         & minor axis & 249 & 1.0 & 1.70 & 60 & 1.19 & 0.17 \cr
}
\tabletext{Column~(3) gives the position angle of the slit in degrees. 
The slit width is given in column~(4). Column~(5) gives the seeing
FWHM $S$, defined as in Section~3.3. The exposure time is
given in column~(6), and the airmass during each exposure in
column~(7). Column~(8) gives the offset of the slit from the centre,
corrected for differential atmospheric refraction.}
\endtable


\section{3 Ground-based spectra}

We obtained high-$S/N$, ground-based, long-slit spectra of NGC~4342
and NGC~4570, with a spatial resolution of $\asim 1''$, using the 4.2m
WHT on La Palma. In this section we describe the observations
(Section~3.1), the reduction (Section~3.2), the seeing PSF of the
observations (Section~3.3), and our observations of a library of
template stars for use in the kinematical analysis (Section~3.4).

\subsection{3.1 Observations}

The observations were done in March 1994 with the WHT/ISIS
spectrograph. With ISIS two spectra are obtained simultaneously, on
the red and blue arms of the spectrograph. On both arms we used the
high resolution gratings with 1200 lines/mm. The red arm was centred
around the Ca II triplet (8498, 8542, 8662{\AA}), while the blue arm
covered the Mg $b$ triplet (5167, 5173, 5184{\AA}).  The blue spectra
were irreparably disturbed by an internal reflection (`ghost') and
will not be further discussed.

On the red arm we used a Tek CCD of $1124\times 1124$ pixels.  Each
pixel measures $0.36''$ by 0.41{\AA}.  All spectra were obtained with
a slit width of $1''$, which is roughly equal to the average
seeing. The instrumental resolution, expressed as the Gaussian
dispersion of spectral lines in the arc lamp frames, was $9 \kms$.

The galaxy exposures were split into consecutive exposures of
typically 20-30 minutes. Before each galaxy exposure we took exposures
of arc lamps to allow accurate wavelength calibration. Bias frames,
tungsten lamp flats and sky flats were taken during twilight.
We also took spectra of a spectro-photometric standard, to allow
correction for the wavelength sensitivity of the CCD (see
Section~3.2). A log of the galaxy observations is given in Table~4.

Guiding during each exposure was done with a TV camera with a Johnson
$V$-band filter. Differential atmospheric refraction can play an
important role, because the $V$-band central wavelength ($\lambda_{\rm
cen} = 5500${\AA}) is offset considerably from the central wavelength
of our red spectra ($\lambda_{\rm cen} = 8580${\AA}). This results in
an offset of the slit from the intended position on the galaxy that
was selected with the TV camera. These offsets can be calculated and
are listed in Table~4. Note that for the `offset exposure' of NGC~4570
we aimed for an intentional offset of $1.5''$, perpendicular to the
major axis. However, due to the atmospheric refraction the real offset
only amounted to $0.94''$. Differences in atmospheric refraction over
the observed spectral range (8390--8770{\AA}) are at most $0.03''$,
and can be neglected.

\subsection{3.2 Reduction}

All spectra where reduced using {\tt IRAF}. The bias level was
determined from the overscan columns and subtracted.  For each night
two flat-field frames were created. The tungsten flats were used to
create one high-S/N flat-field that shows the pixel-to-pixel
sensitivity variations of the CCD. The spectra of the twilight sky
were used to construct a high-S/N flat-field that shows the large scale
illumination pattern due to vignetting of the slit. Both flat-fields
were normalized and divided into all spectra.

Cosmic rays were removed from all frames by interpolating over
high-$\sigma$ deviations, as judged from Poisson statistics and the
known gain and read-out noise of the detector. The wavelength
calibration was done using the arc-lamp frames. Spectra were
rebinned using the resulting wavelength solution, both in the
spatial direction (to align the direction of dispersion with the rows
of the frames), and in logarithmic wavelength. The latter was
done to a scale of $11.076 \kms/{\rm pixel}$, covering the
wavelength range from 8390{\AA} to 8770{\AA}. Subsequently, we
determined the sensitivity as a function of wavelength from the
spectra of the spectro-photometric standards, and corrected all
spectra for these sensitivity variations. All exposures of the
same galaxy and slit position were added. Sky spectra were
determined from the data beyond $90''$ from the centre of the slit,
and were subtracted. Template star frames (see Section~3.4), were
summed along columns to yield one high-S/N spectrum for each star.


\begintable*{5}
\caption{{\bf Table~5.} HST/FOS observations: log and kinematical results}
\halign{#\hfil&\quad \hfil#\hfil\quad& \hfil# \quad& \hfil# \quad&
\hfil# \quad& \hfil# \quad& \hfil# \quad& \hfil# \quad& 
\hfil# \quad& \hfil# \quad& \hfil# \quad& \hfil# \quad\cr
id. & Galaxy & $x_{\rm ap}$ & $y_{\rm ap}$ & $t_{\rm exp}$ & Intensity & 
$\gamma$ & $\Delta\gamma$ & $V_{\rm rot}$ & $\Delta V$ & $\sigma$ & 
$\Delta\sigma$ \cr
 & & (arcsec) & (arcsec) & (sec) & (counts/sec) & & & ($\kms$) & 
($\kms$) & ($\kms$) & ($\kms$) \cr
 A1 & NGC 4342 & $-0.014$ & $-0.010$ & 1000 &$1421.1$ & 1.132 & 0.079 & 
 --0.2 & 30.8 & 418.3 & 32.7 \cr
 A2 &          & $+0.236$ & $-0.010$ & 1190 & $746.3$ & 1.165 & 0.091 &
+188.2 & 25.5 & 321.1 & 33.3 \cr
 A3 &          & $-0.264$ & $-0.010$ & 1200 & $714.4$ & 1.281 & 0.089 & 
--202.1 & 21.1 & 308.2 & 26.8 \cr
 A4 &          & $+0.486$ & $-0.010$ &  990 & $407.6$ & 1.025 & 0.174 & 
+239.8 & 55.0 & 312.7 & 53.5 \cr
 A5 &          & $-0.514$ & $-0.010$ &  900 & $384.3$ & 1.285 & 0.152 &
--253.7 & 25.1 & 237.5$^{*}$ & 23.9 \cr
 A6 &          & $+0.098$ & $+0.214$ &  500 & $456.6$ & 1.447 & 0.185 & 
+111.0 & 49.0 & 327.2 & 50.2 \cr
 A7 &          & $-0.126$ & $-0.234$ &  570 & $521.4$ & 1.152 & 0.177 &
--185.2 & 73.0 & 391.9 & 101.0 \cr
    &          &          &          &      &         &       &       &
       &      &       &      \cr 
 B1 & NGC 4570 & $-0.011$ & $-0.052$ &  810 & $866.5$ & 1.147 & 0.082 &
+35.8 & 25.0 & 249.4 & 31.9 \cr
 B2 &          & $+0.239$ & $-0.052$ &  450 & $434.9$ & 1.335 & 0.158 &
 --4.1 & 31.7 & 217.6 & 38.0 \cr
 B3 &          & $-0.261$ & $-0.052$ &  450 & $446.6$ & 1.323 & 0.162 &
 --119.7 & 41.0 & 205.8 & 38.9 \cr
 B4 &          & $+0.489$ & $-0.052$ &  450 & $273.7$ & 0.990 & 0.168 &
+104.2 & 27.7 & 121.7 & 46.9 \cr
 B5 &          & $-0.511$ & $-0.052$ &  450 & $267.3$ & 1.073 & 0.159 &
--108.9 & 39.9 & 170.8$^{*}$ & 20.6 \cr
 B6 &          & $-0.011$ & $+0.198$ &  450 & $360.8$ & 1.279 & 0.162 &
 +79.3 & 34.2 & 183.3 & 41.5 \cr
 B7 &          & $-0.011$ & $-0.302$ &  450 & $306.3$ & 1.466 & 0.230 &
+102.4 & 78.5 & 332.9 & 78.2 \cr
}
\tabletext{Column~(1) gives the label for the spectrum used in the remainder 
of the paper. Columns~(3) and~(4) give the aperture position in a
$(x,y)$ coordinate system centred on the galaxy, and with $x$ along
the major axis. Column (5) gives the exposure time. In Column (6) we
list the observed intensity, integrated over the wavelength range
covered by the grating. Columns (7)--(12) give the results of the
kinematical analysis and their errors; $\gamma$ is the line strength,
$V_{\rm rot}$ the rotation velocity, and $\sigma$ the velocity
dispersion. Asterisks indicate dispersions that where used as
reference dispersions (see Appendix~A).}
\endtable


\subsection{3.3 Seeing estimates}

$I$-band images of photometric standard fields were taken during
the observing run using the Cassegrain focus Auxiliary Port of the
WHT. The detector used was a EEV CCD with $0.10'' \times 0.10''$
pixels. The images were bias subtracted, flat-fielded, and cleaned of
cosmic rays.

The shape of the seeing PSF was determined using bright stars
on these images. For all images obtained throughout the run, the
shape could be well described by a sum of two Gaussians:
\eqnam\psfshape
$${\rm PSF}(r) = A_1 {\rm e}^{-r^2/2\sigma_1^2} + 
A_2 {\rm e}^{-r^2/2\sigma_2^2},\eqno\new $$
with fixed ratios of the dispersions, $\sigma_2/\sigma_1 =
1.65$, and amplitudes, $A_2/A_1 = 0.22$, of the two components. The
PSF is normalized for $A_1 = 0.09897/\sigma_1^2$, and is fully
specified by its FWHM $S = 2.543\sigma_1$.

The auto-guider camera provides an independent measure of the seeing
FWHM, $S_{\rm auto}$, during each exposure. However, this measure does
not necessarily take the full PSF shape (equation~\psfshape) into
account. In addition, the auto-guider was equipped with a $V$-band
filter, whereas our Ca II triplet spectra fall in the $I$-band. We
therefore calibrated the autoguider FWHM estimates, by comparing them
to the FWHM values, $S_{\rm image}$, inferred from double-Gaussian fits
to the images of bright stars on the $I$-band Auxiliary Port
exposures. This yielded $S_{\rm image}/S_{\rm auto} \approx 1.0$,
indicating that $S_{\rm auto}$ can be used directly as an estimate of
the true $I$-band seeing FWHM. The FWHM values $S$ in Table~4 list the
seeing FWHM thus obtained for each of our spectra.

As a consistency check, we convolved the Lucy-de\-con\-vol\-ved
$I$-band HST image of NGC~4342 with the PSF of equation (\psfshape),
for different test values of the FWHM $S$. We subsequently overlaid a
slit (with the same width and position angle as for the WHT spectra),
and binned into pixels of the appropriate size. This profile was then
compared to the observed intensity profile along the slit for each of
the spectra. In all cases we found the best-fitting values of $S$ to
be consistent with the FWHM values listed in Table~4.

\subsection{3.4 Template spectra}

To infer stellar kinematical quantities from the galaxy spectra, they
must be compared to a template spectrum. It is important to choose a
template that closely matches the `average' spectrum of the stars in
the galaxy (without kinematical Doppler broadening), to minimize
possible systematic errors (e.g., van der Marel \etal 1994). Most of
the visible light of elliptical galaxies comes from stars on the
giant, asymptotic, and horizontal branches, and their spectra are
therefore comparable to those of G and K giants. The spectrum of a
single K giant star generally makes a reasonable template, but not an
ideal one, because galaxies are made up of stars of different stellar
types. We therefore observed 13 template stars with spectral types
ranging from F7 to M0. All these spectra were taken with the same
instrumental setup as the galaxy spectra, and were reduced in the same
way. We used this template library to determine the mix of template
stars that best matches the galaxy spectra. We assume that there are
no strong changes in stellar population over the galaxy, and we
therefore determined only one optimal template spectrum per
galaxy. For this purpose we summed the spectra along the major axes of
NGC~4342 and NGC~4570 inside the inner $1''$, to yield one high-$S/N$
spectrum per galaxy. We then used a method similar to a `biased random
walk', in order to search for the optimal template that consists of a
weighted sum of the stellar spectra in the library (van der Marel
1994).


\beginfigure{7}
\centerline{\psfig{figure=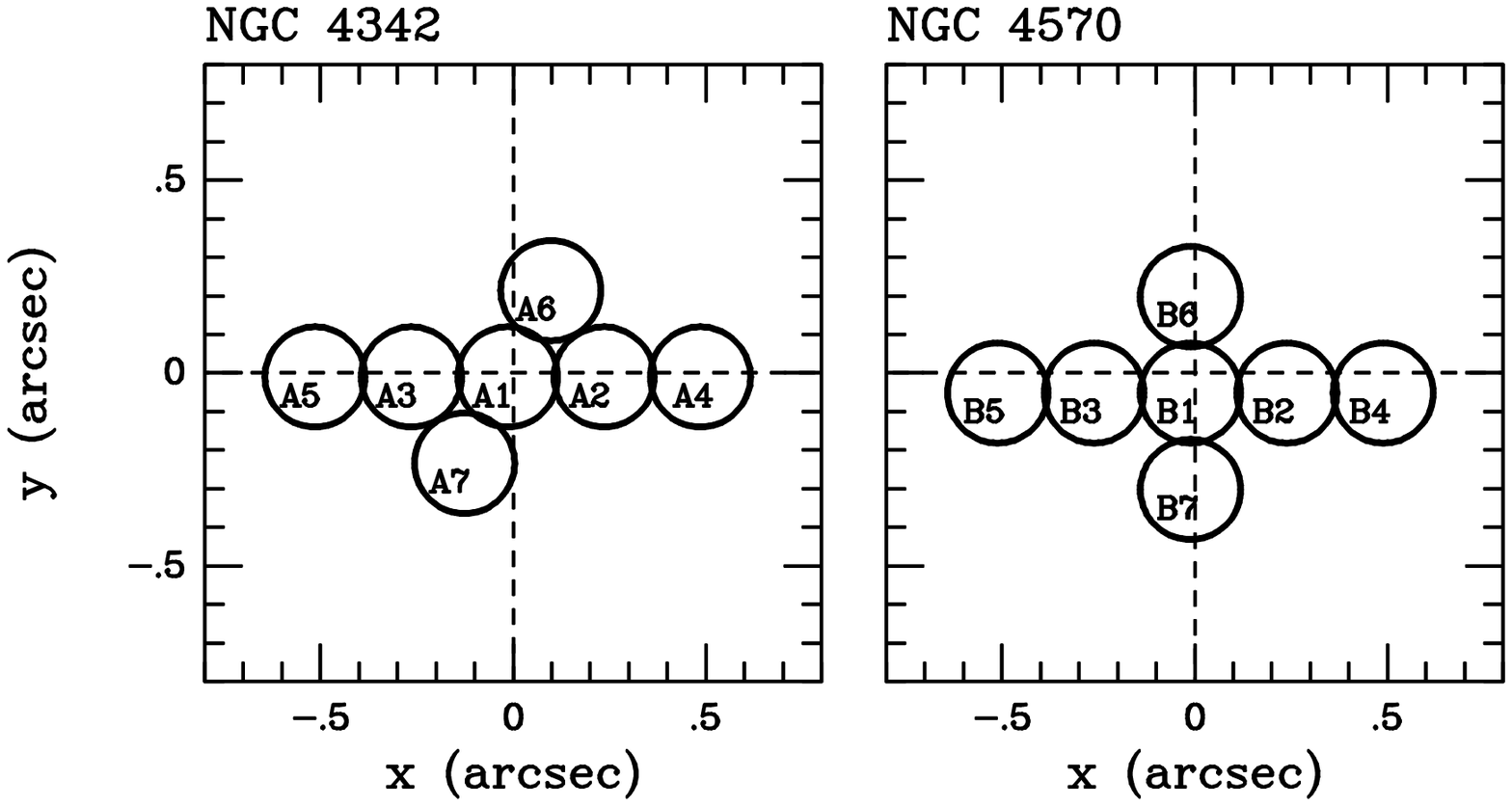,width=\hssize}}\smallskip
\caption{{\bf Figure~7.} Aperture positionings for the HST/FOS spectra. 
The labels are as in Table~5.}
\endfigure


\section{4 HST spectra}

In this section we discuss the high spatial resolution spectra
obtained with the HST/FOS. We describe the observations
(Section~4.1), the target acquisition (Section~4.2), the reduction
(Section~4.3), the wavelength calibration (Section~4.4), and our
choice of template spectrum for use in the kinematical analysis
(Section~4.5). A detailed description of the FOS can be found in
the HST/FOS Instrument Handbook (Keyes \etal 1995).


\beginfigure{8}
\centerline{\psfig{figure=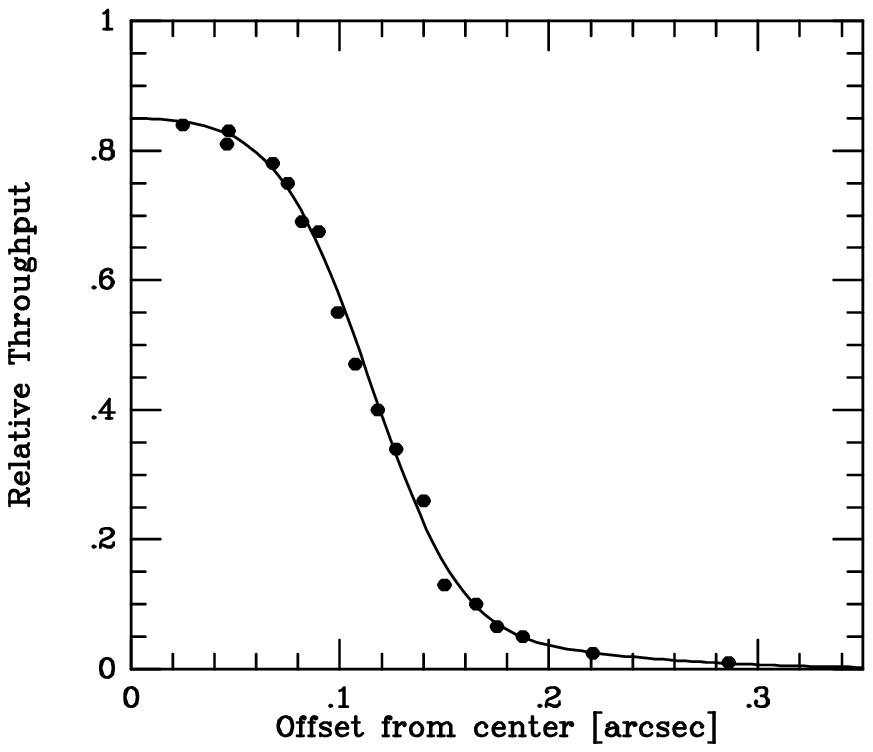,width=\hssize}}\smallskip
\caption{{\bf Figure~8.} The aperture transmission for a star in the
circular $0.26''$ aperture (nominal size), as a function of
its distance from the aperture centre. Data points are from Evans
(1995). The relative throughput is normalized to $1.0$ for a star
centred in the circular $0.86''$ diameter aperture (the FOS {\tt 1.0}
aperture). The solid line is our model fit, which assumes a
PSF that is a sum of three circularly symmetric Gaussians, and an
aperture diameter of $0.238''$.}
\endfigure

 
\subsection{4.1 Observations}

We obtained spectra of NGC~4342 and NGC~4570 with the HST/FOS, using
the circular $0.26''$ diameter aperture (the so-called FOS {\tt 0.3}
aperture) and the G570H grating. This grating covers the wavelength
range from 4569 to 6818{\AA}, and has a dispersion of 4.37{\AA} per
diode. The main absorption lines in this spectral range are the Mg $b$
triplet (5167, 5173, 5184{\AA}) and the Na D lines at 5892{\AA}. The
spectra were quarter-stepped, yielding 2064 $1/4$-diode pixels of
1.09{\AA}. For each galaxy, 7 spectra were taken at different aperture
positions, as illustrated in Figure~7 and summarized in Table~5.
 
\subsection{4.2 Target acquisition}

Some form of target acquisition is required to properly position the
galaxy in the $0.26''$ aperture. We used the `peak-up' acquisition
mode to centre the aperture on the nucleus of each galaxy; spectra at
offset positions were obtained by slewing the telescope from this
position. Pointing drifts during the observations are generally 
not significant ($\lta 0.03''$; Keyes \etal 1995). The
acquisition consists of different stages. In each stage the total flux
through an aperture is measured for a grid of aperture positions. The
telescope is then centred on the aperture position with the highest
flux. In each subsequent stage a smaller aperture is used with a
tighter grid (smaller inter-point spacings), thus increasing the
accuracy of the target positioning. We used different target
acquisition patterns for NGC~4342 and NGC~4570. For NGC~4570, the
final stage consisted of a $3 \times 3$ grid with $0.1''$ inter-point
spacings, using the circular $0.26''$ aperture. This yields an
expected pointing accuracy $\lta 0.08''$ (Keyes \etal 1995). For 
NGC~4342, a $5 \times 5$ grid with $0.052''$ inter-point spacings was 
adopted, again using the $0.26''$ aperture, yielding an expected 
pointing accuracy $\lta 0.04''$.

Precise knowledge of the aperture positions is of great importance
when interpreting the data and comparing it to models. We
therefore modeled the observed fluxes in the final peak-up stage to
verify the success of the target acquisition. We denote the offset
of the true galaxy centre from the grid position that produced the
most counts in the final peak-up stage by
($\Delta_x$,$\Delta_y$). Here $x$ is along the major axis of the
galaxy. We adopt the MGE-model for the $V$-band WFPC2 data to describe
the intrinsic surface brightness distribution of each galaxy. For a
given offset $(\Delta_x,\Delta_y)$, one may calculate the predicted
flux at each grid point in the final peak-up stage, taking into
account the HST/FOS PSF and the aperture size. These predictions can
be compared to the observed fluxes, and the best-fitting offset
$(\Delta_x,\Delta_y)$ can be determined using $\chi^2$-minimization.
The result describes the accuracy of the target acquisition.


\beginfigure{9}
\centerline{\psfig{figure=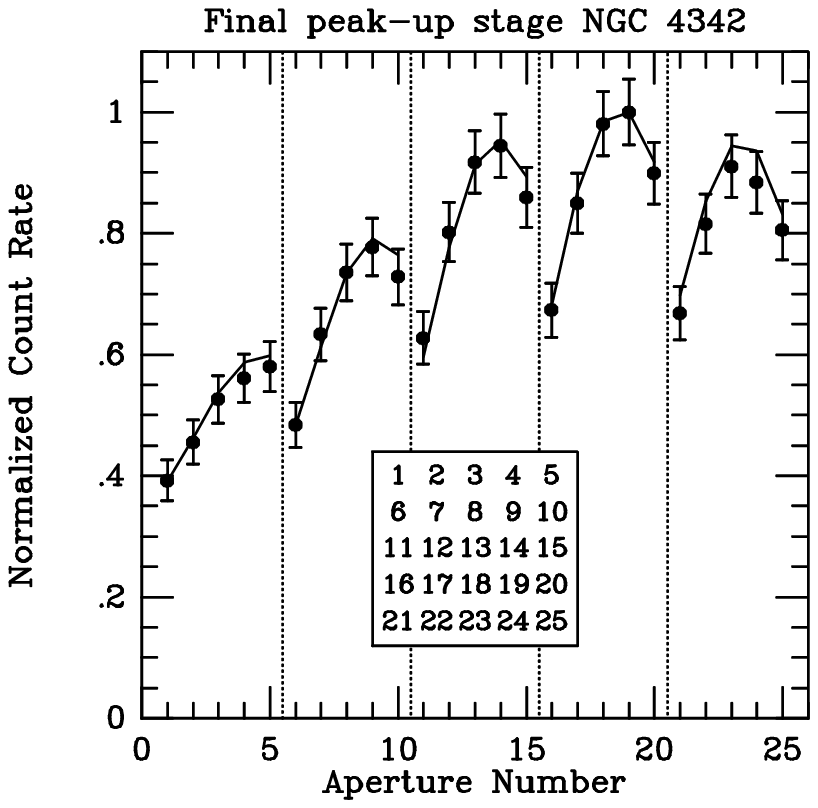,width=\hssize}}\smallskip
\caption{{\bf Figure~9.} Results of the final peak-up stage on the nucleus 
of NGC~4342. The $0.26''$ aperture was placed at the positions of a
$5\times 5$ grid on the sky, as illustrated in the inset. The data
points show the measured counts for each aperture position, normalized
to unity for the position with the maximum count rate (position
\#19). The error bars indicate the Poisson noise on the measurements.
The solid curves connect the predictions of the best-fit model for
these data. This model is based on convolutions of the WFPC2 $V$-band
image, and assumes that the centre of NGC~4342 is offset from the
centre of aperture position \#19 by $-0.014''$ along the major axis,
and $-0.010''$ along the minor axis.}
\endfigure


To properly model the observations one must know the spatial
convolution kernel due to the combined effects of the HST/PSF and the
aperture size, neither of which has been particularly well calibrated
previously. We therefore performed a new calibration of these
quantities, using existing observations obtained by Evans (1995). This
follows the approach of van der Marel \etal (1997), who did the same
for the small {\it square} FOS apertures. Evans measured the
throughput of the $0.26''$ diameter for a star positioned at various
distances from the aperture centre, as shown in Figure~8.  We fitted
these data under the assumption of a purely circular aperture, and a
PSF that can be described as the sum of circularly symmetric
Gaussians. No good fit could be obtained if the aperture diameter was
kept fixed at its nominal value of $0.26''$. This can be due either to
the fact that our model ignores the effects of diffraction at the
aperture edges, or because the aperture does in fact have a different
diameter than its nominal value. An excellent fit to the calibration
observations could in fact be obtained (solid curve in Figure~8) if
the aperture diameter was treated as a free parameter, yielding a
value of $0.238''$. The solid curve in Figure~8 shows the fit under
the assumption that the PSF can be described by the sum of three
Gaussians. The question whether the true aperture diameter is $0.26''$
or $0.238''$ is not relevant here; the diameter enters into the
analysis only through the kernel that describes the combined effect of
the PSF and the aperture size. This kernel is adequately fit by our
model, independent of what the actual aperture size is.

The peak-up data for the final stage of the NGC~4342 target
acquisition are shown in Figure~9. It displays the count rate measured
at each of the 25 positions in the $5\times5$ grid at which the
$0.26''$ aperture was placed. The curve shows the predictions of our
best-fit model, which provides an excellent fit. It has an offset
$(\Delta_x,\Delta_y)$ of only $(+0.014'',+0.010'')$. Similar models
for the NGC~4570 data indicate a somewhat larger offset of
$(+0.011'',+0.052'')$. From a number of experiments we estimate the
errors on our determination of $\Delta_x$ and $\Delta_y$ to be smaller
than $0.005''$. Thus the peak-up acquisitions worked well for both
galaxies, and our models yield the precise positions of the apertures
to high accuracy.

\subsection{4.3 Reduction}
 
The spectra were reduced using the standard pipeline procedure
described by Keyes \etal (1995). The pipeline flat-field was checked
by cross-correlating it with our continuum subtracted galaxy
spectra. A clear cross-correlation peak at zero shift confirmed the
appropriateness of the flat-field. The spectra were converted from
counts to erg cm$^{-2}$ s$^{-1}$ {\AA}$^{-1}$ using the inverse
sensitivity file (IVS) for the circular $0.26''$ aperture. We did not
attempt to correct the calibrated spectra for the PSF dependence on
wavelength; this only affects the continuum slope, which is subtracted
in the stellar kinematical analysis anyway.

\subsection{4.4 Wavelength calibration}

A vacuum wavelength scale is computed by the STScI pipeline, based
upon dispersion coefficients for the given grating and aperture. Due
to non-repeatability of the filter-grating wheel and the aperture
wheel, offset errors in the wavelength scale can occur of up to
several Angstroms. Since we changed neither the aperture nor the
grating during our visits, this will not affect the relative velocity
scale for each galaxy. It may affect the absolute velocity scale, but
that is of little importance.

The FOS suffers from the so-called `geomagnetically induced image
motion problem' (GIMP). Although on-board corrections are applied to
correct for this, residual effects still affect the wavelength
scale considerably ($0.13${\AA} RMS, according to the HST Data
Handbook). Additional wavelength calibration is therefore useful. For
NGC~4570, one arc lamp spectrum was obtained after the acquisition. We
used this spectrum to check the pipeline wavelength calibration, by
comparing the wavelengths of the emission line centres to their actual
vacuum wavelengths. In addition to an offset of $\asim 2${\AA} (due to
the non-repeatability of the wheels), a small non-linearity of the
wavelength scale was found. We therefore recalibrated the wavelength
scale using the arc spectrum, including an additional shift of
-0.769{\AA} to correct for the offset between internal and external
sources (Keyes \etal 1995). For NGC~4342, an arc lamp spectrum was
obtained at the end of each orbit. The wavelength scale of these
spectra was found to vary by 0.35{\AA} during the visit, as a result
of residual GIMP. The arc spectra were used to recalibrate the
wavelength scale of each galaxy spectrum. For this, we used linear
interpolation in time to estimate the wavelength scale for
observations between two arc spectra. After wavelength recalibration,
all spectra were rebinned logarithmically to a scale of $58.539
\kms$/pixel, covering the wavelength range between 4570 and 6817{\AA}.

We estimate the uncertainties in the final wavelength scale for
NGC~4342 to be $\lta 3.5 \kms$. For NGC~4570, we could not correct for
residual GIMP variations from orbit to orbit, because only one arc
lamp spectrum was obtained. If the variations of the absolute
wavelength scale were of the same order as during the NGC~4342 visit
(i.e., 0.35{\AA}), the absolute velocity scales of different NGC~4570
spectra may vary by $\asim 20 \kms$. This may induce systematic errors
in the rotation velocities of the same order (see Section~5.2).

\subsection{4.5 Template spectra}

To facilitate the kinematical analysis it is convenient to have
template spectra that are observed with the same instrumental setup.
However, because of the time consuming target acquisitions, only very
few template stars have been observed with the FOS. From the HST
archive we took a spectrum of the KIII-star F193, which was observed
with the same setup as our galaxy spectra under GO proposal
5744 (PI: H.C. Ford). Unfortunately, after reducing the spectrum, it
was found to provide a poor match to our galaxy spectra.

We therefore decided to use a template library obtained from
ground-based observations. The library consists of 27 stars of
different spectral type, obtained by M. Franx at the 4m telescope of
the KPNO with the RC Spectrograph (see van der Marel \& Franx 1993 for
more details on this template library). The spectra were rebinned
logarithmically to the same scale as the galaxy spectra, and were
shifted to a common velocity. The spectra cover the wavelength range
$4836-5547${\AA}. Although this range is smaller than that covered by
the FOS spectra, it is centred on the Mg b triplet ($\asim
5170${\AA}), which is the most useful wavelength range for stellar
kinematic analysis. The other strong feature in the FOS spectra, the
NaD line, can be influenced by absorption from the interstellar
medium, and is not a good absorption line for kinematic analysis.

Given the relative low $S/N$ of the FOS spectra, we decided
to construct one optimal template spectrum for each galaxy, rather
than for each separate spectrum. For this purpose, we constructed a
grand total spectrum of each galaxy, by summing all spectra at
different aperture positions. We determined the best-fitting stellar
mix using the same method as described in Section~3.4. For both
NGC~4342 and NGC~4570 we found the best-fitting template mix to
consist of giants and dwarfs of spectral types G and K.


\beginfigure*{10}
\centerline{\psfig{figure=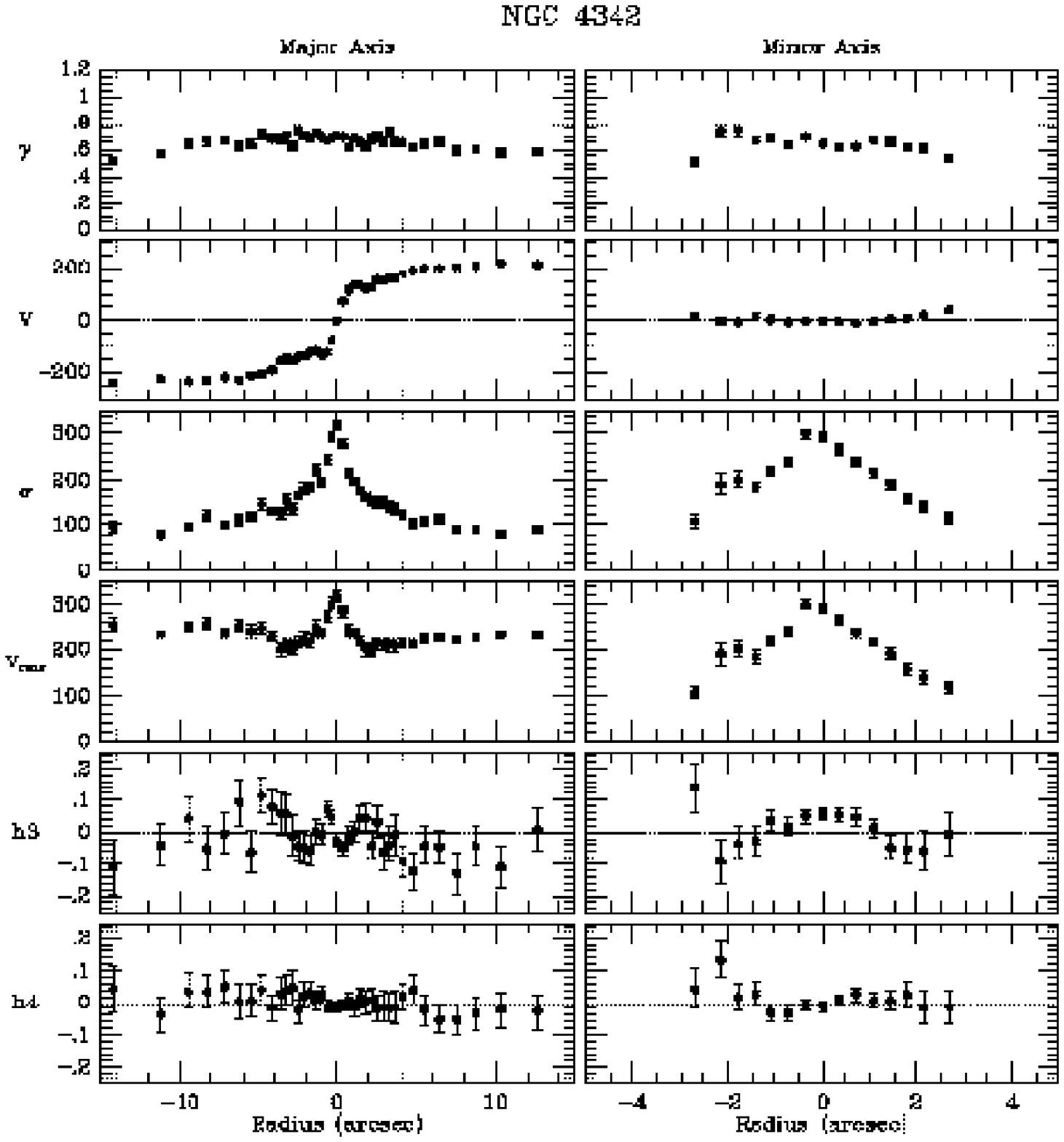,width=0.8\hdsize}}\smallskip
\caption{{\bf Figure~10.} Major and minor axis stellar kinematics of 
NGC~4342 inferred from the WHT spectra. The panels show, from top to
bottom: the line strength $\gamma$, the rotation velocity $V$, the
velocity dispersion $\sigma$, the RMS projected line-of-sight velocity
$\Vrms \equiv \sqrt{V^2+\sigma^2}$, and the Gauss-Hermite
coefficients $h_3$ and $h_4$. All velocities are in units of
$\kms$. The scales on the abscissa are different for the major and
minor axis data, as a result of the strong flattening of NGC~4342.}
\endfigure



\beginfigure*{11}
\centerline{\psfig{figure=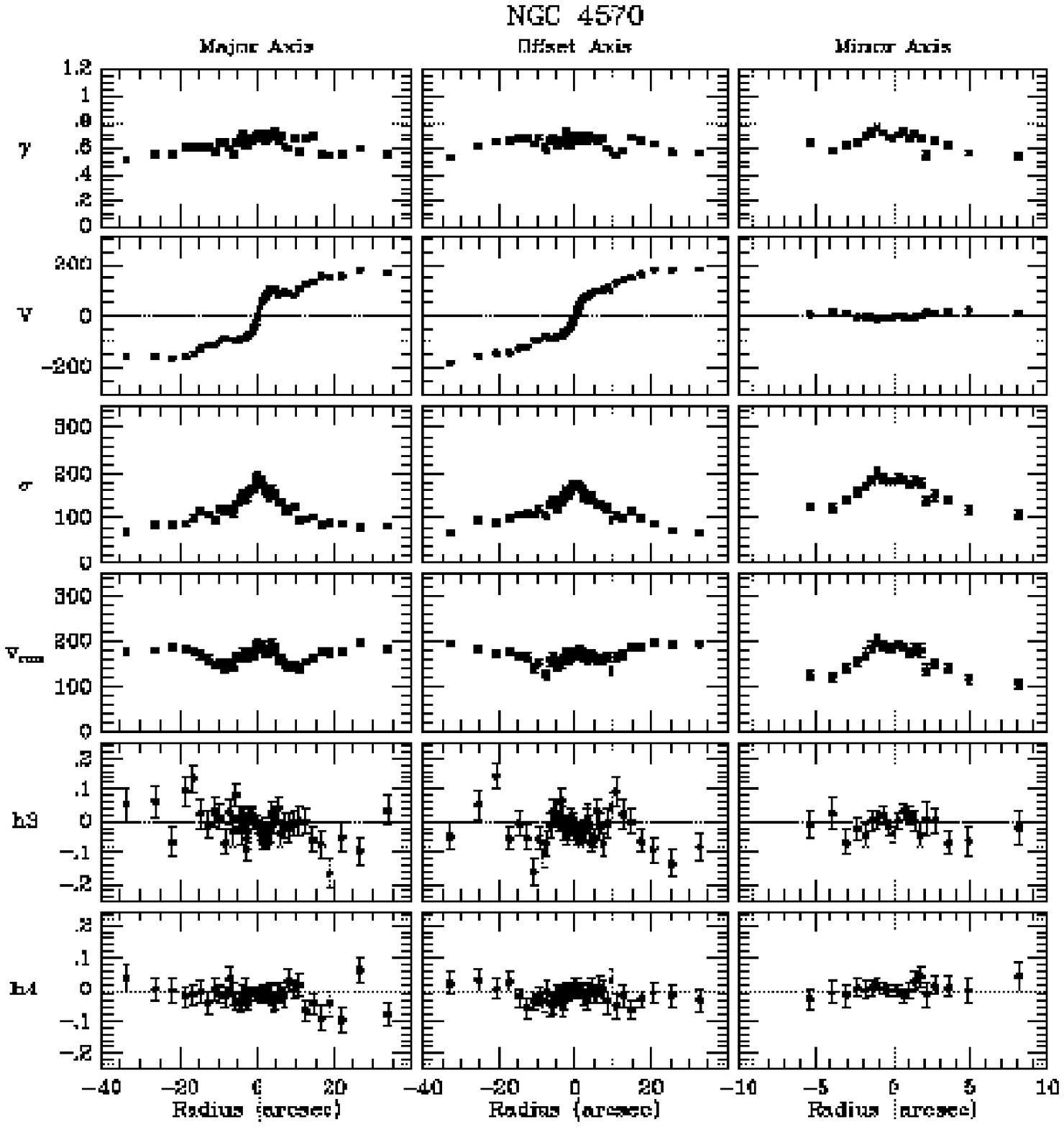,width=0.8\hdsize}}\smallskip
\caption{{\bf Figure~11.} Same as Figure~10, but now for the major axis, 
offset axis (see Table~4), and minor axis of NGC~4570.}
\endfigure



\beginfigure*{12}
\centerline{\psfig{figure=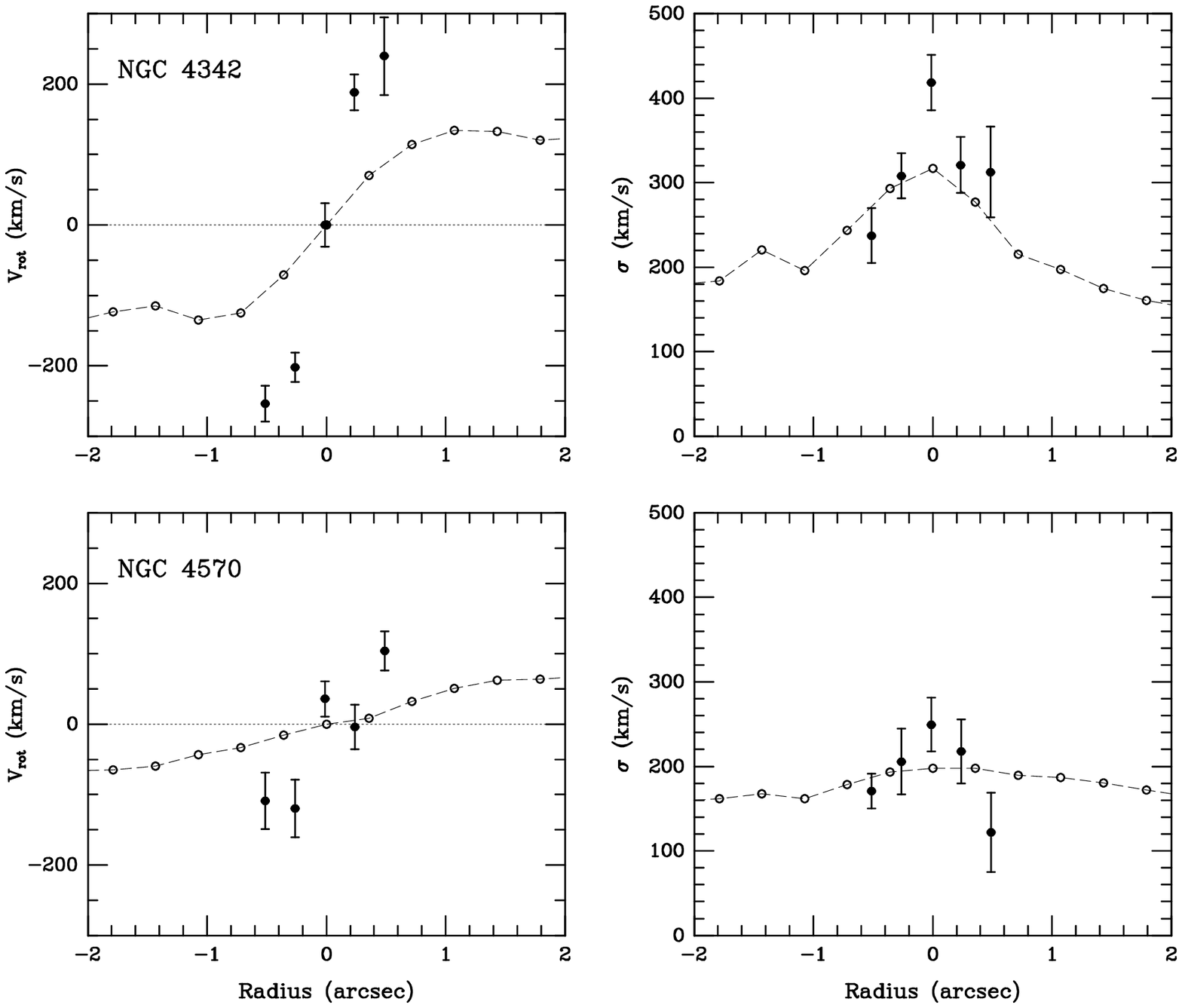,width=0.8\hdsize}}\smallskip
\caption{{\bf Figure~12.} Major axis rotation velocities $V$ (left) and 
velocity dispersions $\sigma$ (right) for the nuclear regions of
NGC~4342 (top) and NGC~4570 (bottom), inferred from the HST/FOS
spectra (data points with error bars). For comparison, the open
circles connected by dashed lines show the lower-spatial resolution
results from the WHT spectra.}
\endfigure


\section{5 Stellar kinematical analysis}

The accessible stellar kinematical information of galaxies is
contained in the stellar line-of-sight velocity profiles (VPs). We
expand each VP in a Gauss-Hermite series, following the approach of
van der Marel \& Franx (1993):
\eqnam\ghseries
$$ {\rm VP}(v) = {\gamma \over \sigma} 
          \alpha(w) \left( 1 + \sum_{j=3}^{N} h_j H_j(w) \right)    
          ,\eqno\new$$
where
$$ \alpha(w) \equiv {1 \over \sqrt{2 \pi}} {\rm e}^{-{1\over2} w^2} 
          , \qquad
   w \equiv (v-V)/\sigma . \eqno\new$$
Here $v$ is the line-of-sight velocity, $H_j$ are the Hermite
polynomials of degree $j$, and $h_j$ are the Gauss-Hermite
coefficients.  The first term in equation (\ghseries) represents a
Gaussian with line strength $\gamma$, mean radial velocity $V$, and
velocity dispersion $\sigma$.  The even Gauss-Hermite coefficients
quantify symmetric deviations of the VP from the best fitting
Gaussian, and the odd coefficients quantify anti-symmetric deviations.

We determined the best-fitting VP parameters for each galaxy spectrum
by $\chi^2$-minimization of the difference between the galaxy spectrum
and a broadened template spectrum (using the Gauss-Hermite series as
the broadening function). The fitting can be done either in Fourier
space (e.g., van der Marel \& Franx 1993) or in pixel space (e.g., van
der Marel 1994). We adopted the latter approach, because it allows
straightforward masking of sky lines (which are especially abundant in
the WHT Ca II triplet spectra) and bad pixel regions (e.g., due to
dead or noisy diodes in the FOS spectra). However, we also performed
Fourier-space fitting for all spectra, as a consistency check, and
found excellent agreement in all cases.

\subsection{5.1 The WHT spectra}

The stellar kinematical results obtained from the WHT spectra are
shown in Figures~10 and~11, for NGC~4342 and NGC~4570, respectively,
and are listed in the tables of Appendix B. Prior to analysis, the
spectra were spatially rebinned along the slit to a S/N $\geq 20$ per
$10 \kms$. As template we used the optimal mix of stellar spectra
determined as described in Section~3.4.

The kinematics of both galaxies are remarkably similar. Each galaxy
shows rapid rotation along the major axis, no measurable rotation
along the minor axis, and a central peak in the velocity dispersion
profile. The central dispersion measured for NGC~4342 is $317\kms$;
that for NGC~4570 is $198\kms$. The central velocity dispersion of
NGC~4342 is extremely high. Only very few galaxies have central
velocity dispersions, measured from the ground, that are larger than,
or comparable to that of NGC~4342; examples are: M87 (van der Marel
1994), NGC~3115 (Kormendy \& Richstone 1992) and NGC~4594 (Kormendy
1988; van der Marel \etal 1994). These galaxies are all strong
candidates for harbouring a massive nuclear black hole. The latter two
galaxies are S0s, and have rather similar kinematics as NGC~4342 and
NGC~4570.

The observed rotation curve shapes are typical for S0 galaxies (e.g.,
Simien, Michard \& Prugniel 1992; Fisher 1997). They are steep in the
centre, show a dip at intermediate radii, and then rise more gradually
out to the last measured point. The dip is also present in the radial
profile of the rms-velocity, $V_{\rm rms} = \sqrt{V^2 +
\sigma^2}$. For NGC~4342, the radial profile of $h_3$ also shows an
interesting feature. Although the profile is somewhat noisy, it
appears that $h_3$ changes its sign in the radial region where the dip
in the rotation curve occurs. This has in fact been observed in a
number of other S0s as well (Fisher 1997). Most likely, all these
kinematical features reflect radial changes in the relative
contributions of the different structural components. The even
Gauss-Hermite coefficient $h_4$, expressing symmetric deviations of
the VP from a Gaussian, is never significantly different from zero.

\subsection{5.2 The HST spectra}

{}From the HST/FOS spectra we determined only the mean velocity and
velocity dispersion of the best-fitting Gaussian VPs. The spectra are
not of sufficient $S/N$ to determine the deviations from a Gaussian
shape. A complication in the kinematical analysis is provided by the
fact that we must use a template spectrum that was obtained with a
different instrument (Section~4.5). This implies that the template and
galaxy spectra do not have the same line-spread-function (LSF; i.e.,
the observed response for a single monochromatic line). The parameters
${\widetilde V}$ and ${\widetilde \sigma}$ obtained from the
kinematical analysis must be corrected for these LSF differences, to
obtain unbiased estimates for the true mean stellar velocity $V$ and
velocity dispersion $\sigma$. The required corrections can be made,
since the LSFs of both the galaxy and template spectra can be measured
and/or calculated. Our approach for this is described in detail in
Appendix~A. The kinematical results obtained after correction for LSF
differences are listed in Table~5. Figure~12 shows the results for the
apertures that were centered on the major axis, as function of major
axis distance. The systemic velocity for each galaxy was estimated as
the mean velocity at $r=0$, obtained by linearly interpolating the
rotation curve between aperture positions~\#4 and~\#5.

For NGC~4342, the FOS spectra show a much higher central velocity
dispersion than the lower spatial resolution WHT spectra, $\sigma_0 =
418 \kms$ vs.~$\sigma_0 = 320 \kms$, respectively. Also, a very steep
central rotation gradient is measured with the FOS, much steeper than
that measured from the ground. The rotation velocity reaches $V_{\rm
rot} \approx 200\kms$ at $0.25''$ from the centre (corresponding to 18
pc at a distance of 15 Mpc). These observations suggest the presence
of a strong central mass concentration in NGC~4342, possibly a massive
black hole. We will address this issue in a forthcoming paper through
detailed dynamical models.
  
For NGC 4570, the FOS results are somewhat more difficult to
interpret. There is certainly much less of a suggestion for a central
mass concentration on the basis of the qualitative features of the
data. The central velocity dispersion is larger than measured from the
ground, but only by a marginal amount. As for NGC~4342, the rotation
curve is steeper than measured from the ground, but the scatter
between neighbouring points (especially observations~B1 and~B2),
suggests the possible presence of small systematic errors. No known
error could be identified, but we cannot exclude GIMP-related
wavelength offsets of several tens of $\kms$ between different spectra
(cf.~Section~4.4). Such offsets are not a problem for the NGC~4342
spectra, for which our wavelength calibration is more accurate.


\beginfigure*{13}
\centerline{\psfig{figure=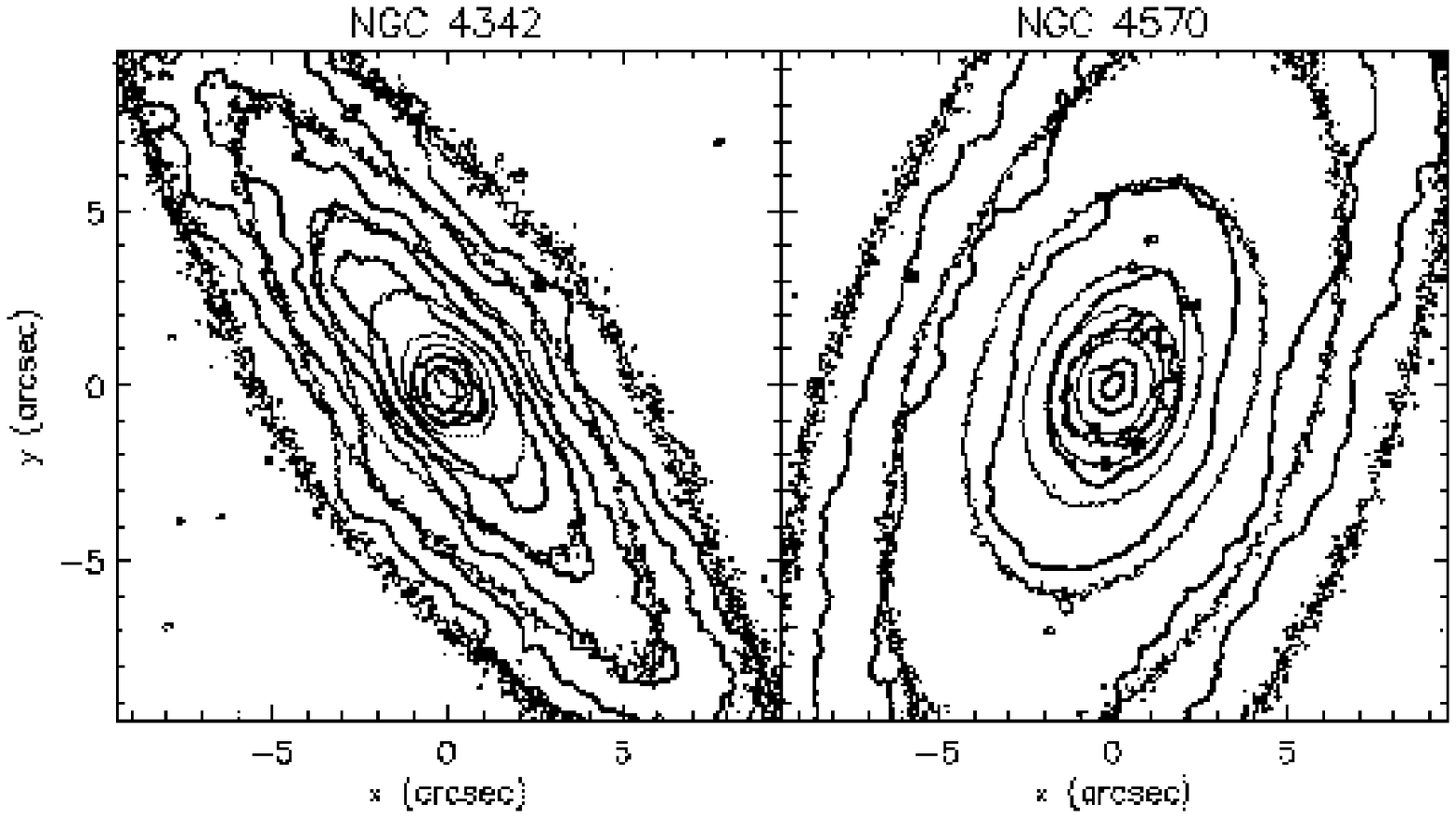,width=\hdsize}}\smallskip
\caption{{\bf Figure~13.} Contour plots of $V-I$ (thick contours),
superimposed on isophotal $V$-band contours (thin contours). In both
galaxies, the outermost colour isophote corresponds to $V-I=1.22$. 
Subsequent contours are 0.02 mag redder. For NGC~4342, the inner four 
contours step by only 0.01 mag, in order to better sample the small 
colour gradient in this galaxy (see Figure~14). The colour images were 
smoothed to suppress noise, while maintaining the information in the 
images. At large radii, the colour contours are flatter than the isophotal
contours. In the central region, the flattening of both contours is
similar. The dents in the $V-I=1.30$ contour of NGC~4570 are caused by 
the relatively blue colours of the two features marked `A' and `B' in 
Figure~6 (see van den Bosch \& Emsellem 1997, for a discussion on the 
nature of these features).}
\endfigure


\section{6 Stellar populations}

In this section we investigate the stellar populations of NGC~4342
and NGC~4570, by studying broad-band colour images (Section~6.1) and
line strengths indices (Section~6.2). The results are compared to
models to study the age and metallicity of the populations
(Section~6.3). This allows us to address the formation of the nuclear
discs, and to discuss the evidence for either coeval or secular
formation.

\subsection{6.1 Broad-band colour images}

Fisher, Franx \& Illingworth (1996) presented $B-R_c$ colour images of
$148'' \times 148''$ for a number of close to edge-on S0s. They found
that the $B-R_c$ contours are flatter than the isophotes;
i.e., whereas the colour gradients along the minor axis decrease
outwards, the major axis colour gradients flatten out towards larger
radii. Similar behaviour was found for the Mg2 line strength
gradients. The H$\beta$ gradients, however, were found to be rather
flat throughout the entire galaxy. These findings suggest that
the (outer) discs of S0s are more metal rich than their bulges,
therewith contradicting formation scenarios in which the bulges are
formed from heated disc material.

We constructed $U-V$ and $V-I$ colour images from the HST/WFPC2 data.
To take into account that the PSFs are significantly different for the
three bands, we convolved the $U$-band image with the $V$-band PSF,
the $I$-band image with the $V$-band PSF, and the $V$-band image with
either the $U$-, or $I$-band PSF (using PSFs constructed with the
TinyTim software package). This approach degrades the spatial
resolution of the colour images somewhat, but provides the safest way
to avoid systematic colour errors near the centre.

Figure~13 presents contour maps of the $V-I$ colour images (thick
contours), superimposed on contour maps of the $V$-band images (thin
contours). The images only go out to $\sim 10''$, and within this
limited extent, both the bulge and the outer disc add
significantly to the projected surface brightness. A determination of
the colours of the individual components is therefore not
straightforward. Nonetheless, our results clearly show that the colour
contours are flatter than the isophotes. Similar results were obtained
for the $U-V$ images. Our results are therefore at least
qualitatively consistent with those of Fisher, Franx \& Illingworth (1996).
 

\beginfigure*{14}
\centerline{\psfig{figure=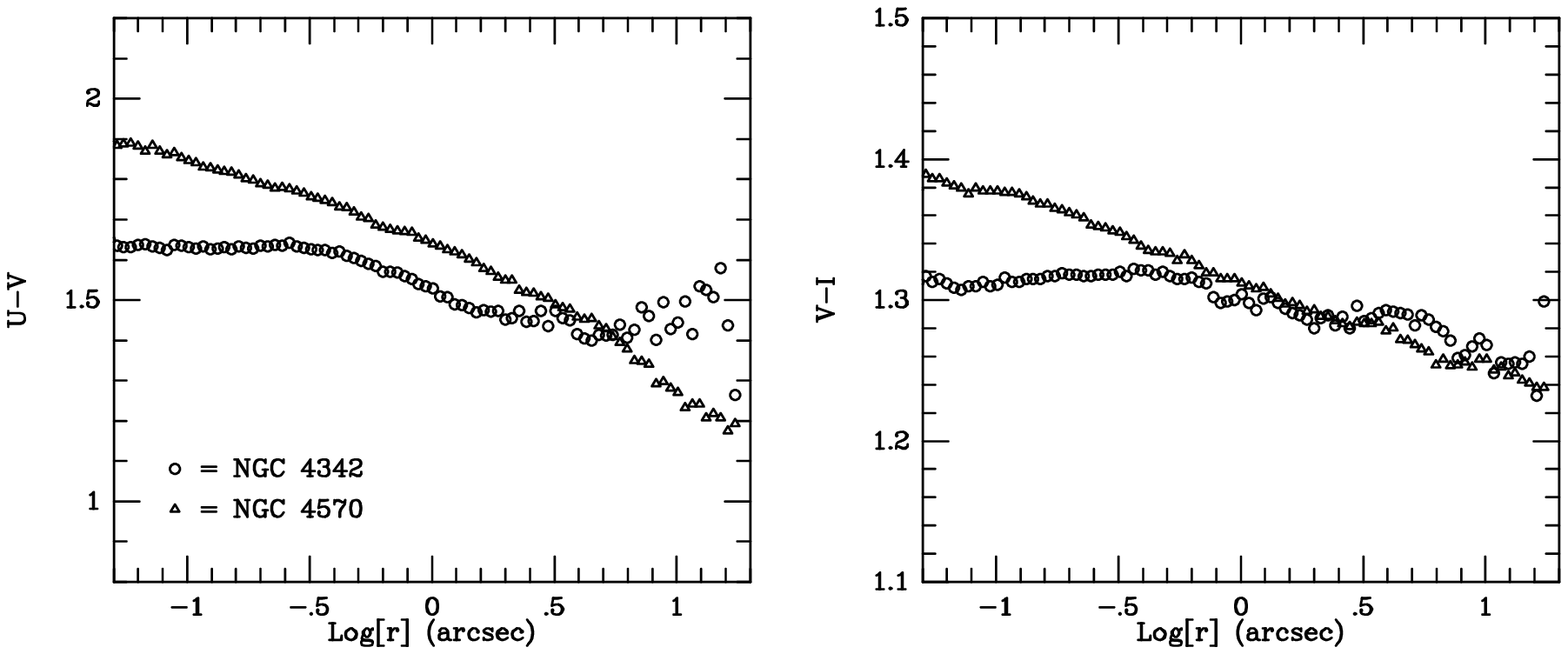,width=\hdsize}}\smallskip
\caption{{\bf Figure~14.} Major-axis $U-V$ and $V-I$ colour gradients, 
as a function of radius, for NGC~4342 and NGC~4570. The behaviour of
the gradients is markedly different for the two galaxies.}
\endfigure


Our high spatial resolution colour images are best suited to study
population differences between the bulges and the {\it nuclear} discs
of both galaxies. If these components had the same, uniform stellar
population, then the flattening of the colour contours and the
isophotes in the central arcsec would have to be similar. This is
exactly what is observed, and contrasts strongly with the results at
larger radii. Thus, the colour images do not suggest a clear
difference in colour between the nuclear discs and the central regions
of the bulges in either of the two galaxies. However, one has to keep
in mind that broad band colours are rather poor population diagnostics
(Worthey 1994), in that different populations can have similar broad
band colours (see e.g., Figure~15 below). Therefore, we cannot rule
out that a more detailed analysis may yet reveal some subtle
population differences. In Section~6.2 below we probe the stellar
populations of the nuclear regions in NGC~4342 and NGC~4570 through 
their absorption line strengths.


\begintable{6}
\caption{{\bf Table~6.} Line strength indices}
\halign{#\hfil&\quad \hfil#\hfil\quad& \hfil#\hfil\quad \cr
& NGC~4342 & NGC~4570 \cr Mg2 & $0.338 \pm 0.087$ & $0.386 \pm 0.015$
\cr H$\beta$ & $1.52 \pm 0.30$ & $2.51 \pm 0.48$ \cr Mgb & $5.11 \pm
0.33$ & $5.32 \pm 0.53$ \cr Fe5270 & $2.94 \pm 0.37$ & $3.38 \pm 0.60$
\cr Fe5335 & $3.29 \pm 0.42$ & $3.40 \pm 0.67$ \cr }
\tabletext{Line strength indices and errors for the central regions of 
NGC~4342 and NGC~4570 (corrected for velocity dispersion broadening
and converted to the Lick-IDS system), derived from grand-total FOS
spectra as described in the text.}
\endtable


Figure~14 shows the $U-V$ and $V-I$ colour gradients as a function of
major axis radius. There is a marked difference between the two
galaxies. NGC~4342 shows no clear colour-gradients inside $\asim
0.3''$ and outside $\asim 3''$. Around $\asim 1''$ we find
$\Delta(U-V)/\Delta \log r = -0.26 \pm 0.02$ and $\Delta(V-I)/\Delta
\log r = -0.05 \pm 0.01$. By contrast, NGC~4570 shows strong gradients
over the entire radial interval studied: $\Delta(U-V)/\Delta \log r$
ranges from $-0.41 \pm 0.02$ at the outside, to $-0.21 \pm 0.02$ in
the centre; $\Delta(V-I)/\Delta \log r$ has more or less a constant
value of $-0.06 \pm 0.01$ from the centre out to $\asim 20''$. The
$U-V$ gradient at the outside of NGC~4570 is extremely large as
compared with those of other early-type galaxies (cf.~Peletier 1989).

The central reddening typical of early-type galaxies is
generally interpreted as due to a metallicity gradient. The best
evidence for this comes from spectroscopic measurements of absorption
lines in ellipticals (e.g., Faber 1977; Burstein \etal 1984; Efstathiou
\& Gorgas 1985; Peletier 1989). Several studies have found a
correlation between colour (and line strength) gradients and
total luminosity. For low-mass galaxies ($M_B > -20.5$), the gradients
increase with the total mass of the galaxy (e.g., Vader \etal 1988;
Carollo, Danziger \& Buson 1993), consistent with the
predictions of simple models of dissipative collapse coupled with
supernovae-induced winds (Larson 1974; Carlberg 1984; Arimoto \& Yoshii
1987; Matteucci \& Tornamb\`e 1987). For NGC~4342 and
NGC~4570, the presence of both an outer and a nuclear disc indeed
indicates that dissipation has played a role during their
formation. The morphology of these galaxies is very similar, but
NGC~4570 is almost 1.6 magnitudes brighter. This indicates that
NGC~4570 is 4.2 times more massive, assuming that both galaxies have a
similar mass-to-light ratio. The finding that the colour gradients
in NGC~4570 are larger than those in NGC~4342 is therefore
qualitatively consistent with dissipative galaxy formation, in which
star formation lasts longer and the onset of a galactic wind starts 
later, in higher mass galaxies. 


\beginfigure*{15}
\centerline{\psfig{figure=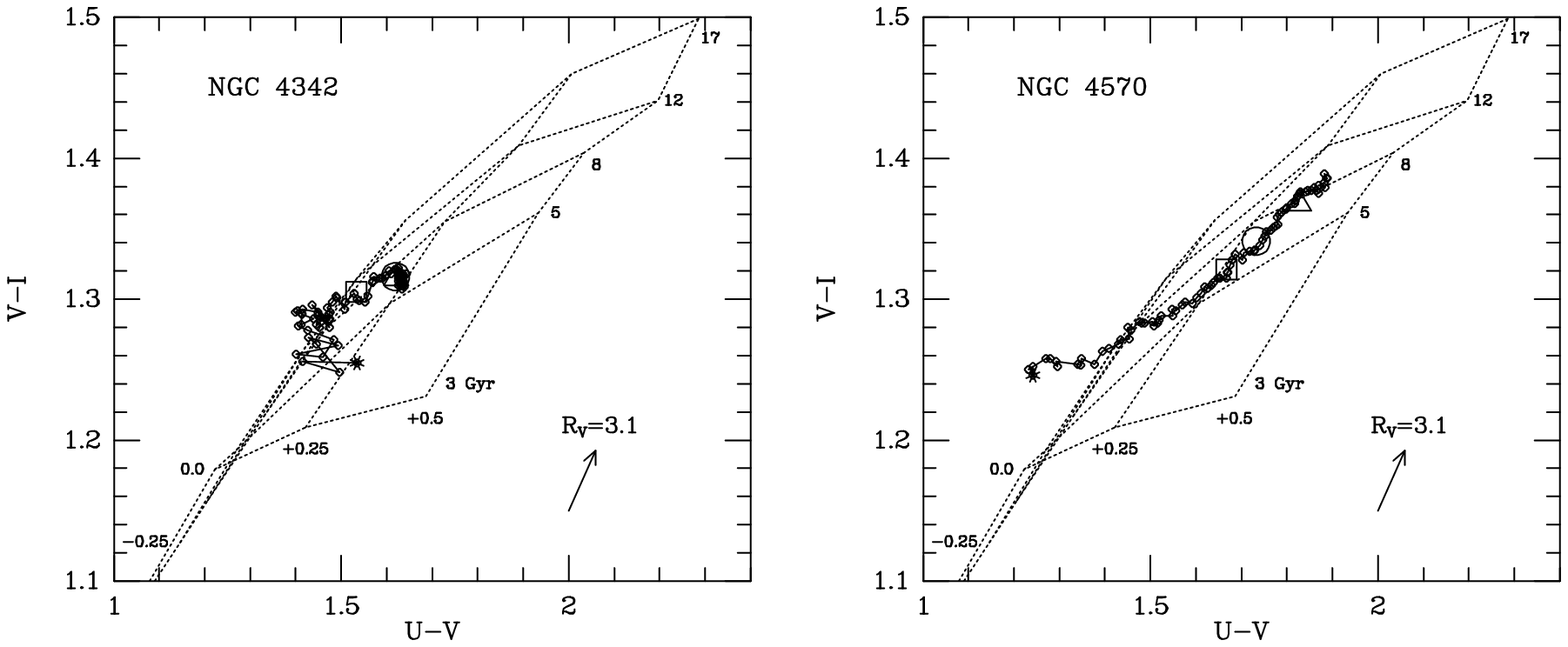,width=\hdsize}}\smallskip
\caption{{\bf Figure~15.} Colour-colour diagrams for NGC~4342 and 
NGC~4570. Small circles connected by solid lines show the colours for
isophotes of different radii; the outermost isophote is indicated by
an asterisk. Large symbols indicate the average colours of the nucleus
(triangle), nuclear disc (circle) and bulge (square). Overplotted as
dotted lines are the predictions of the single-burst stellar
population models of Worthey (1994), for a grid of age (3--17 Gyr) and
metallicity ([Fe/H] $=-0.25$ -- $+0.5$) values. The lines for [Fe/H]
$=-0.25$ and [Fe/H] $=0$ partly overlay each other, indicating a
degeneracy between metallicity and age at low metallicities. The arrow
indicates the slope of dust reddening for $R_V \equiv A(V)/E(B-V) =
3.1$, as typical for Galactic dust.}
\endfigure


\subsection{6.2 Line strengths}

Line strength measurements provide additional information about ages
and metallicities that is complementary to, and often more accurate
than, that provided by broad-band colours. The most commonly used
spectral indices are those of the Lick-IDS system (Faber \etal 1985;
Worthey \etal 1994), in particular H$\beta$, Mg2, Mgb, Fe5270 and
Fe5335. These are all in the wavelength range 4800--5400{\AA}, which
is included in the FOS spectra. The $S/N$ of the FOS spectra was not
sufficient to infer line strengths at each aperture position
individually, and we therefore measured line strength indices from one
grand-total spectrum for each galaxy, obtained by summing the
different spectra available for each galaxy. Two corrections have to
be applied to these indices. First, they must be corrected for the
broadening effect of velocity dispersion, which weakens most of the
lines. We determined empirical correction factors, $C(\sigma) \equiv
{\rm index}(0)/{\rm index}(\sigma)$, for each of the 5 indices listed
above; ${\rm index}(0)$ is the index measured from the template star
K193, $\sigma$ is the velocity dispersion of the grand-total spectrum
derived using K193 as template, and ${\rm index}(\sigma)$ is the index
of the K193 spectrum broadened with a Gaussian of dispersion
$\sigma$. Second, to be able to compare our indices with the Worthey
(1994) stellar population models (see Section~6.3), they must be
converted to the Lick scale. This is necessary to correct for
differences in the spectral resolution and in the spectral response
function between our FOS data and the Lick group data. We compared the
indices derived by us from the FOS spectrum of F193 with those
obtained by the Lick group from observations of the same star (Worthey
\etal 1994). This yields correction factors 
${\rm index(LICK)/index(FOS)}$ for each of the four atomic
indices. For the molecular Mg2 index, we used the difference ${\rm
[index(LICK) - index(FOS)]}$. Table~6 lists the line strength indices
thus obtained for NGC~4342 and NGC~4570, corrected for velocity
dispersion broadening and converted to the Lick scale.

\subsection{6.3 Comparison with stellar population models}

To address the age and metallicity of the stellar populations
in NGC~4342 and NGC~4570, we have compared their colours and line
strengths to the single-burst stellar population models of Worthey
(1994). These models give the fluxes, colours, and line strengths of
an evolving stellar population, as a function of age and
metallicity [Fe/H], calculated from isochrones and model flux
libraries of stars with a Salpeter (1955) initial mass function.

In Figure~15 we plot $V-I$ vs.~$U-V$ for the two galaxies. Small
open circles indicate results for isophotes at different radii;
neighbouring isophotes are connected, and an asterisk indicates the
outermost radius.  Large symbols indicate the average colours of the
nucleus (open triangle), the nuclear disk (open circle) and the bulge
(open square). For the nucleus we used the region inside $0.18''$, for
the nuclear disk we used the region along the major axis between
$0.18''$ and $0.71''$ (which is where the nuclear discs most clearly
stand out from the isophotal analysis, cf.~Figures~4 and~5), and for
the bulge we used the region inside $0.71''$, but offset from the
major axis by $0.5'' - 0.6''$. Overplotted in Figure~15 are the
predictions of Worthey's models for an age-metallicity grid, with ages
between 3 and 17 Gyr, and [Fe/H] between $-0.25$ and $+0.50$. For
[Fe/H] $\lta +0.25$ there is strong degeneracy between age and
metallicity, as the grid points crowd together. However, for high
metal abundances the $V-I$ vs.~$U-V$ diagram is a useful discriminator
between age and metallicity effects.

For NGC~4342, the main body of the galaxy lies close to the 8 Gyr,
solar-metallicity grid point. The nucleus and nuclear disc have
different colours than the main body, but have similar colours to each
other (cf.~Figure~15). Their colours are well fit with a
similar age as the main body, but with a somewhat higher metallicity
([Fe/H] $\approx +0.25$). There is considerable uncertainty in these
results though, since most points fall in the region of the model grid
where there is a strong age-metallicity degeneracy.


\beginfigure{16}
\centerline{\psfig{figure=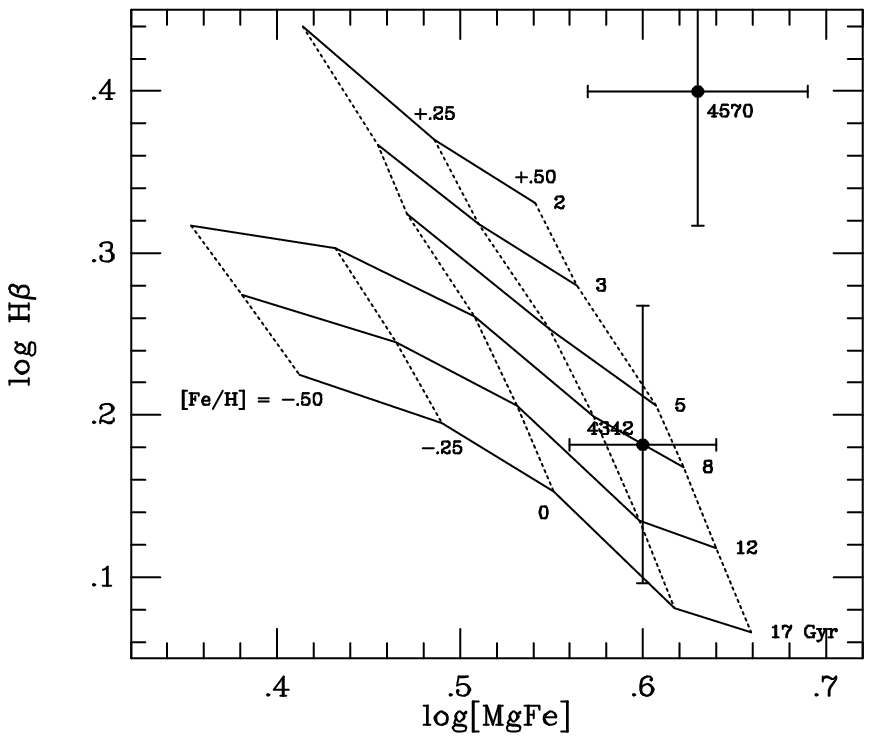,width=\hssize}}\smallskip
\caption{{\bf Figure~16.} Diagram of the age indicator H$\beta$ vs.~the 
metallicity indicator [MgFe]. Data points with error bars indicate the
measurements from the grand-total FOS spectra. Overplotted as dotted
lines are the predictions of the single-burst stellar population
models of Worthey (1994), for a grid of age (3--17 Gyr) and
metallicity ([Fe/H] $=-0.25$--$+0.5$) values.}
\endfigure

 
The colour-colour diagram of NGC~4570 is very different. The
colours change dramatically from the outermost point to the
very nucleus. Whereas the outermost points fall outside the grid of
Worthey's models, the nucleus is consistent with an 8 Gyr old
population with [Fe/H] $\approx +0.35$. The run of the central
isophotes in the colour-colour diagram has a similar slope as would be
expected from reddening due to dust (as indicated by the arrow in
Figure~15), but dust reddening is not likely to be the cause of the
observed gradients: both the isophotes and the colour-images are very
smooth, and no $100\mu{\rm m}$ emission has been detected. It would be
tempting to interpret the observed colour differences between the
nucleus, nuclear disc, and bulge as due to changes in age. However,
the nuclear disc does not show up as a separate component in the
colour images. The colour differences therefore most likely reflect
mere changes in stellar population with radius from the centre, rather
than differences between separate components.

{}From the line indices listed in Table~6, we can calculate the new index
[MgFe], which is defined as $\sqrt{{\rm Mgb} \langle {\rm Fe}
\rangle}$, where $\langle {\rm Fe} \rangle = ({\rm Fe5270} + {\rm
Fe5335})/2$. This index is often used as a metallicity indicator
(e.g., Gonz\'alez 1993). We derive $\log{\rm [MgFe]} = 0.60
\pm 0.04$ and $0.63 \pm 0.06$, for NGC~4342 and NGC~4570,
respectively.  The H$\beta$ index is a sensitive
age-indicator. Figure~16 shows H$\beta$ vs.~[MgFe] for each galaxy,
with overplotted the predictions of Worthey's stellar population
models. The observed indices refer to the central region ($\lta
0.5''$) of each galaxy. The models indicate that this central region
in NGC~4342 has a high metallicity (${\rm [Fe/H]} \approx +0.35$) and
an age of $\asim 8$ Gyr. This is roughly consistent with the age and
metallicity found from the broad band colours (see Figure~15). The
indices for NGC~4570, on the other hand, fall outside the model
grid. The high H$\beta$ index suggests a very young stellar
population, whereas both the colours and the value of [MgFe] suggest a
high metallicity. 

Although these results are suggestive, we do not wish to draw very
strong conclusions. The error bars on the measured line indices are
large, due to the low $S/N$ of the FOS spectra. We have been careful
and conservative in the determination of these errors, but note that
it is very difficult to quantify possible systematic errors. In
addition, the age and metallicity values inferred from the line
strengths refer only to the nuclear region $\lta 0.5''$, and not
necessarily to the entire bulge population. Fisher, Franx \&
Illingworth (1996) have shown that the nuclear regions of S0s are
typically a few Gyr younger than the bulk of the galaxy. Also,
ages measured from the H$\beta$ index have to be interpreted with
care. First, the observed H$\beta$ index could be
dominated by horizontal branch stars, rather than by
main-sequence turnoff stars. It this case it would be more
sensitive to variations in the initial mass function, than to
variations in age. Second, the H$\beta$ index is very
sensitive to small numbers of young stars present. The
relatively young age inferred for the nuclear population in NGC~4570
may therefore reflect merely the most recent generation of
stars, rather than the age of the entire, presumably much older
generation. Indeed, van den Bosch \& Emsellem (1997) present
evidence that the central region of NGC~4570 has had recent
star formation. A more thorough investigation of the stellar
populations of the separate components in NGC~4243 and NGC~4570 than
presented here will have to await additional observations of line
indices, both at larger radii from the centre and with higher
$S/N$.

\section{7 Conclusions and discussion}

Small stellar discs embedded in the nuclei of early-type galaxies are
intriguing: they contain clues about the process of galaxy formation,
and their dynamics provide a useful tool to constrain the central mass
distribution. We have presented high spatial resolution photometric
and spectroscopic data for two E/S0 galaxies in the Virgo cluster,
NGC~4342 and NGC~4570, for which pre-refurbishment HST images showed
nuclear discs in addition to larger outer discs. New HST/WFPC2 images
confirm the existence of the nuclear discs, and dismiss suggestions
that the earlier detections were artifacts of Lucy deconvolution; the
new images clearly show the discs even without deconvolution. The
decomposition of both galaxies in disc and bulge components will be
presented in a forthcoming paper (Scorza \& van den Bosch 1997). Here
we have focussed on using the multi-colour WFPC2 images, WHT spectra
and HST/FOS spectra to do a first study of the stellar populations and
dynamics of the nuclear discs.

Broad-band colour images were constructed from the WFPC2 data,
properly taking into account the different PSFs in the different
bands. For both galaxies we find the colour contours outside $\asim
2''$ to be flatter than the isophotes. We have also determined the
radial colour gradients in both galaxies. The gradients in the more
massive galaxy NGC~4570 are larger than those for NGC~4342, consistent
with the predictions of simple models of dissipative collapse coupled
with supernovae-induced galactic winds. All these results are in good
agreement with those of Fisher, Franx \& Illingworth (1997), who
studied the ages and metallicities of a sample of 20 S0s, and seem to
suggest that the bulges of S0s are not formed out of heated disc
stars, but are most likely the results of dissipational formation.
By contrast, from the colour images and gradients inside
$\asim 2''$ we find no strong evidence for population differences
between the bulges and the {\it nuclear} discs in both galaxies.

We have measured line strength indices for several diagnostic lines
from the summed FOS spectra, to obtain additional stellar population
information about the nuclear ($\lta 0.5''$) components of both
galaxies. Comparison with the single-burst stellar population models
of Worthey (1994) indicates that the central regions of both galaxies
have metallicities [Fe/H] $\approx +0.25$ or larger. NGC~4342 is well
fit with an age of $\asim 8$ Gyr. NGC 4570 has an unusually large
H$\beta$ line strength, which may be suggestive of recent star
formation. In van den Bosch \& Emsellem (1997) evidence is presented
that the central region of NGC~4570 has experienced bar induced
secular evolution. Unfortunately, neither the broad band colours nor
the line strength measurements presented here, place important
constraints on possible formation scenarios for the nuclear
discs. Additional observations of line indices, both at larger radii
from the centre and with higher $S/N$, are required for a more
thorough investigation of the stellar populations of the separate
components in these galaxies.

The WHT and FOS spectra were used to determine the nuclear stellar
kinematics of NGC~4342 and NGC~4570. The dynamical structure of both
galaxies is found to be remarkably similar to that of other
well-studied S0s, such as NGC~3115 (Kormendy \& Richstone 1992),
NGC~4026 and NGC~4111 (Simien, Michard \& Prugniel 1993; Fisher
1997). The long-slit WHT spectra have high $S/N$ and a large radial
extent. They reveal a very centrally peaked velocity dispersion
profile in both galaxies. The rotation curves clearly show the
different dynamical properties of the structural components identified
photometrically. The single-aperture FOS spectra of the nuclear
regions of both galaxies have lower $S/N$ than the WHT spectra, but
have four times higher spatial resolution. The FOS data of NGC~4342
show significantly higher velocities than the WHT data. The observed
central velocity dispersion is $\asim 420\kms$, compared to `only'
$\asim 320\kms$ as measured from the WHT spectra. The rotation
velocity reaches $\asim 200 \kms$ at $0.25''$, implying a much steeper
rotation gradient than inferred from the WHT data. These observations
indicate a high central mass density. Detailed dynamical models to be
presented in a forthcoming paper (Cretton \& van den Bosch 1997, see
also van den Bosch \& Jaffe 1997) provide strong evidence for the
presence of a few times $10^8 \Msun$ black hole in NGC~4342. The FOS
data for NGC~4570 are more difficult to interpret. The rotation curve
is considerably steeper than measured from the ground, but there is
some indication for possible systematic errors. The central velocity
dispersion is larger than measured from the ground, but only
marginally so. Although these data certainly do not exclude the
possible presence of a black hole in NGC~4570, the qualitative
features of the data do not suggest such a black hole as strongly as
they do for NGC~4342.

\section*{Acknowledgments}

\tx The observations presented in this paper were obtained with the 
NASA/ESA Hubble Space Telescope and with the William Herschel
Telescope.  HST data are obtained at the Space Telescope Science
Institute, which is operated by AURA, Inc., under NASA contract NAS
5-26555. The WHT is operated on the island of La Palma by the Royal
Greenwich Observatory in the Spanish Observatorio del Roque de los
Muchachos of the Instituto de Astrofisica de Canarias. We are grateful
to Eric Emsellem for his help with the MGE analysis, and to Marijn
Franx for providing a ground-based template library. RPvdM was
supported by a Hubble Fellowship, \#HF-1065.01-94A, awarded by STScI.

\section*{Appendix A: LSF corrections for the HST/FOS spectra}

\subsection{A.1 Basic equations}

For each galaxy there are seven HST/FOS galaxy spectra, $G_i$
($i=1,..,7$). Each spectrum is the convolution of the stellar velocity
profile, ${\rm VP}_i$, the characteristic stellar spectrum of the
population of the galaxy, $S_G$, and the line-spread-function of the
observation, ${\rm LSF}_i$:
$$ G_i = {\rm VP}_i \otimes S_G \otimes {\rm LSF}_i . \eqno (A1) $$
We assume that each ${\rm VP}_i$ is a Gaussian with mean velocity
$V_i$ and velocity dispersion $\sigma_i$. The template spectrum, $T$,
is the convolution of the stellar spectral mix, $S_T$, and the
line-spread-function of the template star observations, ${\rm LSF}_T$:
$$ T = S_T \otimes {\rm LSF}_T . \eqno (A2) $$
The stellar kinematical analysis minimizes
\eqnam\chisq
$$ \chi^2 = \int [G_i - (T \otimes {\widetilde {\rm VP}}_i)]^2 , \eqno (A3) $$
to find the parameters ${\widetilde V}_i$ and ${\widetilde \sigma}_i$ of
the best-fitting Gaussian broadening function ${\widetilde {\rm
VP}}_i$. It is generally assumed that there is no template mismatch,
i.e., $S_T = S_G$. 

If the LSFs of the galaxy and template spectra are identical, ${\rm
LSF}_i \equiv {\rm LSF}_T$, as is usually assumed if all observations
are obtained with the same instrument, then ${\widetilde V}_i$ and
${\widetilde \sigma}_i$ are unbiased estimates of the true mean velocity
$V_i$ and velocity dispersion $\sigma_i$. In our case the galaxy and
template spectra were not obtained with the same instrument, and hence
the LSFs cannot be assumed to be identical. Thus, ${\widetilde V}_i$ and
${\widetilde \sigma}_i$ must be corrected for the LSF differences
between the galaxy and template spectra, to obtain proper estimates
for $V_i$ and~$\sigma_i$.

\subsection{A.2 The HST line-spread-functions}

For HST/FOS observations one may assume (van der Marel 1997) that the
LSF for observation $i$, ${\rm LSF}_i$, is the convolution of the
normalized intensity distribution of the light that falls onto the
grating, $A_i$, and the instrumental line-broadening function due to the
grating and the detector resolution, $H$:
$$ {\rm LSF}_i = A_i \otimes H , \eqno (A4) $$
The illumination function $A_i$ is a function of the unconvolved,
projected light distribution of the galaxy, the aperture position, and
the kernel function that describes the HST/FOS PSF and the aperture
geometry. These are all known: the MGE fits in Table~3 describe the
unconvolved light distribution, the aperture positions $(x_{\rm
ap},y_{\rm ap})$ for all observations are listed in Table~5, and the
PSF+aperture kernel was derived in Section~4.2. The functions $A_i$
can therefore be calculated explicitly for all galaxy observations.

One may write for any pair of observations at positions $i$ and $j$
within the same galaxy, at least formally,
\eqnam\aplumconv
$$ A_j = z_{ji} \otimes A_i \eqno (A5) $$
We found, as did van der Marel (1997), that the functions $z_{ji}$ can be
well approximated by Gaussians with mean velocity ${\widehat V}_{ji}$
and velocity dispersion ${\widehat \sigma}_{ji}$. 

\subsection{A.3 Kinematical corrections}

The functions $H$ and ${\rm LSF}_T$ influence the kinematical analysis
for each galaxy spectrum in exactly the same way. Their shape and
properties therefore do not enter into the {\it differences} between
the stellar kinematics inferred from observations at different
positions $i$ and $j$ in the same galaxy. Since the convolution of two
Gaussians is again a Gaussian, it is straightforward to show that
\eqnam\fosvel
$$ V_j - V_i = {\widetilde V}_{j} - {\widetilde V}_{i} + {\widehat V}_{ji}
      \eqno (A6) $$
and
\eqnam\fossig
$$ \sigma^2_j - \sigma^2_i = 
      {\widetilde \sigma}^2_j - {\widetilde \sigma}^2_i + 
      {\widehat \sigma}^2_{ji} . \eqno (A7) $$
So if the stellar kinematics for any galaxy spectrum $j$ are known
{\it a priori}, then the kinematics at any other position $i$ can be
calculated without knowledge of either of the functions $H$ and ${\rm
LSF}_T$. 

\subsection{A.4 Reference kinematics}

Position \#5 for each galaxy is located at $\asim 0.5''$ from the
galaxy center (cf.~see~Table 5). This position is beyond the region
most affected by seeing in ground-based observations, and can
therefore be used as a `reference position' for use with
equations~(A6) and~(A7). One possibility is to assume that the
velocity dispersions for spectra~A5 and~B5 are known a priori, from
the results of the ground-based WHT observations. This yields
$\sigma_5 = 260.7 \kms$ for NGC 4342, and $\sigma_5 = 173.1 \kms$ for
NGC 4570.  More accurate estimates for $\sigma_5$ can be obtained by
modeling the small residual effect of seeing on the ground-based
observations. For this purpose we calculated predicted kinematical
quantities from $f(E,L_z)$ Jeans equation models for the ground-based
kinematics, based on the MGE-fitted density distributions (see Cretton
\& van den Bosch 1997). We then calculated predictions for 
$\sigma_5$, by convolving the unconvolved model kinematics with the
$0.26''$ aperture kernel at the appropriate positions. This yielded
$\sigma_5 = 237.5 \kms$ and $\sigma_5 = 170.8 \kms$ for NGC 4342 and
NGC 4570 respectively. Comparison with the directly observed values
confirms that the effect of seeing convolution at a galactocentric
distance of $0.5''$ is only modest.

The LSF corrected stellar kinematics listed in Table~5 were
obtained using equations~(A6) and~(A7), with the seeing corrected
$\sigma_5$ as reference value, and with ${\widehat V}_{i5}$ and
${\widehat \sigma}_{i5}$ chosen so as to best fit equation~(A5). The
systemic velocity was independently estimated and subtracted from the
velocities as described in Section~5.2. Hence, any arbitrary value may
be used for the reference velocity $V_5$, because it only enters into
equation~(A6) as an additive constant.


\begintable{7}
\caption{{\bf Table~A1.} HST/FOS stellar kinematics: a consistency check}
\halign{#\hfil&\quad \hfil#\hfil\quad& \hfil#\hfil\quad& 
\hfil#\hfil\quad& \hfil#\hfil\quad& \hfil#\hfil\quad& \hfil#\hfil\quad \cr
\multispan3\quad\hfil NGC~4342 \hfil & & 
\multispan3\quad\hfil NGC~4570 \hfil \cr
id. & $\delta_V$ & $\delta_{\sigma}$ & & 
id. & $\delta_V$ & $\delta_{\sigma}$ \cr
A1 & $-0.73$ & $-0.41$ & & B1 & $-1.29$ & $+0.52$ \cr
A2 & $+0.41$ & $-0.39$ & & B2 & $-0.14$ & $-0.38$ \cr
A3 & $-0.55$ & $+0.93$ & & B3 & $+0.06$ & $+0.59$ \cr
A4 & $-0.32$ & $+0.02$ & & B4 & $-2.01$ & $+0.26$ \cr
A5 & $+0.68$ & $+0.76$ & & B5 & $-2.85$ & $-1.24$ \cr
A6 & $-0.26$ & $-0.50$ & & B6 & $-0.54$ & $-0.28$ \cr
A7 & $-0.32$ & $+0.33$ & & B7 & $-1.51$ & $-1.06$ \cr
}
\tabletext{The quantities $\delta_V$ and $\delta_{\sigma}$ are the
differences between the kinematical results obtained from the HST/FOS
galaxy spectra with: (a) a ground-based template that provides a good
match to the galaxy spectra; and (b) an HST template that has
significant template mismatch. The differences are expressed in units
of the formal error bars.}
\endtable


\subsection{A.5 Consistency check}

As a consistency check on the {\it assumed} values of $\sigma_5$, we
also tried to {\it calculate} $\sigma_5$ from the actual FOS
observations. This can be done under the assumption that the LSFs of
both the galaxy and template spectrum are Gaussian, with dispersions
$\sigma_G$ and $\sigma_T$, respectively. This is not a very accurate
approximation, but provides a useful consistency check. It yields:
\eqnam\testsigma
$$ \sigma_5^2 = {\widetilde \sigma}_5^2 + \sigma_T^2 - \sigma_G^2 
      . \eqno (A8)$$
The dispersion of the template spectrum is $\sigma_T = 71 \kms$ (at
5170{\AA}), whereas $\sigma_G = 100 \pm 2 \kms$ (Keyes \etal 1995).
The dispersions ${\widetilde \sigma}_5$ measured directly from the
$\chi^2$ minimization~(A3) are $240.6 \kms$ and $171.5 \kms$, for
NGC~4342 and NGC~4570, respectively. This yields with equation~(A8)
that $\sigma_5 = 230.1\kms$ for NGC~4342, and $\sigma_5 = 156.4 \kms$
for NGC~4570, in good agreement (within the error bars of the FOS
data) with the values derived from the Jeans modeling. The reference
values assumed on the basis of modeling of the ground-based data are
therefore confirmed by the FOS data.

\subsection{A.6 Analysis with an HST template}

As yet another test, we also analyzed the FOS spectra with an actual
FOS template spectrum: the KIII star F193, observed with the same
instrumental setup as our galaxy spectra. There is definite template
mismatch between this stellar spectrum and our galaxy spectra, but
the results should nonetheless be roughly consistent with those
derived using the more appropriate ground-based template. 

In this case, the instrumental broadening function $H$ (equation~[A4])
is the same for the galaxy and template spectra. However, the template
has a different illumination function $A_T$ than the galaxy
spectra. We calculated $A_T$ assuming that the star was properly
centred in the $0.26''$ aperture. In analogy with equation~(A5), we
subsequently assumed that each $A_i$ can be approximated as a
convolution of $A_T$ with a Gaussian. For each galaxy spectrum we
found the best Gaussian, and convolved it with the template spectrum.
This results in a different template spectrum for each galaxy
spectrum, that has the same LSF as the galaxy spectrum. We refer to
the stellar kinematics determined with these templates as $V_i^{*}$
and $\sigma_i^{*}$. In Table~A1 we list the relative differences
$\delta_V \equiv (V_i^{*} - V_i)/\Delta V_i$ and $\delta_{\sigma}
\equiv (\sigma_i^{*} - \sigma_i)/\Delta \sigma_i$, between these results
and those listed in Table~5. The agreement is satisfactory: most
residuals are smaller than the error bars listed in Table 5 (i.e.,
$\delta < 1$). The results in Table~5 are the more accurate ones,
because they suffer less from template mismatch. 

The results obtained with an HST template are consistent with those
derived using a ground-based template. This confirms that there are no
large systematic errors in our analysis.

\section*{Appendix B: WHT kinematics}

The tables in this Appendix list the stellar kinematics inferred
from the WHT spectra. The quantities $V$ and $\sigma$ (in $\kms$) are
the mean and dispersion of best-fitting Gaussian VPs, and $h_3$ and
$h_4$ are the lowest-order Gauss-Hermite moments of the VPs. The
kinematics inferred from the HST/FOS spectra are listed in
Table~5.


\begintable{8}
\caption{{\bf Table~B1.} NGC~4342 major axis}
\halign{\hfil#&\quad \hfil#\quad& \hfil#\quad& \hfil#\quad& \hfil#\quad&
\hfil#\quad& \hfil#\quad& \hfil#\quad& \hfil# \cr
R & $V$ & $\Delta V$ & $\sigma$ & $\Delta\sigma$ & 
$h_3$ & $\Delta h_3$ & $h_4$ & $\Delta h_4$ \cr
-14.235&-238.5&   9.4&  94.6&  12.1&-0.111& 0.088& 0.041& 0.070\cr
-11.218&-223.6&   6.3&  78.6&   7.7&-0.040& 0.068&-0.038& 0.056\cr
 -9.463&-233.1&   7.0&  95.8&   9.6& 0.042& 0.068& 0.037& 0.054\cr
 -8.222&-231.1&   8.8& 118.0&  11.6&-0.052& 0.070& 0.035& 0.055\cr
 -7.147&-216.7&   6.5& 100.2&   9.1&-0.003& 0.061& 0.051& 0.048\cr
 -6.259&-227.8&   8.4& 109.8&  11.0& 0.093& 0.072& 0.003& 0.057\cr
 -5.542&-212.2&   8.3& 117.9&  10.6&-0.062& 0.066& 0.007& 0.052\cr
 -4.826&-204.2&   8.3& 144.1&  11.9& 0.114& 0.056& 0.042& 0.044\cr
 -4.109&-190.2&   7.3& 129.4&   8.8& 0.078& 0.050&-0.016& 0.040\cr
 -3.580&-155.8&   9.3& 126.5&  11.9& 0.059& 0.065& 0.029& 0.053\cr
 -3.222&-145.9&  10.3& 152.3&  13.8& 0.055& 0.061& 0.038& 0.050\cr
 -2.864&-152.3&   9.5& 134.9&  12.7&-0.011& 0.063& 0.046& 0.052\cr
 -2.506&-141.1&   9.5& 165.3&  11.1&-0.043& 0.049&-0.025& 0.039\cr
 -2.148&-137.6&  10.0& 180.0&  13.4&-0.049& 0.050& 0.024& 0.040\cr
 -1.790&-123.6&   9.4& 183.8&  12.8&-0.057& 0.046& 0.031& 0.036\cr
 -1.432&-114.9&  10.1& 220.5&  12.8& 0.001& 0.038& 0.010& 0.031\cr
 -1.074&-135.0&   6.9& 196.3&   9.2&-0.008& 0.031& 0.025& 0.025\cr
 -0.716&-124.6&   7.4& 243.5&   9.8& 0.071& 0.026&-0.011& 0.021\cr
 -0.358& -71.2&   7.7& 292.9&  10.6& 0.046& 0.022&-0.012& 0.018\cr
  0.000&   0.0&   7.3& 317.1&  10.6&-0.026& 0.021&-0.015& 0.016\cr
  0.358&  70.5&   7.0& 276.8&   9.5&-0.050& 0.021&-0.007& 0.017\cr
  0.716& 114.2&   7.1& 215.3&   9.5&-0.011& 0.027& 0.006& 0.022\cr
  1.074& 134.1&   7.3& 197.1&   9.3& 0.002& 0.030&-0.016& 0.025\cr
  1.432& 132.8&   8.0& 175.1&  10.9& 0.045& 0.038& 0.017& 0.032\cr
  1.790& 120.4&   8.8& 160.5&  11.4& 0.041& 0.046& 0.002& 0.038\cr
  2.148& 124.2&   8.7& 152.2&  11.1&-0.042& 0.047& 0.006& 0.040\cr
  2.506& 154.8&   8.9& 147.9&  11.1& 0.032& 0.050&-0.020& 0.042\cr
  2.864& 153.3&  10.1& 147.0&  12.2&-0.062& 0.056&-0.012& 0.047\cr
  3.222& 158.6&   9.3& 139.1&  11.2&-0.039& 0.055&-0.016& 0.045\cr
  3.580& 158.3&  10.1& 135.9&  12.5&-0.008& 0.063&-0.014& 0.052\cr
  4.109& 178.7&   6.8& 121.0&   9.1&-0.094& 0.050& 0.020& 0.041\cr
  4.826& 189.7&   6.6& 101.2&   9.2&-0.123& 0.059& 0.040& 0.047\cr
  5.542& 197.2&   7.6& 108.1&   9.3&-0.041& 0.060&-0.024& 0.048\cr
  6.429& 197.8&   7.1& 113.1&   8.1&-0.045& 0.052&-0.055& 0.042\cr
  7.505& 203.9&   6.8&  90.0&   8.0&-0.130& 0.063&-0.053& 0.051\cr
  8.749& 207.4&   6.3&  90.4&   7.9&-0.043& 0.061&-0.034& 0.049\cr
 10.344& 219.8&   6.5&  81.4&   7.9&-0.112& 0.066&-0.024& 0.055\cr
 12.614& 214.5&   7.1&  90.0&   8.8& 0.010& 0.067&-0.028& 0.055\cr
}
\endtable



\begintable{9}
\caption{{\bf Table~B2.} NGC~4342 minor axis}
\halign{\hfil#&\quad \hfil#\quad& \hfil#\quad& \hfil#\quad& \hfil#\quad&
\hfil#\quad& \hfil#\quad& \hfil#\quad& \hfil# \cr
R & $V$ & $\Delta V$ & $\sigma$ & $\Delta\sigma$ & 
$h_3$ & $\Delta h_3$ & $h_4$ & $\Delta h_4$ \cr
 -2.661&  15.2&   9.9& 107.4&  13.3& 0.138& 0.079& 0.044& 0.064\cr
 -2.148&  -3.7&  15.1& 190.4&  25.5&-0.092& 0.068& 0.133& 0.056\cr
 -1.790&  -6.2&  12.3& 200.4&  16.4&-0.035& 0.052& 0.019& 0.043\cr
 -1.432&  16.8&   9.1& 183.8&  13.0&-0.027& 0.044& 0.028& 0.036\cr
 -1.074&   5.0&   9.3& 219.5&  10.9& 0.036& 0.033&-0.031& 0.028\cr
 -0.716&  -5.8&   7.1& 240.9&   9.6& 0.017& 0.027&-0.034& 0.022\cr
 -0.358&  -0.8&   6.9& 299.0&  10.1& 0.048& 0.021&-0.008& 0.016\cr
  0.000&   0.0&   6.2& 293.3&   9.5& 0.056& 0.020&-0.012& 0.016\cr
  0.358&  -3.7&   6.6& 265.5&   9.5& 0.054& 0.022& 0.009& 0.018\cr
  0.716& -10.7&   6.9& 239.4&  10.1& 0.047& 0.025& 0.028& 0.021\cr
  1.074&  -1.7&   7.1& 216.7&   9.9& 0.013& 0.029& 0.007& 0.024\cr
  1.432&   6.4&   8.6& 191.2&  11.0&-0.048& 0.037& 0.006& 0.031\cr
  1.790&   9.0&   9.1& 156.1&  12.0&-0.051& 0.049& 0.025& 0.041\cr
  2.148&  20.7&   9.8& 138.5&  12.3&-0.059& 0.060&-0.013& 0.050\cr
  2.658&  37.9&   8.4& 114.3&  10.8&-0.007& 0.065&-0.011& 0.052\cr
}
\endtable



\begintable{10}
\caption{{\bf Table~B3.} NGC~4570 major axis}
\halign{\hfil#&\quad \hfil#\quad& \hfil#\quad& \hfil#\quad& \hfil#\quad&
\hfil#\quad& \hfil#\quad& \hfil#\quad& \hfil# \cr
R & $V$ & $\Delta V$ & $\sigma$ & $\Delta\sigma$ & 
$h_3$ & $\Delta h_3$ & $h_4$ & $\Delta h_4$ \cr
-33.516&-161.3&   4.4&  71.4&   6.2& 0.051& 0.051& 0.037& 0.044\cr
-26.363&-157.4&   4.7&  86.6&   6.6& 0.061& 0.048& 0.001& 0.039\cr
-21.951&-165.5&   4.3&  84.3&   6.6&-0.065& 0.048&-0.003& 0.039\cr
-18.749&-158.6&   4.5&  88.2&   5.9& 0.093& 0.043&-0.024& 0.034\cr
-16.259&-148.2&   5.0&  99.2&   6.7& 0.133& 0.044&-0.016& 0.035\cr
-14.308&-127.7&   6.1& 113.8&   7.9& 0.025& 0.045&-0.006& 0.036\cr
-12.518&-118.5&   5.5& 108.1&   6.6&-0.012& 0.042&-0.045& 0.033\cr
-10.907&-114.0&   5.1&  95.1&   7.1& 0.029& 0.047& 0.003& 0.038\cr
 -9.472& -96.3&   5.4& 118.7&   7.2& 0.014& 0.039&-0.016& 0.032\cr
 -8.222& -83.5&   5.5& 113.9&   7.1&-0.066& 0.041&-0.025& 0.034\cr
 -7.144& -91.2&   5.3& 118.2&   8.2& 0.029& 0.041& 0.037& 0.034\cr
 -6.258& -90.6&   6.1& 112.9&   8.2&-0.030& 0.047&-0.017& 0.038\cr
 -5.542& -96.7&   5.7& 131.7&   7.2& 0.081& 0.036&-0.036& 0.030\cr
 -4.825& -95.4&   5.2& 130.8&   6.5& 0.014& 0.033&-0.038& 0.028\cr
 -4.296& -81.6&   7.2& 142.8&   9.4&-0.028& 0.042&-0.036& 0.035\cr
 -3.938& -85.8&   7.0& 149.1&   9.4& 0.010& 0.038&-0.012& 0.032\cr
 -3.580& -88.5&   6.7& 149.1&   9.0& 0.005& 0.037&-0.019& 0.031\cr
 -3.222& -86.3&   6.5& 141.2&   9.0&-0.018& 0.039& 0.009& 0.033\cr
 -2.864& -74.9&   6.4& 147.5&   8.4&-0.090& 0.036&-0.033& 0.030\cr
 -2.506& -75.1&   5.8& 157.1&   7.4&-0.032& 0.030&-0.041& 0.025\cr
 -2.148& -66.1&   5.5& 158.6&   7.4& 0.017& 0.028&-0.016& 0.024\cr
 -1.790& -64.6&   5.1& 161.8&   7.0& 0.018& 0.026&-0.014& 0.022\cr
 -1.432& -59.2&   4.6& 167.8&   6.7& 0.018& 0.023&-0.007& 0.019\cr
 -1.074& -43.2&   4.0& 161.9&   5.6&-0.003& 0.021&-0.006& 0.017\cr
 -0.716& -33.6&   4.1& 178.4&   5.8& 0.003& 0.019&-0.011& 0.015\cr
 -0.358& -15.7&   4.0& 193.0&   5.9&-0.001& 0.017&-0.012& 0.014\cr
  0.000&   0.0&   3.9& 197.8&   5.8&-0.019& 0.016&-0.018& 0.013\cr
  0.358&   8.6&   4.0& 197.6&   5.8&-0.018& 0.017&-0.030& 0.014\cr
  0.716&  32.3&   4.1& 189.3&   6.0&-0.039& 0.018&-0.018& 0.015\cr
  1.074&  50.9&   4.5& 186.9&   6.2&-0.047& 0.020&-0.020& 0.016\cr
  1.432&  62.6&   4.8& 180.3&   6.9&-0.069& 0.022&-0.006& 0.018\cr
  1.790&  63.8&   5.2& 171.8&   7.2&-0.014& 0.025&-0.002& 0.020\cr
  2.148&  68.1&   5.6& 164.2&   8.0&-0.045& 0.029&-0.007& 0.024\cr
  2.506&  77.8&   6.2& 159.6&   8.4&-0.061& 0.032&-0.015& 0.026\cr
  2.864&  90.7&   5.8& 152.1&   7.8&-0.039& 0.032&-0.012& 0.027\cr
  3.222& 100.0&   6.5& 147.4&   8.3&-0.049& 0.036&-0.012& 0.030\cr
  3.580&  91.3&   7.5& 164.0&  10.2&-0.024& 0.037&-0.009& 0.031\cr
  3.938& 101.5&   7.8& 161.2&  10.3& 0.024& 0.039&-0.018& 0.033\cr
  4.296& 101.4&   7.7& 153.6&   9.7& 0.006& 0.040&-0.041& 0.034\cr
  4.825& 100.0&   5.6& 154.5&   7.1& 0.019& 0.029&-0.039& 0.024\cr
  5.541&  82.6&   6.0& 135.8&   7.9& 0.040& 0.037&-0.010& 0.031\cr
  6.258&  88.4&   6.2& 133.5&   8.1&-0.037& 0.038&-0.005& 0.032\cr
  7.145&  97.4&   5.2& 115.9&   6.7&-0.013& 0.039&-0.024& 0.032\cr
  8.223&  85.1&   6.0& 117.6&   8.8&-0.019& 0.046& 0.027& 0.037\cr
  9.470&  80.6&   5.4& 122.7&   7.4&-0.005& 0.038&-0.002& 0.031\cr
 10.909& 104.2&   5.3&  95.0&   7.3& 0.002& 0.048& 0.017& 0.038\cr
 12.517& 119.8&   4.5&  98.7&   5.2& 0.000& 0.037&-0.069& 0.029\cr
 14.303& 128.3&   4.9& 102.3&   5.7&-0.057& 0.039&-0.046& 0.031\cr
 16.266& 152.4&   4.9&  86.5&   5.5&-0.072& 0.046&-0.091& 0.037\cr
 18.755& 148.4&   5.2&  90.6&   6.2&-0.167& 0.047&-0.045& 0.038\cr
 21.948& 152.1&   5.2&  88.1&   5.7&-0.050& 0.046&-0.099& 0.037\cr
 26.375& 176.4&   4.3&  83.1&   7.1&-0.095& 0.046& 0.064& 0.038\cr
 33.529& 161.8&   4.6&  83.4&   5.5& 0.034& 0.045&-0.078& 0.037\cr
}
\endtable



\begintable{11}
\caption{{\bf Table~B4.} NGC~4570 offset axis}
\halign{\hfil#&\quad \hfil#\quad& \hfil#\quad& \hfil#\quad& \hfil#\quad&
\hfil#\quad& \hfil#\quad& \hfil#\quad& \hfil# \cr
R & $V$ & $\Delta V$ & $\sigma$ & $\Delta\sigma$ & 
$h_3$ & $\Delta h_3$ & $h_4$ & $\Delta h_4$ \cr
-32.931&-182.6&   4.6&  68.5&   5.5&-0.046& 0.046& 0.022& 0.040\cr
-25.097&-155.7&   5.0&  96.9&   6.9& 0.051& 0.043& 0.034& 0.035\cr
-20.515&-147.1&   4.3&  90.5&   5.8& 0.142& 0.041& 0.002& 0.033\cr
-17.318&-145.0&   4.3&  99.9&   6.5&-0.054& 0.040& 0.027& 0.032\cr
-14.835&-128.6&   4.6& 107.3&   6.1&-0.006& 0.039&-0.025& 0.031\cr
-12.874&-121.6&   5.2& 111.3&   6.0&-0.051& 0.039&-0.060& 0.031\cr
-11.080& -92.7&   5.1& 106.5&   6.0&-0.162& 0.039&-0.038& 0.031\cr
 -9.471& -92.7&   5.0& 120.9&   6.8&-0.055& 0.038&-0.020& 0.031\cr
 -8.222& -71.1&   5.7& 107.2&   6.8&-0.104& 0.045&-0.037& 0.036\cr
 -7.146& -73.8&   5.1& 102.8&   6.8&-0.072& 0.043&-0.006& 0.034\cr
 -6.259& -90.8&   6.6& 127.4&   7.6&-0.012& 0.041&-0.050& 0.034\cr
 -5.542& -85.9&   5.9& 138.3&   7.6& 0.030& 0.036&-0.051& 0.030\cr
 -4.825& -76.3&   5.3& 120.6&   6.8& 0.023& 0.037&-0.013& 0.030\cr
 -4.109& -78.4&   4.8& 132.3&   6.4& 0.000& 0.031&-0.010& 0.026\cr
 -3.580& -71.8&   6.8& 135.6&   8.9& 0.063& 0.042&-0.017& 0.035\cr
 -3.222& -60.1&   7.0& 147.2&   8.4& 0.007& 0.038&-0.054& 0.032\cr
 -2.864& -62.4&   6.6& 147.3&   8.4&-0.022& 0.037&-0.036& 0.031\cr
 -2.506& -60.5&   6.5& 161.9&   8.9&-0.023& 0.033& 0.000& 0.027\cr
 -2.148& -58.5&   6.5& 163.3&   9.1&-0.038& 0.033&-0.007& 0.027\cr
 -1.790& -47.3&   5.7& 149.8&   8.0&-0.026& 0.033&-0.012& 0.028\cr
 -1.432& -40.1&   5.8& 159.6&   8.3&-0.025& 0.031&-0.003& 0.026\cr
 -1.074& -30.7&   5.5& 165.8&   7.8& 0.013& 0.028&-0.006& 0.023\cr
 -0.716& -23.3&   5.4& 175.2&   7.5& 0.002& 0.025&-0.012& 0.021\cr
 -0.358&  -5.7&   5.4& 176.9&   7.9&-0.015& 0.026& 0.000& 0.021\cr
  0.000&   0.0&   5.4& 172.2&   8.1&-0.033& 0.026& 0.019& 0.022\cr
  0.358&  16.0&   5.0& 165.9&   7.6&-0.047& 0.026& 0.013& 0.022\cr
  0.716&  26.2&   5.4& 175.7&   7.9&-0.048& 0.026& 0.001& 0.021\cr
  1.074&  34.9&   5.6& 174.3&   8.2&-0.028& 0.027& 0.001& 0.022\cr
  1.432&  45.9&   5.9& 173.5&   8.3&-0.039& 0.028&-0.005& 0.023\cr
  1.790&  62.2&   6.0& 163.4&   8.3&-0.048& 0.030&-0.003& 0.025\cr
  2.148&  70.7&   6.0& 154.0&   8.4&-0.028& 0.033&-0.005& 0.028\cr
  2.506&  72.6&   6.0& 145.8&   7.9&-0.029& 0.034&-0.022& 0.029\cr
  2.864&  78.4&   6.6& 148.3&   8.5& 0.001& 0.037&-0.036& 0.031\cr
  3.222&  76.1&   6.5& 141.4&   8.7& 0.012& 0.039&-0.015& 0.033\cr
  3.580&  81.7&   6.9& 143.7&   9.1&-0.019& 0.040&-0.013& 0.033\cr
  4.109&  89.0&   5.1& 146.6&   7.4&-0.068& 0.031& 0.013& 0.026\cr
  4.826&  94.5&   5.5& 136.4&   7.1&-0.055& 0.034&-0.013& 0.028\cr
  5.542&  92.6&   5.6& 127.1&   7.3& 0.031& 0.038&-0.013& 0.031\cr
  6.259&  94.5&   6.5& 129.3&   8.7&-0.018& 0.042& 0.010& 0.035\cr
  7.146&  99.7&   5.6& 126.9&   7.4&-0.074& 0.037& 0.003& 0.031\cr
  8.223& 108.9&   6.2& 121.3&   7.9&-0.005& 0.045&-0.040& 0.037\cr
  9.472&  97.2&   5.0&  96.2&   7.2& 0.044& 0.046& 0.030& 0.037\cr
 11.080& 126.1&   5.6& 106.1&   6.7& 0.091& 0.045&-0.055& 0.036\cr
 12.870& 137.5&   5.3&  99.6&   6.9& 0.022& 0.046&-0.022& 0.037\cr
 14.841& 144.7&   5.1& 115.5&   6.1&-0.001& 0.038&-0.067& 0.030\cr
 17.316& 158.0&   4.7& 100.6&   5.7&-0.061& 0.038&-0.031& 0.031\cr
 20.516& 176.3&   4.2&  87.7&   5.8&-0.091& 0.041&-0.008& 0.033\cr
 25.086& 176.7&   3.4&  73.9&   5.1&-0.136& 0.043&-0.020& 0.036\cr
 32.870& 179.2&   3.6&  70.0&   4.7&-0.082& 0.043&-0.035& 0.037\cr
}
\endtable



\begintable{12}
\caption{{\bf Table~B5.} NGC~4570 minor axis}
\halign{\hfil#&\quad \hfil#\quad& \hfil#\quad& \hfil#\quad& \hfil#\quad&
\hfil#\quad& \hfil#\quad& \hfil#\quad& \hfil# \cr
R & $V$ & $\Delta V$ & $\sigma$ & $\Delta\sigma$ & 
$h_3$ & $\Delta h_3$ & $h_4$ & $\Delta h_4$ \cr
 -5.430&   8.4&   6.5& 125.1&   8.0&-0.010& 0.043&-0.031& 0.035\cr
 -3.903&  18.4&   6.7& 121.5&   8.9& 0.025& 0.048&-0.010& 0.039\cr
 -3.027&  10.5&   7.1& 138.6&   9.4&-0.066& 0.043&-0.020& 0.036\cr
 -2.308&  -2.1&   6.5& 156.6&   9.1&-0.022& 0.034& 0.007& 0.029\cr
 -1.790&  -1.8&   7.8& 170.5&  11.1&-0.043& 0.037&-0.002& 0.031\cr
 -1.432&  -1.3&   7.3& 185.9&  10.9& 0.003& 0.032& 0.006& 0.027\cr
 -1.074& -10.7&   6.6& 201.8&  10.5& 0.006& 0.027& 0.015& 0.022\cr
 -0.716&  -5.5&   5.8& 189.5&   8.8& 0.018& 0.025& 0.003& 0.021\cr
 -0.358&  -4.7&   5.6& 184.3&   8.3&-0.016& 0.025&-0.004& 0.021\cr
  0.000&   0.0&   5.7& 184.0&   8.4&-0.039& 0.025&-0.004& 0.021\cr
  0.358&   3.1&   5.8& 190.6&   8.7& 0.003& 0.025&-0.005& 0.021\cr
  0.716&  -0.2&   5.8& 187.0&   8.1& 0.032& 0.025&-0.020& 0.021\cr
  1.074&  -3.5&   6.3& 176.6&   9.1& 0.019& 0.030&-0.003& 0.025\cr
  1.432&  -1.8&   7.1& 183.6&  11.7& 0.001& 0.033& 0.029& 0.027\cr
  1.790&   3.5&   8.1& 178.4&  14.0&-0.041& 0.039& 0.043& 0.032\cr
  2.148&  13.6&   8.0& 136.0&  11.1& 0.006& 0.053&-0.017& 0.044\cr
  2.668&  12.0&   6.5& 149.0&   9.6& 0.004& 0.037& 0.014& 0.032\cr
  3.543&  18.5&   6.5& 138.3&   9.2&-0.066& 0.040& 0.009& 0.034\cr
  4.922&  22.2&   6.2& 116.2&   8.6&-0.062& 0.048&-0.003& 0.039\cr
  8.059&   9.6&   6.6& 106.5&   9.7&-0.019& 0.054& 0.043& 0.043\cr
}
\endtable



\section*{References}

\beginrefs

\bibitem Arimoto N., Yoshii Y., 1987, A\&A, 173, 23

\bibitem Bender R., D\"obereiner S., M\"ollenhoff C., 1988, A\&AS, 74, 385

\bibitem Burrows C.J. et al., 1995, Hubble Space Telescope Wide Field 
	 and Planetary Camera 2 Instrument Handbook, Version 3.0. 
         STScI, Baltimore

\bibitem Burstein D., 1979, ApJ, 234, 435

\bibitem Burstein D., Faber S.M., Gaskell C.M., Krumm N., 1984, ApJ, 287, 586

\bibitem Carlberg R.C., 1984, ApJ, 286, 403

\bibitem Carollo C.M., Danziger I.J., Buson L., 1993, MNRAS, 265, 553

\bibitem Cretton N., van den Bosch F.C., 1997, in preparation 

\bibitem de Vaucouleurs G., de Vaucouleurs A., Corwin H.C., 1976, 
	 Second Reference Catalogue of Bright Galaxies. University of Texas, 
         Austin (RC2)

\bibitem Dressler A., 1984, ApJ, 286, 97

\bibitem Efstathiou G., Gorgas J., 1985, MNRAS, 215, 37p
 
\bibitem Emsellem E., Monnet G., Bacon R., 1994, A\&A, 285, 723

\bibitem Evans I.N., 1995, FOS Instrument Science Report CAL/FOS-140, 
	 `Post-COSTAR FOS small aperture relative throughputs derived from 
	 SMOV data'. STScI, Baltimore

\bibitem Faber S.M., 1977, in Tinsley B.T., Larson R.B., eds., 
         The Evolution of Galaxies and Stellar Populations,.
	 Yale University Press, New Haven, p.~157

\bibitem Faber S.M., Friel E., Burstein D., Gaskell C.M., 1985, ApJS,
	 57, 711

\bibitem Fisher D., Franx M., Illingworth G.D., 1996, ApJ, 459, 110

\bibitem Fisher D., 1997, AJ, 113, 950

\bibitem Forbes D.A., 1994, AJ, 107, 2017

\bibitem Gebhardt K. et al., 1996, AJ, 112, 105

\bibitem Gonz\'alez J.J., 1993, PhD Thesis. University of California, 
	 Santa Cruz
 
\bibitem Holtzman J.A. et al., 1995a, PASP, 107, 156

\bibitem Holtzman J.A. et al., 1995b, PASP, 107, 1065

\bibitem Jacoby G., Ciardullo R., Ford H.C., 1990, ApJ, 356, 332

\bibitem Jaffe W., Ford H.C., Ferrarese L., van den Bosch F.C., O'Connell R.W.,
	 1994, AJ, 108, 1567

\bibitem Jedrzejewski R.I., 1987, MNRAS, 226, 747

\bibitem Keyes T. et al., 1995, Hubble Space Telescope Faint Object 
         Spectrograph Instrument Handbook, Version 6.0. STScI, Baltimore

\bibitem Knapp G.R, Guhathakurta P., Kim D.-W., Jura M., 1989, ApJS, 70, 329

\bibitem Kormendy J., 1988, ApJ, 335, 40
 
\bibitem Kormendy J., Richstone D.O., 1992, ApJ, 393, 559

\bibitem Kuijken K., Merrifield M.R., 1993, MNRAS, 264, 712

\bibitem Larson R.B., 1974, MNRAS, 166, 385

\bibitem Lauer T.R., 1985, MNRAS, 216, 429

\bibitem Lauer T.R. et al., 1995, AJ, 110, 2622 

\bibitem Lucy L.B., 1974, AJ, 74, 745

\bibitem Matteucci F., Tornamb\`e A., 1987, A\&A, 185, 51

\bibitem Michard R., 1996, A\&AS, 117, 583

\bibitem Nieto J.-L., Bender R., Arnaud J., Surma P., 1991, A\&A, 244, L25

\bibitem Peletier R.F., 1989, PhD Thesis. University of Groningen

\bibitem Rix H.-W., White S.D.M., 1992, MNRAS, 254, 389

\bibitem Sargent W.L.W., Young P.J., Boksenberg A., Shortridge K., Lynds C.R., 
	 Hartwick F.D.A., 1978, ApJ, 221, 731

\bibitem Scorza C., Bender R., 1990, A\&A, 235, 49

\bibitem Scorza C., Bender R., 1995, A\&A, 293, 20

\bibitem Scorza C., van den Bosch F.C., 1997, in preparation

\bibitem Simien F., Michard R., Prugniel P., 1993, in
         Danziger I.J., Zeilinger W.W., Kj\"ar K., eds., 
         Structure, Dynamics and Chemical Evolution of Elliptical Galaxies.
         ESO Conference and Workshop Proceedings, Garching bei M\"unchen,
         p.~211

\bibitem Tonry J.L., Davis M., 1979, AJ, 84, 1511

\bibitem Vader J.P., Vigroux L., Lachi\`eze-Rey M., Souviron J., 1988,
	 A\&A, 203, 217

\bibitem van den Bosch F.C., Ferrarese L., Jaffe W., Ford H.C., O'Connell R.W.,
         1994, AJ, 108, 1579

\bibitem van den Bosch F.C., de Zeeuw, P.T., 1996, MNRAS, 283, 381

\bibitem van den Bosch F.C., Jaffe W., 1997, in Arnaboldi M., Da Costa G.S.,
	 Saha P., eds., The Nature of Elliptical Galaxies, A.S.P Conference
 	 Series Volume 116, p. 142

\bibitem van den Bosch F.C., Emsellem E., 1997, MNRAS, submitted

\bibitem van der Marel R.P., 1991, MNRAS, 253, 710

\bibitem van der Marel R.P., 1994, MNRAS, 270, 271

\bibitem van der Marel R.P., Franx M., 1993, ApJ, 407, 525

\bibitem van der Marel R.P., Rix H.-W., Carter D., Franx M., White S.D.M.,
	 de Zeeuw P.T., 1994, MNRAS, 268, 521

\bibitem van der Marel R.P., de Zeeuw P.T., Rix H.-W., 1997, ApJ, in press

\bibitem Worthey G., 1994, ApJS, 95, 107

\bibitem Worthey G., Faber S.M., Gonz\'alez J.J., Burstein D., 1994, ApJS, 94,
         687

\bibitem Wrobel J.M., 1991, AJ, 101, 127

\endrefs
\bye